\documentclass[11pt]{article}

\usepackage{graphicx}
\usepackage{amsmath}
\usepackage{amssymb}
\usepackage{hyperref}
\usepackage{cite}
\usepackage{times}
\usepackage{fullpage}
\usepackage{ifthen}
\usepackage{deluxetable}
\usepackage{color}

\newcommand{\spacing}[1]{\renewcommand{\baselinestretch}{#1}\large\normalsize}
\spacing{1}

\def\@maketitle{%
  \newpage\spacing{1}\setlength{\parskip}{12pt}%
    {\Large\bfseries\noindent\sloppy \textsf{\@title} \par}%
    {\noindent\sloppy \@author}%
}

\newcommand{\simgt}{\lower.5ex\hbox{$\; \buildrel > \over \sim \;$}}
\newcommand{\simlt}{\lower.5ex\hbox{$\; \buildrel < \over \sim \;$}}

\def\aap{Astron.\ Astrophys.}
\def\apj{Astrophys.\ J.}

\def\apjl{Astrophys.\ J.\ Lett.}
\def\apjs{Astrophys.\ J.\ Suppl.\ S.}
\def\mnras{Mon.\ Not.\ R.\ Astron.\ Soc.}

\def\pasj{Pub.\ Astron.\ Soc.\ Jap.}

\def\physrep{Phys.\ Rep.}
\def\prd{Phys.\ Rev.\ D}

\def\aaps{Astron.\ Astrophys.\ Supp.}

\def\procspie{Proceedings of the SPIE}
\def\sovast{Sov.\ Astron.}
\def \prl{Phys.\ Rev.\ Lett.}
\def \prd{Phys.\ Rev.\ D}

\newcommand{\dl}{d}
\newcommand{\ds}{d_{\rm s}}
\newcommand{\drm}{{\rm d}}
\newcommand{\vv}{\mathbf{v}}

\newcommand{\RNum}[1]{\uppercase\expandafter{\romannumeral #1\relax}}

\title{Microlensing constraints on
primordial black holes with the Subaru/HSC Andromeda observation}

\date{}

\begin{document}

\maketitle

\newenvironment{affiliations}{%
    \setcounter{enumi}{1}%
    \setlength{\parindent}{0in}%
    \slshape\sloppy%
    \begin{list}{\upshape$^{\arabic{enumi}}$}{%
        \usecounter{enumi}%
        \setlength{\leftmargin}{0in}%
        \setlength{\topsep}{0in}%
        \setlength{\labelsep}{0in}%
        \setlength{\labelwidth}{0in}%
        \setlength{\listparindent}{0in}%
        \setlength{\itemsep}{0ex}%
        \setlength{\parsep}{0in}%
        }
    }{\end{list}\par\vspace{12pt}}

\renewenvironment{abstract}{%
    \setlength{\parindent}{0in}%
    \setlength{\parskip}{0in}%
    \bfseries%
    }{\par\vspace{0pt}}


{\bf
{\noindent \author{Hiroko~Niikura$^{1,2}$,
Masahiro~Takada$^{1}$,
Naoki~Yasuda$^{1}$,
Robert~H.~Lupton$^{3}$,
Takahiro~Sumi$^{4}$,
Surhud~More$^{1}$,
Toshiki~Kurita$^{1,2}$,
Sunao~Sugiyama$^{1,2}$,
Anupreeta~More$^{1}$,
Masamune~Oguri$^{1,2,5}$, 
Masashi~Chiba$^{6}$}}
}


\begin{affiliations}
\item {Kavli Institute for the Physics and Mathematics of the
Universe (WPI),
The University of Tokyo Institutes for Advanced Study (UTIAS),
The
University of Tokyo, Chiba, 277-8583, Japan}
\item {Physics Department,
The University of Tokyo, Bunkyo, Tokyo 113-0031, Japan}
\item {Department
of Astrophysical Sciences, Princeton University, Peyton Hall, Princeton
NJ 08544 USA}
\item {Department of Earth and Space Science, Graduate
School of Science, Osaka University, Toyonaka, Osaka 560-0043, Japan}
\item {Research Center for the Early Universe, University of Tokyo, Tokyo 113-0033, Japan}
\item {Astronomical Institute, Tohoku University, Aoba-ku, Sendai
980-8578, Japan}
\end{affiliations}

\begin{abstract}
Primordial black holes (PBHs) have long been suggested as a viable candidate for the elusive dark matter (DM).
The abundance of such PBHs has been constrained using a number of astrophysical observations, except for a hitherto unexplored mass window of 
$M_{\rm PBH}=[10^{-14},10^{-9}]M_\odot$. 
Here we carry out a dense-cadence (2~min sampling rate), 7 hour-long observation of
the Andromeda galaxy (M31) with the Subaru Hyper Suprime-Cam to search for
microlensing of stars in M31 by PBHs lying in the halo regions of the Milky Way
(MW) and M31.  Given our simultaneous monitoring of more than tens of millions
of stars in M31, if such light PBHs 
make up a significant fraction of DM, we expect
to find many microlensing events for the PBH DM scenario. However, we 
identify only a single candidate event, which translates into the most
stringent upper bounds on the abundance of PBHs in the mass range $M_{\rm
PBH}\simeq [10^{-11}, 10^{-6}]M_\odot$. 
\end{abstract}

The nature of dark matter (DM) remains 
one of the most important problems in
physics. Previous studies have suggested that DM is non-baryonic,
non-relativistic, and interacts with ordinary matter only via gravity
\cite{Davisetal:85,Cloweetal:06,DodelsonLiguori:06}.  Currently, unknown stable
particle(s) beyond the Standard Model of Particle Physics, such as Weakly
Interacting Massive Particles (WIMPs), are considered to be viable candidates
\cite{Jungmanetal:96}. However such particles have so far evaded detection in
either elastic scattering experiments, indirect experiments or collider
experiments \cite{Klasenetal:15}. Primordial black holes (PBH),  which can be
formed during the early universe, are also viable candidates for the elusive DM
\cite{Zel'dovichNovikov:67,Hawking:71,CarrHawking:74}.

\begin{figure}[h]
\centering
\includegraphics[width=0.7\textwidth]{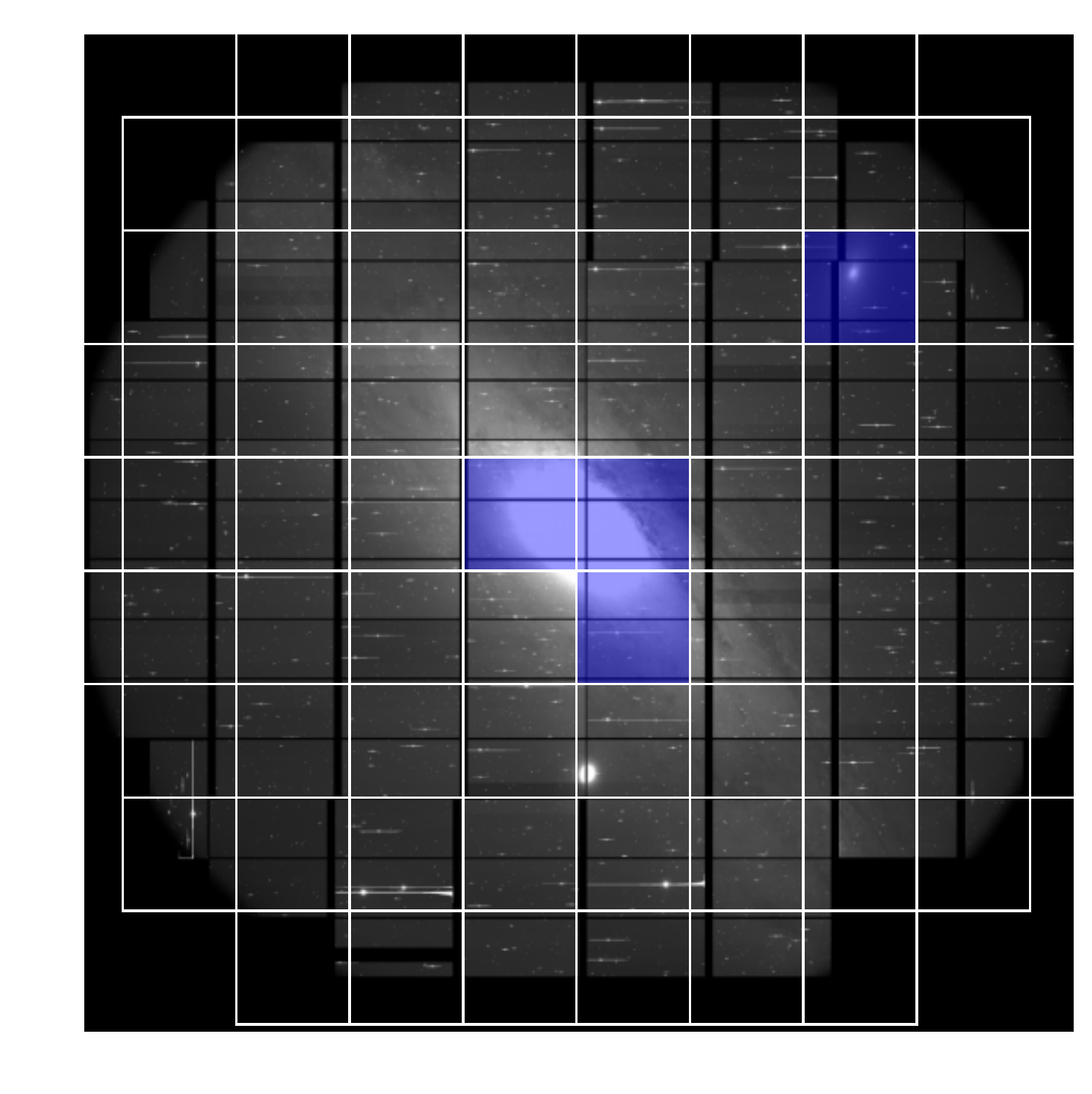}
\caption{The background shows the HSC image of M31 as seen by the 104 CCD chips
of the Subaru/HSC camera. The white-colored grid represents a predefined
iso-latitude tessellation grid, called the HSC ``patch'' (approximately 12 arcmin
on a side).  Our data analysis including the image subtraction is performed on
individual patches. We exclude those patches which are marked in dark-blue color from our
analysis as the dense star fields in these patches result in a saturation of
the CCDs.
\label{fig:m31_image}}
\end{figure}
\begin{figure}[h]
 \centering
 \includegraphics[width=0.7\textwidth]{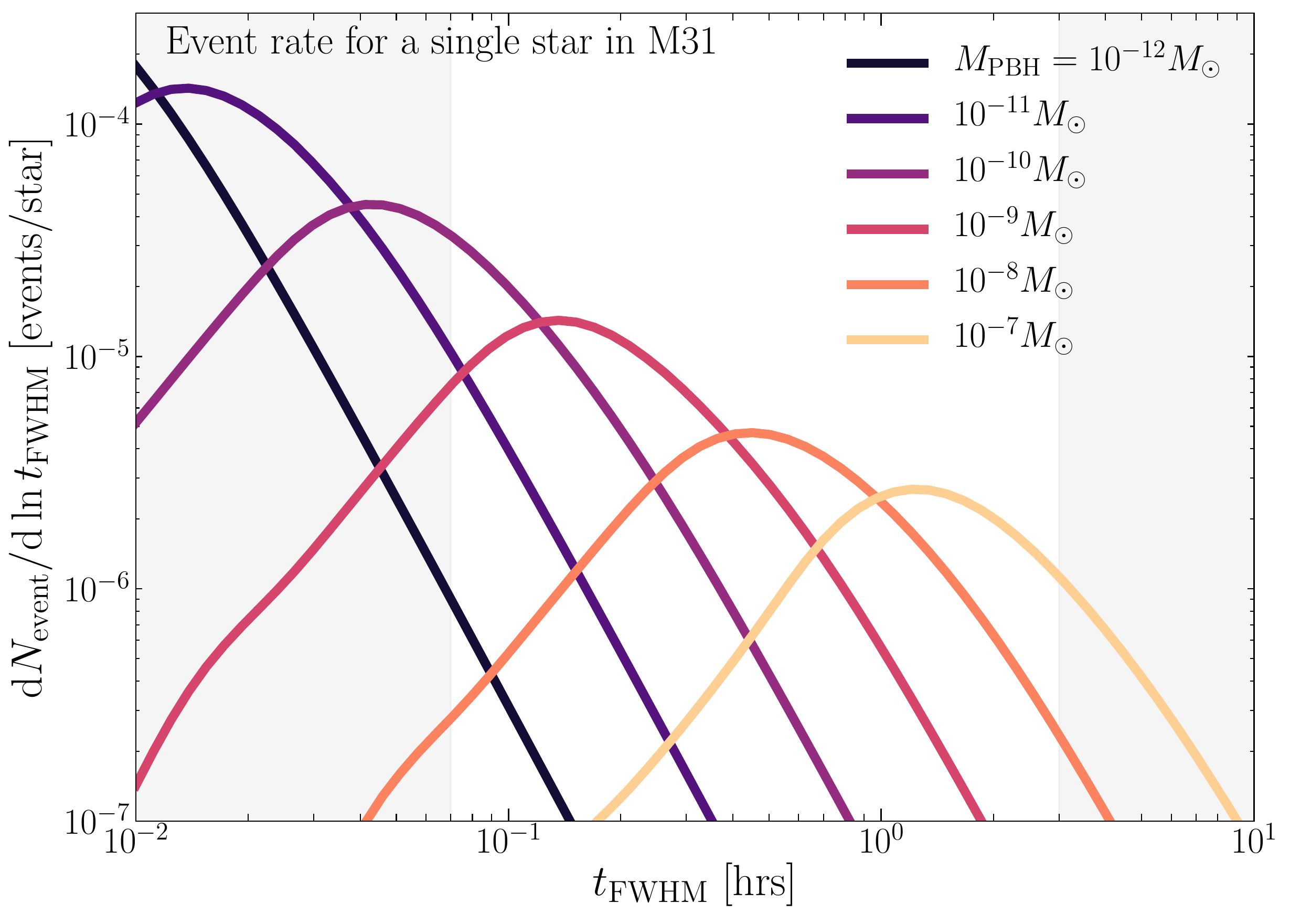}
 \caption{The expected differential number of PBH microlensing events per logarithmic
interval of the full-width-at-half-maximum (FWHM) microlensing timescale ${t_{\rm FWHM}}$,
for a {\it single} star in M31. Each solid line corresponds to a monochromatic
PBH DM scenario and assumes that all the dark matter consists of such PBHs.
We adopt DM halo models for the MW and M31 halos which reproduce their
individual rotation curves.  The event rate calculation includes distributions
of impact parameters and velocities of PBHs relative to a source star. Given
the cadence, our data has the highest sensitivity to measure lightcurves with
$t_{\rm FWHM}\simeq [0.07,3]~{\rm hours}$ shown by the unshaded regions. 
 \label{fig:eventrate}}
\end{figure}
The abundance of PBHs of different mass scales is already constrained by
various observations except for a mass window of $M_{\rm PBH}\simeq
[10^{19},10^{24}]$g or equivalently $[10^{-14},10^{-9}]M_\odot$
\cite{Carretal:16}. 
The existing constraints based on
the capture of neutron stars and white dwarfs \cite{Capelaetal:13b} in this
mass regime are based on uncertain assumptions about the presence of DM in a
globular cluster \cite{Laneetal:09}. 
Thus it is of critical importance to
further explore observational constraints on the PBH abundance for this mass
window.

Gravitational microlensing is a powerful method to probe DM in the Milky Way
(MW) \cite{Paczynski:86, Griestetal:91}.
Microlensing causes a time-varying magnification of a background star when a
lensing object crosses the line-of-sight to the star at close proximity.
The microlensing experiments, MACHO \cite{Alcocketal:00} and EROS
\cite{EROS:07}, have previously monitored large number of stars in the Large Magellanic
Cloud (LMC) with roughly a 24 hour cadence. They have ruled out massive compact
halo objects (MACHOs) such as brown dwarfs with mass scales $[10^{-7},
10]M_\odot$ as DM candidates.
We also note that if PBHs with mass around $10M_\odot$ comprise even 1\% of the DM and form binaries, 
then their merger rate could be larger than the LIGO event rate
\cite{Sasakietal:16,Ali-Haimoudetal:17}.
Microlensing searches on time scales of 15 or 30 minutes have also been carried out
using the public 2-year Kepler data to constrain the abundance of $10^{-8}
M_\odot$ PBHs \cite{Griestetal:14}.
With the aim of constraining the abundance of PBH on even smaller mass scales,
we carried out a dense cadence observation of the Andromeda galaxy (M31),
with the Subaru Hyper Suprime-Cam (HSC). We search for microlensing event(s) of
M31 stars by intervening PBHs in both the halo regions of MW and M31. M31 is the
MW's largest neighboring spiral galaxy, at a distance of 770~kpc (the distance
modulus $\mu\simeq 24.4~$mag). Even a single night of HSC/Subaru yields an
ideal dataset to search for the PBH microlensing events for the following reasons. First, the 1.5 degree
diameter field-of-view of HSC \cite{2018PASJ...70S...1M}
allows us to cover the
entire region of M31 (the bulge, disk and halo regions) with a single pointing,
as shown in Fig.~\ref{fig:m31_image}.  Secondly, the 8.2m large aperture and
its superb image quality (typically $0.6^{\prime\prime}$
seeing)\cite{Aiharaetal:17} allow us to detect fluxes from M31 stars down to
$m_r\simeq 26$ even with a short exposure of 90~sec.  These two facts allow us to
simultaneously monitor a sufficiently large number of stars in M31. Thirdly,
the 90~sec exposure and a short camera readout of $\sim$35~sec enable us to
take data at an unprecedented cadence of 2~min. Thus, 
we can search for microlensing events with PBH mass scales smaller than those probed by
Ref.~\cite{Griestetal:14}. Finally, the huge volume between M31 and the
Earth, leads to a large optical depth of PBH microlensing to each star in M31,
which allows us to put meaningful constraints on the PBH DM scenario.

In Fig.~\ref{fig:eventrate} we show the differential number of PBH microlensing
events for a single star in M31 per logarithmic timescale, for our 7 hour-long
HSC
observation, assuming that PBHs of a single mass scale make up all DM in
the halo regions of MW and M31. Here we consider a monochromatic PBH mass-scale for 
illustrative purposes. However, our limits will apply to a general scenario of
PBH DM with an arbitrary mass spectrum.
To model the DM distribution, we adopt the halo model for the MW and M31 from
Ref.~\cite{Klypinetal:02}, with model parameters constrained by the observed galaxy
rotation curves. We assume $M_{\rm vir}=10^{12}M_\odot$ and $1.6\times
10^{12}M_\odot$ as the virial mass of the MW and the M31 halo, respectively. To
be conservative, we ignore any further DM contribution arising in the
intervening space between MW and M31, e.g. due to a possible filamentary
structure between the two galaxies.  Once the halo model parameters, the
distance to the lensing PBH, the impact parameter and the tangential velocity
relative to the source star on the sky, are specified, we can predict the
microlensing light curve. Different combinations of the model parameters
could produce a similar timescale for the light curve. The expected event
number in Fig.~\ref{fig:eventrate} takes into account variations of the
parameters,  by integrating the differential event rate over the ranges of
model parameters for a fixed PBH mass and a fixed microlensing timescale.
Throughout this paper we characterize each microlensing light curve
by its full-width-at-half-maximum (FWHM) timescale, $t_{\rm FWHM}$.
The constraint on the abundance of PBHs can be obtained by integrating the
expected microlensing events over  all possible light-curve timescales
accessible to our observations. Our dense HSC data is
most sensitive to light curves with timescales ranging from a few minutes to a
few hours. 
The expected number of PBH microlensing is quite high, up
to $\mathrm{d}N_{\rm exp}/\mathrm{d}\ln t_{\rm FWHM}\sim 10^{-5} $ for a
light curve timescale of $t_{\rm FWHM}\sim 0.1~$hours.
Hence, if we monitor $10^8$ stars in M31 with each exposure (visit) as we will
describe below, we can expect to observe $\sim 10^3$ events
if such PBHs constitute 
most DM in the MW and M31 halo regions.  Since a PBH in the halo region
has a typical motion of 200~${\rm km}/{\rm s}$ as implied by the rotation curve
irrespectively of PBH mass, a lighter PBH will result in an event with a
shorter timescale, owing to its smaller Einstein radius. This makes
high-cadence observations of M31 ideally suited for the search of microlensing
events arising from lighter PBHs.

\begin{figure}
\centering
 \includegraphics[width=0.32\textwidth,clip]{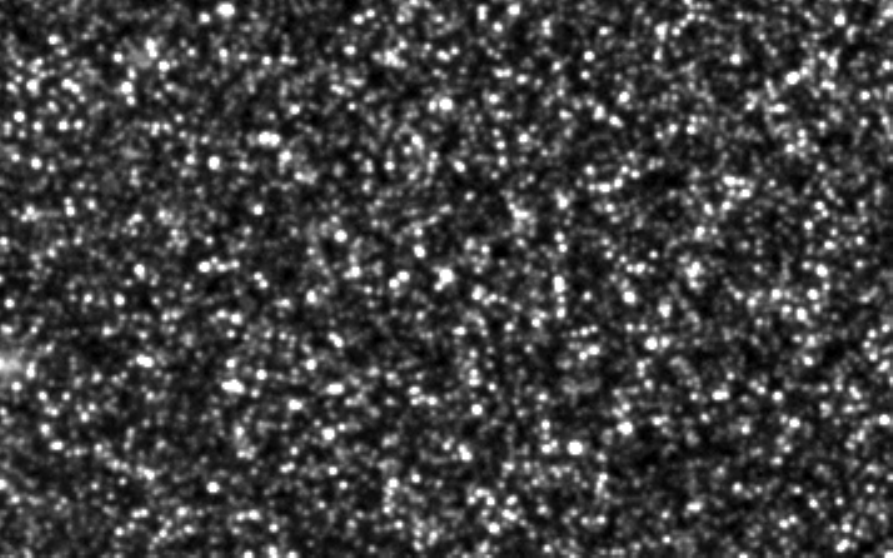}
 \includegraphics[width=0.32\textwidth,clip]{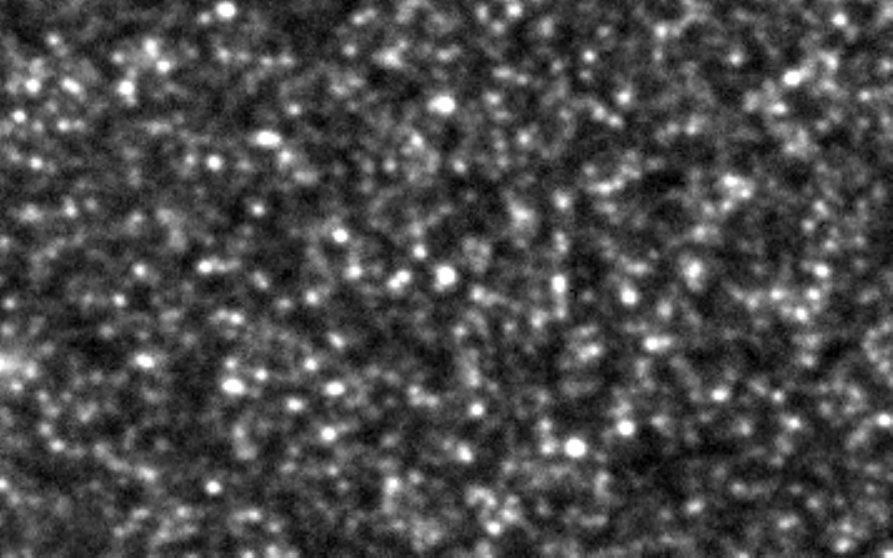}
 \includegraphics[width=0.32\textwidth,clip]{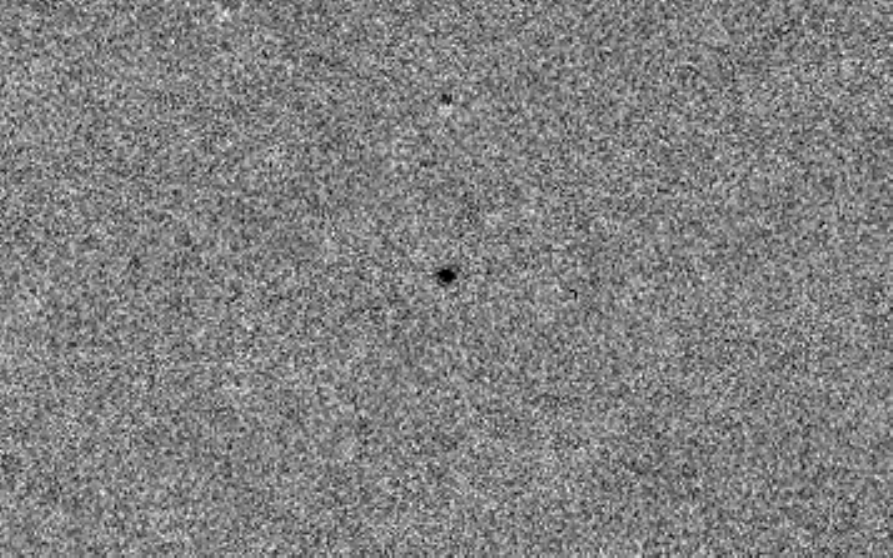}
 \caption{An example of the image subtraction technique we use for the
 analysis in this paper. The left panel shows the reference image
 which was constructed by co-adding the images of 10 best-seeing epochs, with a
 typical seeing of $0.45^{\prime\prime}$. The size of the image is
 $222\times 356$ pixels (corresponding to about 0.63 sq. arcmin), 
 and corresponds to the disk region in M31.  The middle panel shows a target
 image (coadded image using 3 sequential exposures) with seeing size of
 $0.8^{\prime\prime}$.  The right panel shows the difference image generated by
 our image subtraction pipeline properly accounting for the different seeing of
 the target and reference images even in such a densely populated stellar field.
 A variable star candidate shows up in the difference image at the center.
In this case, the candidate object appears as a negative flux
 in the difference image, because the object was fainter in the
 target image than in the reference image.  \label{fig:ex_diffimage}}
\end{figure}
Motivated by these considerations, we carried out a dense-cadence HSC
observation of M31 in the $r$-band on the night of November 23,
2014. The HSC camera has 104 science detectors with a pixel scale of
$0.168^{\prime \prime}$ \cite{2018PASJ...70S...1M,HSCOverView:17} (see Fig.~\ref{fig:m31_image}).  
The pointing was centered at the coordinates of the M31 central region: (RA,
dec) = (00h 42m 44.420s,+41d 16m 10.1s). We did not adopt any dithering (moving the pointing directions) between
different exposures 
in order to keep the same stars in the same CCD chip. We
carried out the observations with a cadence of 2~minutes,
and acquired 194 exposures for M31 during 7 hours within the same night until
the elevation of M31 fell below about $30$ degrees. The last 6 of the
exposures at the end of our observations suffered from bad seeing, $\simgt
1.2^{\prime\prime}$. Therefore in our analysis, we use 188 exposures in total.
These data yield a densely-sampled light curve for every variable star
candidate in the field with a 2-min cadence. 

The analysis of M31 time domain data presents a formidable challenge, as it is
a dense stellar field.  We are in the pixel lensing regime, where we need to
detect the microlensing of a single {\em unresolved} star among many stars that
contribute photons to each CCD pixel \cite{Crotts:92,Baillonetal:93,Gould:96}.
All of the previous work on M31 microlensing (e.g., see \cite{CalchiNovati:10}) has been carried out using smaller
aperture telescopes, which can only be sensitive to microlensing of relatively
bright stars such as red giants
\cite{Auriereetal:01}. 
In addition the image quality of HSC
corresponds to a significant step ahead, especially given the typical seeing size $\sim
0.6^{\prime\prime}$. In order to search for pixel lensing, we used the image
subtraction technique described in Alard \& Lupton \cite{AlardLupton:98}. This
technique has been integrated into the standard HSC data reduction pipeline,
$\mathtt{hscPipe}$ \cite{Boschetal:17}. The pipeline subtracts a reference
image (constructed from the 10 epochs with the best seeing data) from a target
image for M31 taken at a different epoch, and catalogs variable star candidates
that 
are identified 
in the difference image.  In Fig.~\ref{fig:ex_diffimage}, we
demonstrate an example of the image difference technique successfully performed
by our pipeline in a typically dense stellar field in M31. A variable star
candidate, which undergoes a flux change between the reference and target
epochs, appears in the difference image, as shown in the right panel. We
exclude the core bulge of M31 and the region including M101 from our analysis
where there are many saturated stars as there is no hope of recovering
microlensing events buried in saturated pixels even with an image difference
technique.

\begin{figure*}[t]
 \centering
 \includegraphics[width=0.7\textwidth]{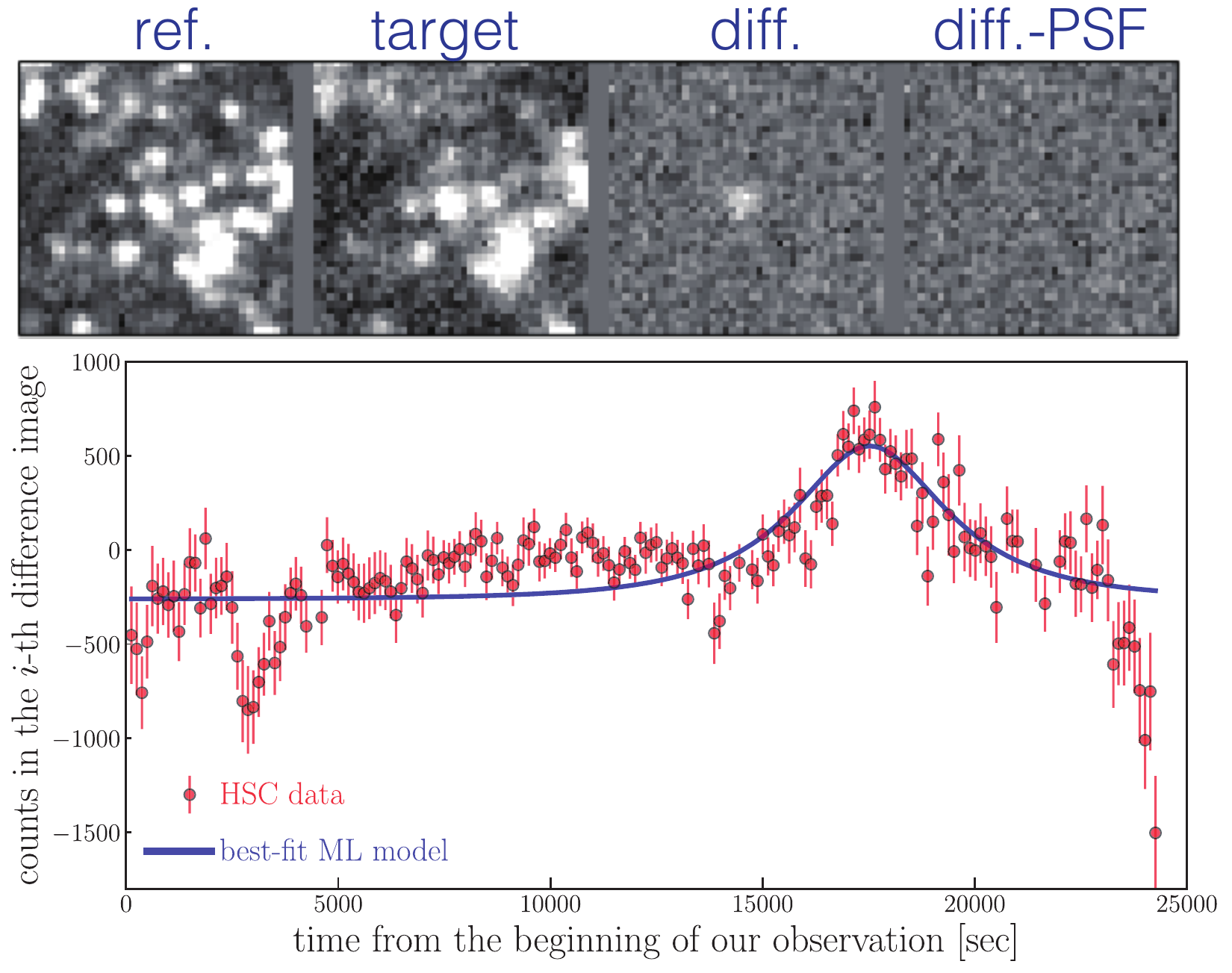}
 \caption{The single remaining candidate that passed all the criteria
 we impose to select microlensing events. The images in the upper panels show the
 postage-stamped images around the candidate: the reference image, the
 target image, the difference image and the residual image after
 subtracting the best-fit PSF image, respectively. The lower panel shows
 that the best-fit microlensing model (blue curve) gives an acceptable fit to the
 measured light curve. The error bars denote photometric errors in the 
 brightness measurement in the different image at each epoch.
  \label{fig:one_remained}}
\end{figure*}
We extract 15571 candidate variable stars, from the difference images
constructed by subtracting the reference image from each of the 188 target
images. All of these candidates satisfy our basic selection criteria -- (a) at
least a $5\sigma$ detection of flux in any of the 188 difference images, and
(b) the difference image of the candidate is consistent with the Point Spread
Function (PSF). Subsequently, we perform PSF photometry at the center of each
candidate in all of the difference images. This allows us to measure the light
curve of the candidate as a function of time, sampled every 2~min, through our
7~hour-long observation period. Our candidates include many secure detections
of variable events such as stellar flares, eclipsing contact binaries and
Cepheid variables. However, our photometry comes with an important caveat. The
photometry in the difference image measures only the flux {\it change} between
the reference and target images. Although we also use the PSF photometry at the
candidate position in the reference image to {\it infer} the intrinsic
flux of the candidate star, this photometry could be contaminated by 
fluxes of neighboring stars. Among the 15,571 variables we detect, about 3000 are
brighter than $m_r=25$, while the rest extend down to $m_r\sim 26$.  Thus, even
with a 90~sec short exposure, the 8.2~m aperture and excellent image quality
enables us to detect variables stars down to $m_r\sim  25 $--$26$, which
clearly shows the power of HSC/Subaru for time domain astronomy.

We then search for microlensing events from our master catalog of 15,571
variable star candidates.  As the optical depth of microlensing towards a
single star is $\tau\ll 1$, the probability to have multiple lensing events for
the same source star is negligible. Therefore, we impose a level 1 requirement
that a candidate should have a single ``bump'' feature in the light curve,
defined by 3 time-consecutive flux changes each with a significance greater
than $5\sigma$ in the difference image. This selection leaves us with $11,703$
candidates. Then we fit the parameters of a microlensing model with each
measured light curve. The microlensing light curve in the difference images is
characterized by 3 parameters: the impact parameter, the light-curve FWHM
timescale and the intrinsic flux (more exactly the intrinsic ADU counts of the
candidate in the difference image). To perform a $\chi^2$-fit to the data, we
estimate the rms noise of PSF photometry in each of the difference images by
estimating the PSF photometry at 1,000 random points in each HSC patch region
(see Fig.~\ref{fig:m31_image}).
We keep only those candidates which yield a best fit reduced $\chi$-squared
value, $\chi^2_{\rm best-fit}/185<3.5$ (the degrees of freedom are $185
=188-3$). This criterion is sufficiently conservative (the P-value is $\sim
10^{-5}$) for us not to miss a real microlensing candidate, if it exists. We further
impose the condition that the light curve has a symmetric shape around the
peak. These selections leave us with a total of 66 candidates.

Finally we perform a visual inspection of each of the remaining candidates. We
found various impostors that are not removed by the above automated criteria.
Most of them are a result of imperfect image subtraction; in most cases the
difference image has significant residuals near the edges of CCD chip or around
a bright star. In particular, bright stars cause a spiky residual in the
difference image, which result in impostors with a microlensing-like light
curve if the PSF flux is measured at a fixed position. We found 44 such
impostors which were a result of such spike-like images around bright stars.
Of the remaining, 20 impostors were located at the edges of the CCDs. 
We also identified 1 impostor event caused by a moving object, an asteroid. If
the light curve is measured at a fixed position where the asteroid passes, it
results in a light curve which mimics microlensing. In summary, the visual
inspection left us with a single candidate  which passed all our cuts and
visual checks. The candidate position is $({\rm RA},
{\rm dec})=(00{\rm h}~45{\rm m}~33.413{\rm s}, +41{\rm d}~07{\rm
m}~53.03{\rm s})$. Fig.~\ref{fig:one_remained} shows the images and the
light curve for the remaining candidate. Although the light curve looks noisy,
it is consistent with the microlensing prediction. The magnitude of the star
inferred from the reference image $m_r\sim 24.5$. 
Unfortunately, the candidate is placed just outside the
survey regions of the Panchromatic Hubble Andromeda Treasury (PHAT)
catalog in Refs.~\cite{Williamsetal:14,Dalcantonetal:12}\footnote{\url{https://archive.stsci.edu/prepds/phat/}}, 
so the HST image at
this location is not available.  To address whether the candidate is a variable
star, we looked into the $r$-band data that was taken during the HSC
commissioning run in 2013, a different epoch from our observing night. If our
candidate is a variable star, it would display a time variability at the
different epoch. 
However, the $r$-band commissioning image was unfortunately taken with a seeing
of about $1.2^{\prime\prime}$, so the difference image at the candidate
position appears noisy.  Similarly, we also analyzed the $g$-band images taken
during the HSC commissioning run. However, due to the short duration of the
data itself ($\sim 15~$min), it is difficult to judge whether this candidate
has any time variability between the $g$ band images. Therefore, we cannot
conclusively infer the nature of this candidate. In what follows, we derive an
upper bound on the abundance of PBHs as a constituent of DM assuming that this
remaining candidate is real.

\begin{figure}
 \centering
 \includegraphics[width=0.95\textwidth]{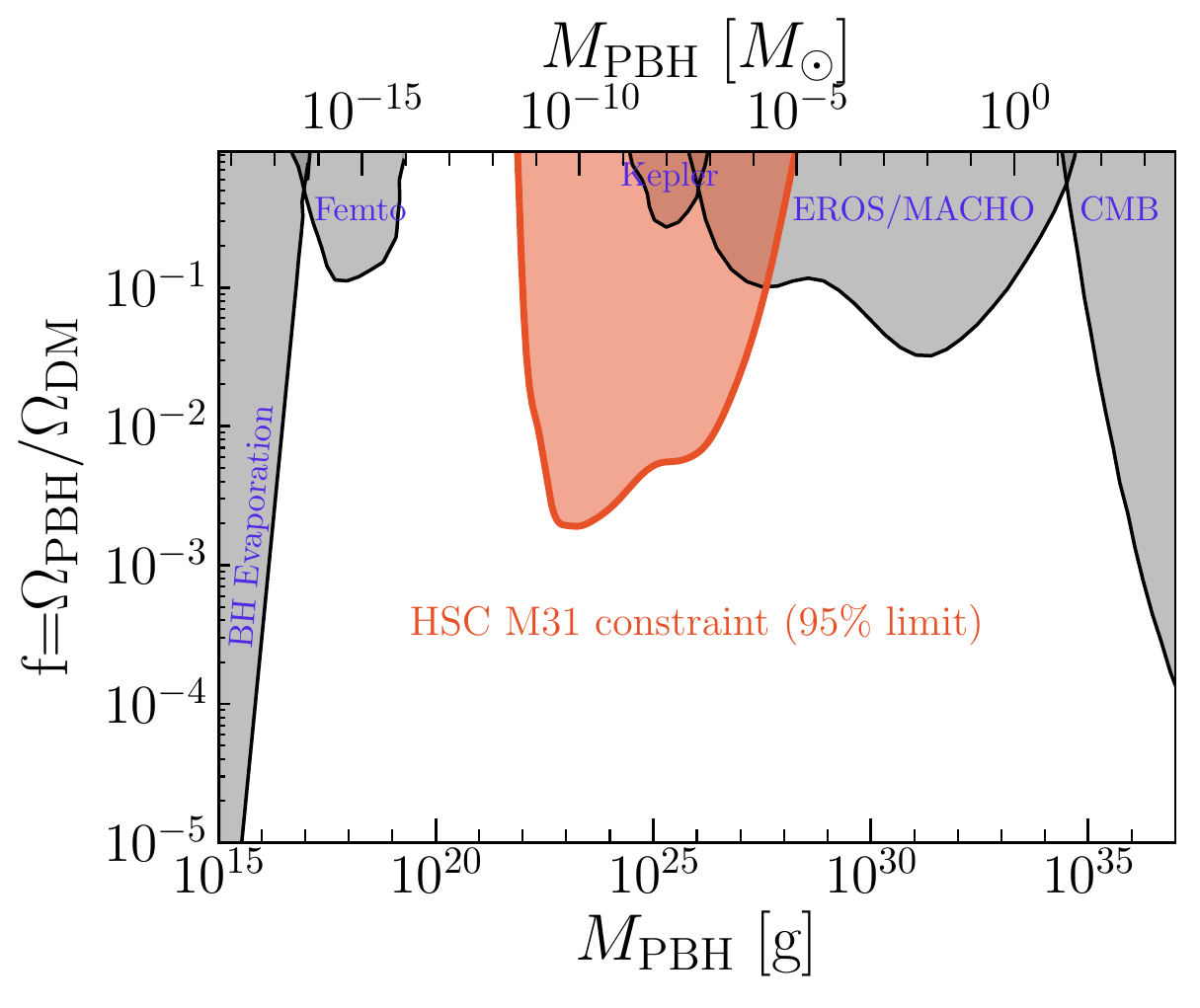}
 \caption{The red shaded region corresponds to the 95\% C.L. upper bound on
 the PBH mass fraction to DM in the halo regions of MW and M31, derived
  from our search for microlensing of M31 stars based on the ``single-night''
HSC/Subaru data and fills a large gap in the existing constraints by closing
the PBH DM window around lunar mass scale. 
 To derive this constraint, we took into 
 account the effect of finite source size, assuming that all source stars in M31 have a solar radius, 
 as well as the effect of wave optics in the HSC $r$-band filter on the microlensing event
 (see text for details). The effects weaken the upper bounds 
 at $M\simlt 10^{-7}M_\odot$, and give no constraint on PBH at $M\simlt 10^{-11}M_\odot$. 
 Our constraint can be
 compared with other observational constraints as shown by the gray
 shaded regions: extragalactic $\gamma$-rays from PBH evaporation
 \cite{Carretal:10}, femtolensing of $\gamma$-ray burst (``Femto'')
 \cite{Barnackaetal:12}, microlensing search of stars from the satellite
 2-years Kepler data (``Kepler'') \cite{Griestetal:14}, MACHO/EROS/OGLE
 microlensing of stars (``EROS/MACHO'') \cite{EROS:07}, and the accretion
 effects on the CMB observables (``CMB'') \cite{Ali-HaimoudKamionkowski:17}, updated
 from the earlier estimate
 \cite{Ricottietal:08}. \label{fig:upper_bound}}
\end{figure}
Now we use the results of our  microlensing search to constrain the abundance
of PBHs in the halo regions of MW and M31. The expected number of PBH
microlensing events in our HSC data is given by
\begin{eqnarray}
 N_{\rm exp}\!\left(M_{\rm
 PBH},\frac{\Omega_{\rm PBH}}{\Omega_{\rm DM}}\right)
 =\frac{\Omega_{\rm PBH}}{\Omega_{\rm DM}}\int_0^{t_{\rm obs}}\!\!\frac{\mathrm{d}t_{\rm FWHM}}{t_{\rm FWHM}}~ 
 \int\!\!\mathrm{d}m_{r}~
\frac{\mathrm{d}N_{\rm event}}{\mathrm{d}{\ln t}_{\rm FWHM}}
\frac{\mathrm{d}N_{s}}{\mathrm{d}m_{r}}\epsilon({t}_{\rm FWHM}
,m_r),
\label{eq:expN}
\end{eqnarray}
where $\mathrm{d}N_{\rm exp}/\mathrm{d}t_{\rm FWHM}$ is the differential event
rate for a single star (Fig.~\ref{fig:eventrate}) per logarithmic timescale,
$\mathrm{d}{N}/\mathrm{d}m_r$ is the luminosity function of source stars in the
$r$-band magnitude range $[m_r,m_r+\mathrm{d}m_r]$, and $\epsilon(m_{\rm
FWHM},m_r)$ is the detection efficiency quantifying the probability that a
microlensing event for a star with magnitude $m_r$ and the light curve
timescale $t_{\rm FWHM}$ is detected by our selection procedures. The event
rate depends on the mass fraction of PBHs to the total DM mass in the halo
regions, $\Omega_{\rm PBH}/\Omega_{\rm DM}$. Note that we have assumed a
parametric model for the total matter content of the MW and M31 halos
constrained by their respective rotation curves (see the explanation for
Fig.~\ref{fig:eventrate}). The PBH DM mass fraction does not depend on the
cosmological matter parameter, $\Omega_{\rm m0}$, that is relevant for the
cosmic expansion. 

We use the following procedure to estimate $\mathrm{d}N_s/\mathrm{d}m_r$ and
$\epsilon$ in Eq.~(\ref{eq:expN}). Since individual stars are not resolved in
the HSC data, especially in the disk region of M31, it is not straightforward
to estimate the number of source stars from the HSC data alone. This
constitutes a significant uncertainty in our results.  One way to
estimate the number of source stars from the HSC data itself is using the
number of ``detected peaks'' in the reference image (the best-seeing co-added
image). This estimate is very conservative as it misses the numerous faint
stars that do not produce prominent peaks. To overcome this difficulty, we use
the HST PHAT star catalog as follows. Our HSC data has an overlap with the HST
PHAT survey for the M31 disk region, where individual stars are resolved thanks
to the high angular resolution of the ACS/HST data. We found that the number
counts of peaks in the HSC image fairly well agrees with the counts in the PHAT
catalog down to $m_r\sim 23$, after applying a color transformation between the HSC
and HST filters, but indeed misses stars at the fainter magnitudes. 
For the overlapping regions, we used the PHAT star counts down
to $m_r\sim26$.
For the non-overlapping regions in
the M31 disk, we infer the luminosity function by extrapolating the number
counts of HSC peaks at $m_r=23$ down to $m_r=26$ based on the PHAT luminosity
function of stars at a similar distance from the M31 center. For our default
analysis, we used about $8.7\times 10^7$ stars down to $m_r=26~$mag over the
entire region of M31, which is a factor 14 more number of stars than that of
HSC peaks.
The large number
of source stars in the M31 region 
can be compared with those
in previous studies, e.g., \cite{Griestetal:14} used $\sim 1.5\times 10^5$
source stars for the microlensing search in Kepler data.

For an estimation of the detection efficiency $\epsilon(t_{\rm FWHM},
m_r)$ in Eq.~(\ref{eq:expN}), we carry out Monte Carlo simulations
of microlensing light curves adopting random combinations of the model
parameters (the impact parameter, $t_{\rm FWHM}$, and the intrinsic
flux) and adding the statistical noise based on the photometry errors
in each HSC-patch region. These simulations allow us to estimate the
fraction of simulated light curves that can be recovered by our
selection procedures. As a cross-check, we also use SynPipe
\cite{Huangetal:17}, a pipeline to inject a synthesized microlensed star into every
individual HSC exposure image at a randomly-selected location. We then test
whether our microlensing search can identify such synthesized events from these
images in order to estimate the detection efficiency. The two methods give
similar estimates for the detection efficiency. Our results indicate that our
pipeline can recover about 70--60\% of microlensing events for stars with
intrinsic magnitude $m_r=23$--$24~$mag, if the timescale is in the range
$t_{\rm FWHM}\simeq [0.1,3]~{\rm hours}$. For fainter stars with
$m_r=25~$--26~mag, the efficiency is reduced to about 30--20\%.

Next we combine the estimates of $\mathrm{dN}_{\rm event}/\mathrm{d}\ln t_{\rm
FWHM}$, $\mathrm{d}N_s/\mathrm{d}m_r$ and $\epsilon(t_{\rm FWHM},m_r)$ in
Eq.~(\ref{eq:expN}) to constrain the abundance of PBHs. Assuming the number of
microlensing events follow a Poisson distribution, the probability to observe a
given number of such events, $N_{\rm obs}$, is given by $P(k=N_{\rm obs}|N_{\rm
exp})= \left[\left(N_{\rm exp}\right)^k/k!\right]\exp[-N_{\rm exp}]$. Hence the 
95\%~C.L. interval is estimated as $P(k=0)+P(k=1)\ge 0.05$, leading to $N_{\rm
exp}\le 4.74$ assuming that the candidate in Fig.~\ref{fig:one_remained} is real.
Fig.~\ref{fig:upper_bound} shows our result in comparison with other
observational constraints on the abundance of PBHs on different mass scales.
In the results, we took into account the effect of finite source star size \cite{WittMao:94}
as well as the effect of wave optics on the microlensing cross section \cite{Gould:92,Nakamura:98}.  
The finite-source size results in the magnification of only a small part of the
star and hence affects the detectability of the event.
The effect modifies our constraints on mass scales,
$M_{\rm PBH}\simlt
10^{-7}M_\odot$ where the Einstein radii of the PBHs become comparable to or
smaller than the size of the source stars. We caution that we may have
underestimated the impact somewhat as we have assumed a solar radius for all
stars in M31, while some of the stars would likely be giants. The wave effect arises from the fact that
the Schwarzschild
radii of light PBHs with $M\simlt 10^{-11}M_\odot$ become comparable to
the wavelength of the HSC $r$-band filter (centered around 600~nm). In this
regime, the wave nature of light becomes important and can further lower the
maximum magnification of the microlensing light curve. This results in a lower
event rate for a given detection threshold. 
These effects
need to be further studied and carefully accounted for. Nevertheless the
figure shows that a single night of HSC data on M31 results in a tight upper
bound on the mass fraction of PBHs to DM, $\Omega_{\rm PBH}/\Omega_{\rm DM}$.
The origin of the constraint can be easily understood. Given that we monitor
about $10^8$ stars, we expected to observe about 1,000 microlensing events 
if PBHs of a single mass scale $M_{\rm PBH}\sim 10^{-9}M_\odot$ make up all
DM in the MW and M31 halo regions (see Fig.~\ref{fig:eventrate}), and yet we
could identify only a single event. In other words, only a small mass fraction
of PBHs such as $\Omega_{\rm PBH}/\Omega_{\rm DM}\simeq 0.001$ is allowed in
order to reconcile the PBH DM scenario with our M31 data. Our results
constrain PBHs in an open window of PBH masses, $M_{\rm PBHs}\simeq
[10^{-11},10^{-9}]M_\odot$, as well as give tighter constraints than 
those reported by
previous work 
in the range of $M_{\rm PBH}\simeq[10^{-9},10^{-6}]M_\odot $.
In particular, our constraint is tighter than the constraint from the 2-year
Kepler data that had monitored an open cluster containing $10^5$ stars, with
about 15 or 30~min cadence over 2 years \cite{Griestetal:14}. 

More generally, theoretical models for formation of PBHs in the early universe
scenario \cite{Muscoetal:08,Kuhneletal:15,Kawasakietal:16,Kawasakietal:16b,Inomataetal:16,KuhnelFreese:17}
predict a mass spectrum.
Some theoretical models even predict mass spectrum extending up to a
$10M_\odot$ scale, the mass scale of LIGO binary black holes. All such models with a
non-monochromatic mass spectrum must reconcile with our constraints (see the discussion around Eq.~\ref{eq:expN_extended} in the Supplementary Information and also see Refs.~\cite{Carretal:16,Inomataetal:17,Carretal:17}).
We expect the observational constraints to be improved in the future. By simply
carrying out observations of M31 for more HSC nights, the bounds on the PBH
abundances could be tightened. For example, an additional monitoring of M31 for
10 clear nights could tighten the upper bounds by a factor of $10$. 
We also expect our constraints to be 
extended to heavier mass scales by monitoring M31 over a
longer timescale from months to years. Repeated observations of M31 every few
months over years, e.g., 10 minutes of monitoring during each observation
run, should be able to provide important constraints on heavier mass scales
including those at LIGO BH mass scales of $10M_\odot$.  Since M31 is the most
suitable  target in the northern hemisphere for HSC, this is a valuable
opportunity, waiting to be exploited.



Correspondence and requests for material in relation to this work should be sent to Hiroko Niikura
(niikura@hep.phys.s.u-tokyo.ac.jp) as well as Masahiro Takada (masahiro.takada@ipmu.jp).


\section*{Acknowledgments}

We would like to dedicate this paper to the memory of Prof.~Arlin
Crotts, a pioneer of pixel lensing.  
We would like to thank the anonymous referees for their comments/suggestions 
that help to improve this paper.
We thank
Sergey Blinnikov,
 Andrew Gould, Bhuvnesh
Jain, Masahiro Kawasaki, Alex Kusenko, Chien-Hsiu Lee, Hitoshi Murayama, 
David Spergel and Masaomi Tanaka for useful discussion.
We thank Nick Kaiser and Misao Sasaki for pointing out the importance of wave optics effect
in our microlensing constraints when M.T. gave a talk at the seminar of YITP, Kyoto University. 
This work was
supported by World Premier International Research Center Initiative (WPI
Initiative), MEXT, Japan, by the FIRST program ``Subaru Measurements of
Images and Redshifts (SuMIRe)'', CSTP, Japan, Grant-in-Aid for
Scientific Research from the JSPS Promotion of Science (No.~23340061,
26610058, and 15H03654), MEXT Grant-in-Aid for Scientific Research on Innovative Areas
(No.~15H05887, 15H05892, 15H05893, 15K21733) and JSPS Program for Advancing
Strategic International Networks to Accelerate the Circulation of
Talented Researchers.

The Hyper Suprime-Cam (HSC) collaboration includes the astronomical
communities of Japan and Taiwan, and Princeton University.  The HSC
instrumentation and software were developed by the National Astronomical
Observatory of Japan (NAOJ), the Kavli Institute for the Physics and
Mathematics of the Universe (Kavli IPMU), the University of Tokyo, the
High Energy Accelerator Research Organization (KEK), the Academia Sinica
Institute for Astronomy and Astrophysics in Taiwan (ASIAA), and
Princeton University.  Funding was contributed by the FIRST program from
Japanese Cabinet Office, the Ministry of Education, Culture, Sports,
Science and Technology (MEXT), the Japan Society for the Promotion of
Science (JSPS), Japan Science and Technology Agency (JST), the Toray
Science Foundation, NAOJ, Kavli IPMU, KEK, ASIAA, and Princeton
University.

The Pan-STARRS1 Surveys (PS1) have been made possible through
contributions of the Institute for Astronomy, the University of Hawaii,
the Pan-STARRS Project Office, the Max-Planck Society and its
participating institutes, the Max Planck Institute for Astronomy,
Heidelberg and the Max Planck Institute for Extraterrestrial Physics,
Garching, The Johns Hopkins University, Durham University, the
University of Edinburgh, Queen's University Belfast, the
Harvard-Smithsonian Center for Astrophysics, the Las Cumbres Observatory
Global Telescope Network Incorporated, the National Central University
of Taiwan, the Space Telescope Science Institute, the National
Aeronautics and Space Administration under Grant No. NNX08AR22G issued
through the Planetary Science Division of the NASA Science Mission
Directorate, the National Science Foundation under Grant
No. AST-1238877, the University of Maryland, and E\"otv\"os Lor\'and
University (ELTE).

Based [in part] on data collected at the Subaru Telescope and retrieved
from the HSC data archive system, which is operated by Subaru Telescope
and Astronomy Data Center at National Astronomical Observatory of Japan.

This paper makes use of software developed for the Large Synoptic Survey
Telescope. We thank the LSST Project for making their code available as
free software at http://dm.lsstcorp.org.

\section*{Author contributions}

All the authors discussed the results and commented on the manuscript. M.T., H.N. and S.M. wrote the paper. H.N. performed most of the data analysis, the calculation of microlensing event rates and the model fitting. M.T. proposed the idea, and M.T and T.S. prepared the observation plan and strategy for the HSC/Subaru observation of M31. N.Y. and R.H.L. provided advice about the use of the HSC data analysis pipeline, especially the image difference method. 
T.K. and S.S. carefully estimated the effect of finite source size and the wave optics effect on microlensing event rates for PBH
at $\simlt 10^{-9}M_\odot$, and we were
able to obtain a more accurate estimation of the upper bounds on the abundance of such PBHs. 
All the authors commented on the draft text.

\clearpage
\newpage
\begin{center}
{\bf \large Supplementary Information}
\end{center}

\section{Event rate of PBH microlensing for M31 stars}
\label{sec:pointlens}

In this section we estimate event rates of PBH microlensing for a star
in M31. We extend the formulation in previous studies
\cite{Griestetal:91,Alcocketal:96,Kerinsetal:01,Riffeseretal:06} to
microlensing effect on a star in M31 due to PBHs assuming the PBHs exist in the halo regions of MW and M31.

\subsection{Microlensing basics}
\label{sec:lens}

If a star in M31\footnote{Throughout this paper we assume that a source star is
in M31, not in the MW halo region, because of the higher number density on the
sky.} and a foreground PBH are almost perfectly aligned along the line-of-sight
to an observer, the star is multiply imaged due to strong gravitational
lensing. In case these multiple images are unresolved, the flux from the star
appears magnified. When the source star and the lensing PBH are separated by an
angle $\beta$ on the sky, the total lensing magnification, i.e. the sum of the
magnification of the two images, is
\begin{eqnarray}
 A=A_{1}+A_{2}=\frac{u^2+2}{u\sqrt{u^2+4}},
\label{eq:cano}
\end{eqnarray}
where $u\equiv (\dl \times \beta)/R_E$, and $\dl$ is the distance to a
lensing PBH.  The Einstein radius $R_E$ is defined as
\begin{equation}
R^2_E=\frac{4GM_{\rm PBH}D}{c^2},
\label{eq:RE} 
\end{equation}
where $M_{\rm PBH}$ is the PBH mass. $D$ is the lensing weighted
distance,
$D\equiv \dl (1-\dl /\ds)$, where $\ds$ is the distance to a
source star in M31, and $d$ is the distance to the PBH. By plugging typical
values of the parameters, we can find the typical Einstein radius:
\begin{equation}
 \theta_E\equiv \frac{R_E}{\dl}\simeq 3\times 10^{-8}~{\rm arcsec}
  \left(\frac{M_{\rm PBH}}{10^{-8}M_\odot}\right)^{1/2}
  \left(
   \frac{\dl}{100~{\rm
   kpc}}\right)^{-1/2}
  \label{eq:thetaE}
\end{equation}
where we assumed $\ds=770~{\rm kpc}$ for distance to a star in M31 and
we assumed $D\sim d$ for simplicity, and employed $M_{\rm
PBH}=10^{-8}M_\odot$ as a working example for the sake of comparison
with Ref.~\cite{Griestetal:14}. In the following analysis we will consider a
wide range of PBH mass scales. 
The PBH lensing phenomena we search for are in the microlensing
regime; we cannot resolve two lensed images with angular resolution of an
optical telescope, and we can measure only the total magnification.  A size of
a star in M31 is viewed as
\begin{equation}
 \theta_s\simeq \frac{R_s}{d_{\rm s}}\simeq 5.8\times 10^{-9}~{\rm
  arcsec}\ ,
  \label{eq:theta_star}
\end{equation}
if the source star has a similar size to the solar radius ($R_\odot\simeq
6.96\times 10^{10}~$cm).
Comparing with Eq.~(\ref{eq:thetaE}) we find that the Einstein radius becomes
smaller than the source size if PBH mass $M_{\rm PBH}\simlt 10^{-10}M_\odot$
corresponding to $M_{\rm PBH}\simlt 10^{23}~$g.  We will  later
discuss such lighter PBHs, where we will take into account the effect of finite
source size on the microlensing
\cite{WittMao:94,CieplakGriest:13,Griestetal:14}.

Since the PBH and the source star move relative to each other on the sky, the
lensing magnification varies with time, allowing us to identify the star as a
variable source in a difference image from the cadence observation. The
microlensing light curve has a characteristic timescale that is needed for a
lensing PBH to move across the Einstein radius:
\begin{equation}
\label{eq:etime}
 t_E\equiv \frac{R_E}{v},
\end{equation}
where $v$ is the relative velocity. Assuming fiducial values for these
parameters, we can estimate the typical timescale as
\begin{equation}
 t_E\simeq 34~{\rm min}\left(\frac{M_{\rm
               PBH}}{10^{-8}M_\odot}\right)^{1/2}
 \left(\frac{\dl}{100~{\rm kpc}}\right)^{1/2}
\left(\frac{v}{200~{\rm km/s}}\right)^{-1}~ ,
\end{equation}
where we assumed $v=200~{\rm km/s}$ for the typical relative velocity.
Thus the microlensing light curve is expected to vary over several tens of
minutes, and should be well sampled by our HSC observation. It should
also be noted that a PBH closer to the Earth gives a longer timescale
light curve for a fixed velocity. 
Since we can safely
assume that the relative velocity stays constant during the Einstein radius
crossing, the light curve should have a symmetric shape around the peak, which
we will use to eliminate impostors.

\subsection{Microlensing event rate}
\label{sec:mlrate}

Here we estimate expected microlensing event rates from PBHs assuming that they
consist of a significant fraction of DM in the MW and M31 halo regions.

We first need to assume a model for the spatial distribution of DM (therefore
PBHs) between M31 and us (the Earth). Here we simply assume that the DM
distribution in each halo region of MW or M31 follows the NFW profile
\cite{NFW97}:
\begin{equation}
\rho_\mathrm{NFW}(r)= \frac{\rho_{c}}{(r/r_{s})(1+r/r_{s})^2},
\label{eq:rho_nfwm}
\end{equation}
where $r$ is the radius from the MW center or the M31 center, $r_s$ is
the scale radius and $\rho_c$ is the central density parameter. In this
paper we adopt the halo model in Ref.~\cite{Klypinetal:02}: $M_{\rm
vir}=10^{12}M_\odot$, $\rho_c=4.88\times 10^{6}~{M_\odot/{\rm kpc}^3}$,
and $r_s=21.5~{\rm kpc}$ for MW, taken from Table~2 in the paper, while
$M_{\rm vir}=1.6\times 10^{12}M_\odot$, $\rho_c=4.96\times
10^6~{M_\odot/{\rm kpc}^3}$, and $r_s=25~{\rm kpc}$ for M31, taken from
Table~3. Thus we assume a slightly larger DM content for the M31 halo
than the MW halo.  Dark matter profiles with these parameters have been
shown to fairly well reproduce the observed rotation curves for MW and
M31, respectively. There might be an extra DM contribution in the
intervening space between MW and M31, e.g. due to a filamentary
structure bridging MW and M31.  However, we do not consider such an
unknown contribution.

Consider a PBH at a distance $\dl$ (kpc) from the Earth and in the angular
direction to M31, $(l,b) = (121.2^{\circ},
-21.6^{\circ})$ in the Galactic coordinate system. Assuming that the
Earth is placed at distance $R_\oplus=8.5~{\rm kpc}$ from the MW center,
we can express the separation to the PBH from the MW center, $r_{\rm
MW-PBH}$, in terms of the distance from the Earth, $d$, as
\begin{equation}
r_{\rm MW-PBH}(\dl)=\sqrt{R_\oplus^2-2R_\oplus \dl\cos(l)\cos(b)+\dl^2}.
\label{eq:dislens}
\end{equation}
If we ignore the angular extent of M31 on the sky (which is restricted to
1.5~degree in diameter for our study), the distance to the PBH from the M31
center, $r_{\rm M31-PBH}$, is approximately given by,
\begin{equation}
 r_{\rm M31-PBH}(d)\simeq \ds-\dl,
\label{eq:dislens_m31}
\end{equation}
where we approximated the distance to a source star in M31 to be the same as
the distance to the center of M31, $D_{\rm M31}\simeq \ds$, which we assume to
be equal to $\ds =770~{\rm kpc}$ throughout this paper.
\begin{figure}
\centering
\includegraphics[width=0.48\textwidth]{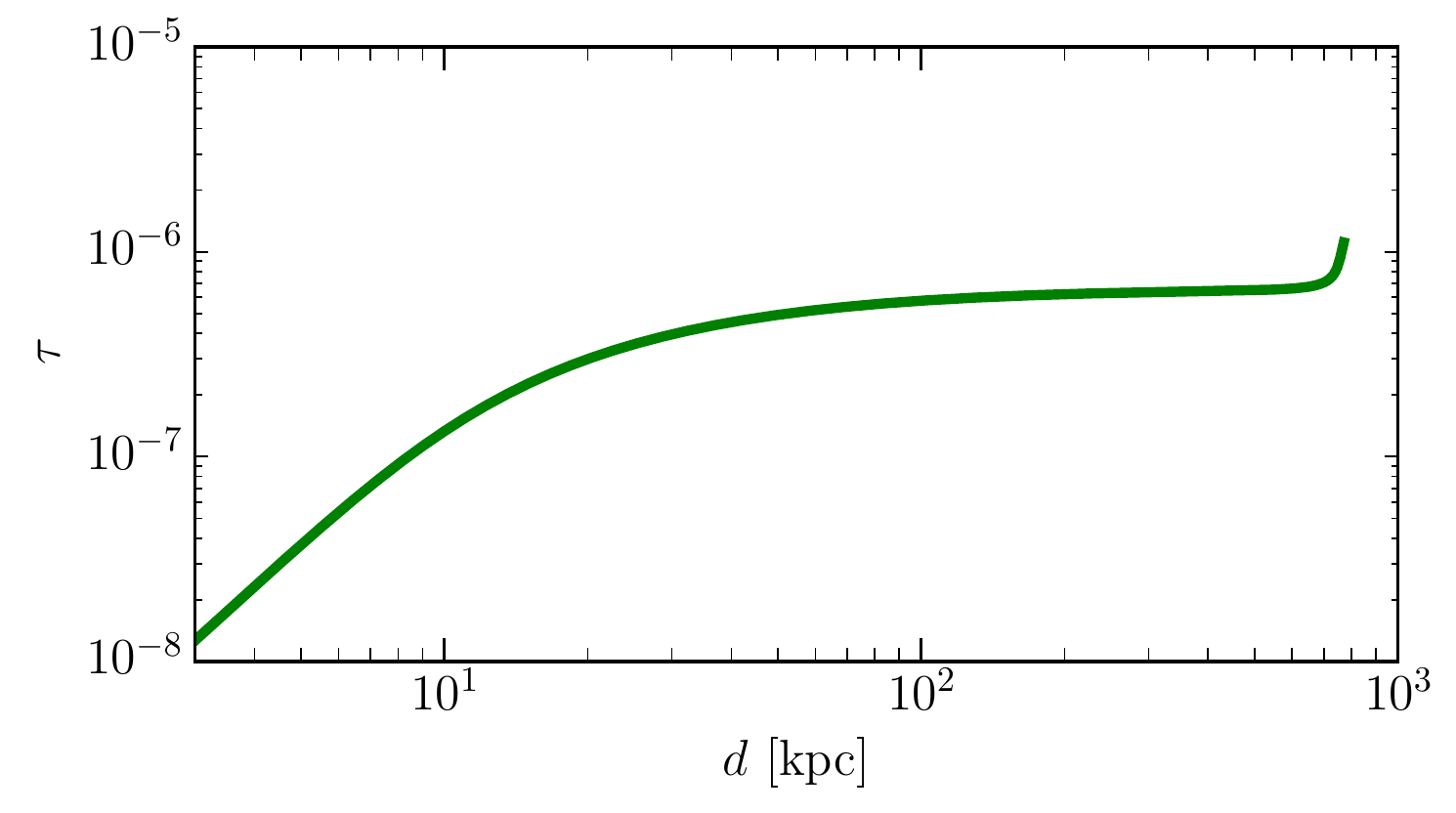}
\includegraphics[width=0.48\textwidth]{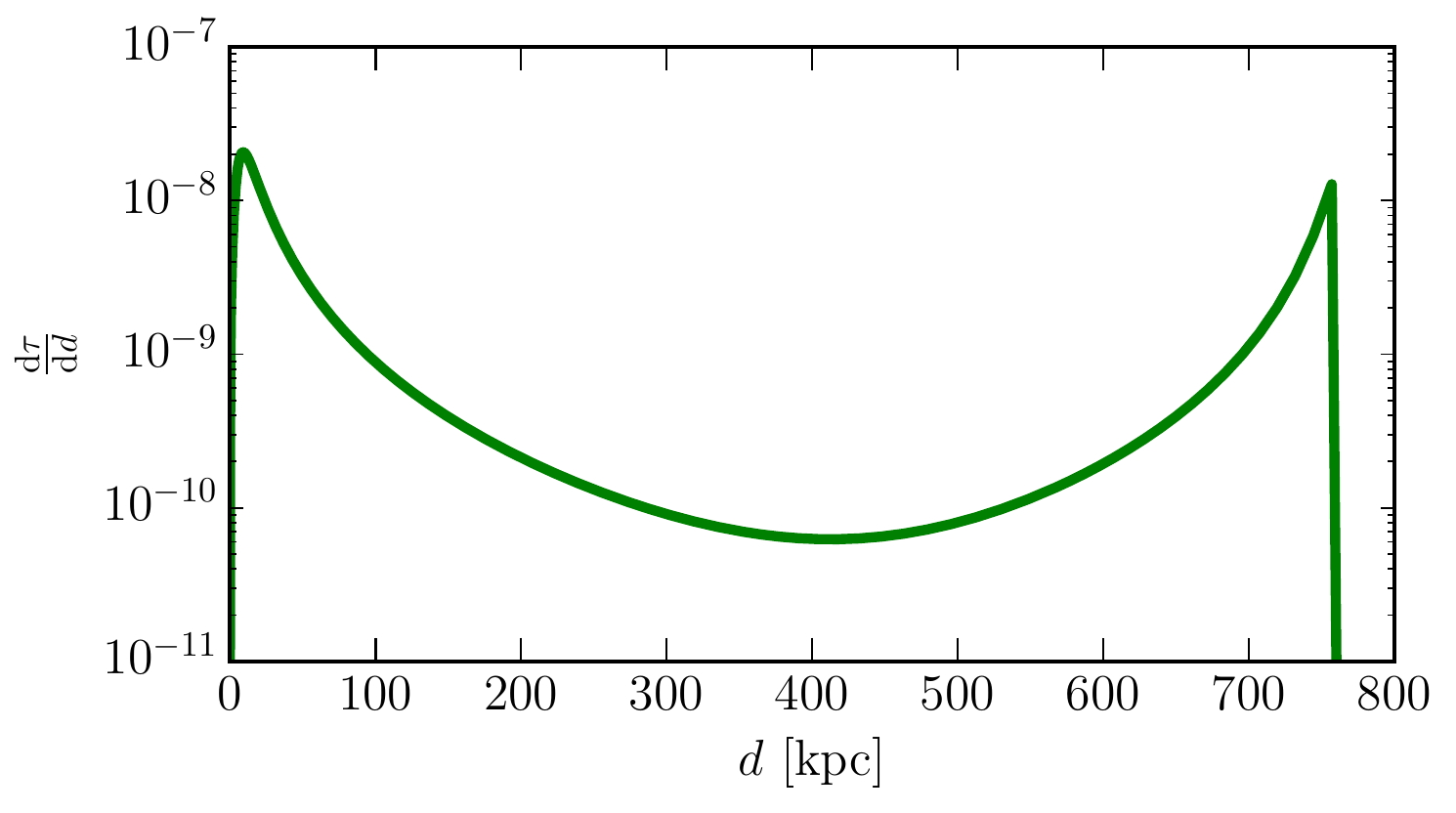}
 %
\caption{{\it Upper}: The optical depth of PBH microlensing effect on a
 single star in M31 as a function of the distance to PBH, $\dl$, which
 can be obtained by integrating the integrand in Eq.~(\ref{eq:tau}) over
 $[0,\dl]$, rather than $[0,\ds]$. The optical depth is independent of
 PBH mass, and we assumed NFW parameters to model the DM distribution in
 each of the MW and M31 halo regions, where we determined the NFW parameters
 so as to reproduce their rotation curves (see text for details). {\it
 Lower}: Similar plot, but the fractional contribution of PBHs at the
 distance, $\dl$,
 to the optical depth. Note that $\dl$ in the $x$-axis is in
 linear scale. The area under this curve up to $\dl$ gives the optical
 depth to $\dl$ in the upper plot.}  \label{fig:tau}
\end{figure}
By using Eqs.~(\ref{eq:rho_nfwm})-(\ref{eq:dislens_m31}), we can compute
the DM density, contributed from both the MW and M31 halos, as a
function of the distance to PBH, $\dl$.

Assuming that PBHs make up the DM content by a a fraction of $\Omega_{\rm
PBH}/\Omega_{\rm DM}$, we can compute the optical depth $\tau$ for the
microlensing of PBHs with mass $M_{\rm PBH}$ for a {\it single} star in
M31.  The optical depth is defined as the probability for a source star
to be inside the Einstein radius of a foreground PBH on the sky or
equivalently the probability for the magnification of source flux to be
greater than that at the Einstein radius, $A\ge 1.34$
\cite{Paczynski:86}:
\begin{eqnarray}
 \tau(d; M_{\rm PBH})= \frac{\Omega_{\rm PBH}}{\Omega_{\rm DM}}
  \int^{\ds}_0\!\!\mathrm{d}d~ 
  \frac{\rho_{\rm DM}(d)}{M_{\rm PBH}}\pi R_E^2(d,M_{\rm PBH}).
  \label{eq:tau}
\end{eqnarray}
Here the mass density field of DM is given by the sum of NFW profiles
for the MW and M31 halos: $\rho_{\rm DM}(d)=\rho_{\rm NFW, MW}(d)+\rho_{\rm
NFW, M31}(d)$. Note that, because of $R_E^2\propto M_{\rm PBH}$, the
optical depth is independent of PBH mass.

In Fig.~\ref{fig:tau}, we show the optical depth of PBH microlensing for a
single star in M31, calculated using the above equation. Here we have assumed
that all the DM in the halo regions of MW and M31 is composed of PBHs, i.e.,
$\Omega_{\rm PBH}/\Omega_{\rm DM}=1$.  The optical depth for microlensing,
$\tau\sim10^{-6}$, is larger compared to that to LMC or a star cluster in MW
($\tau\sim 10^{-7}$) by an order of magnitude, due to the enormous volume and
large mass content between the Earth and M31. The PBHs in each of the MW and M31
halos result in a roughly equal contribution to the optical depth to an M31 star.
Although there is an uncertainty in the DM density in the inner region of MW or
M31 (at radii $\simlt 10$~kpc) due to poorly-understood baryonic effects, the
contribution is not large.

\begin{figure}
 \centering
 \includegraphics[width=0.35\textwidth]{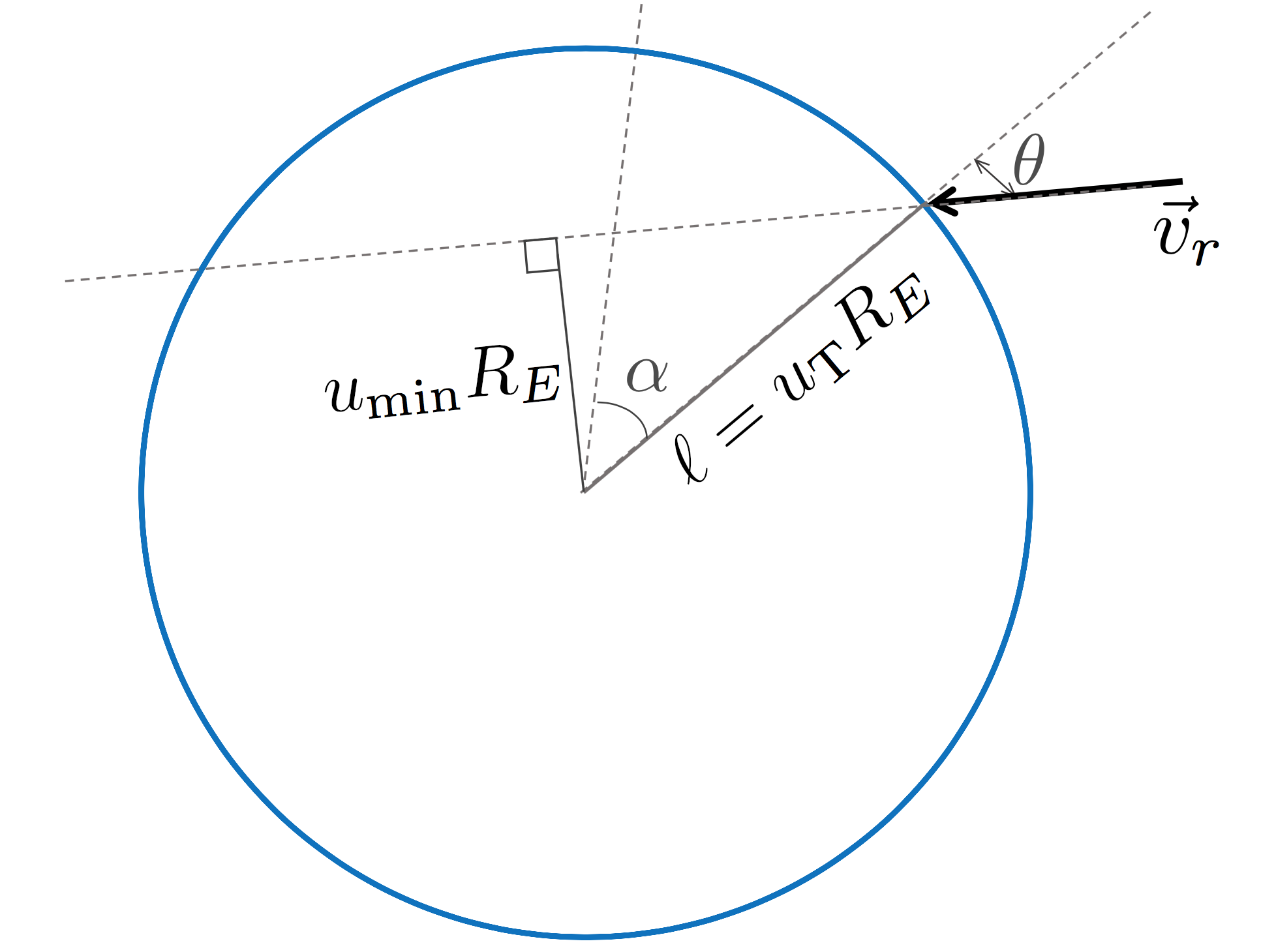}
 \caption{A schematic illustration of configurations of a lensing PBH
 and a source star in M31 in the lens plane, following Fig.~4 of
 Ref.~\cite{Griestetal:91}. The orbit of a lensing PBH, around a source star
 in M31 (placed at the origin in this figure), is parameterized as in
 the figure, which is used to derive the microlensing event rate (see
 text for details). 
 \label{fig:eventrate_illust}}
\end{figure}
Next we estimate the rate for microlensing events with a given timescale for its
light curve. First we model the velocity distribution of DM in the halo regions. We simply
assume an isotropic Maxwellian velocity distribution for DM particles (e.g., 
\cite{Jungmanetal:96}):
\begin{equation}
 f(\vv; r)\mathrm{d}^3\vv=\frac{1}{\pi^{3/2}v_{\rm c}(r)^3}
  \exp\left[-\frac{|\vv|^2}{v_{\rm c}(r)^2}\right]\mathrm{d}^3\vv
\label{eq:bolvelocity}
\end{equation}
where $V_{\rm halo}(r)$ is the velocity dispersion at radius $r$ from
the MW or M31 center.
For $V_{\rm halo}(r)$, we assume that it is given as
\begin{equation}
 v_{\rm c}(r)=\sqrt{\frac{GM_{\rm NFW}(<r)}{r}},
\label{eq:halovelocity}
\end{equation}
where $M_{\rm NFW}(<r)$ is the interior mass within radius $r$ from the
halo center, defined as
$ M_{\rm NFW}(<r)=4\pi\rho_sr^3_s\left[\ln(1+c)-{c}/({1+c})\right]$,
where $c=r/r_s$ for each of the MW and M31 halos.

We start from the geometry and variables shown in Fig.~4 of
Ref.~\cite{Griestetal:91} and their Eq.~(10) (see
Fig.~\ref{fig:eventrate_illust}), which gives the rate
$\mathrm{d}\Gamma$ of PBHs entering a volume element along the
line-of-sight where they can cause microlensing for a single star in
M31:
\begin{equation}
 \drm\Gamma =\frac{\Omega_{\rm PBH}}{\Omega_{\rm DM}}\frac{\rho_{\rm
  DM}(\dl)}{M_{\rm PBH}}
  \frac{u_{\rm T} R_{\rm E}}{\pi v_{\rm c}^2}
  \exp\left[-\frac{v_{\rm r}^2}{v_{\rm c}^2}\right]\, v_{\rm r}^2
  \cos\theta\, \drm v_{\rm r}\, \drm\theta\, \drm\dl\,\drm\alpha.
\end{equation}
Here $n_{\rm PBH}(\dl)$ is the number density of PBHs at the distance
$\dl$ from the Earth, $v_{\rm r}$ is the velocity of the PBH in the lens
plane, $\theta$ is the angle at which the PBH enters the volume element,
and $\alpha$ is an angle with respect to an arbitrary direction in the
lens plane, as shown in Fig.~\ref{fig:eventrate_illust}.  Microlensing
events are identified if they have a given threshold magnification
$A_{\rm T}$ at peak.  This threshold magnification defines a threshold
impact parameter with respect to the Einstein radius of a PBH, $u_{\rm
T}=R_{\rm T}/R_{\rm E}$.
Compared to Ref.~\cite{Griestetal:91}, we have further ignored motions of
source stars for simplicity, i.e. $v_{\rm t}=0$. The parameters vary in
the range of $\theta\in[-\pi/2,\pi/2]$, $\alpha\in[0, 2\pi]$, $v_{\rm
r}=[0,\infty)$.

The timescale for the microlensing event described by the above
geometry is given by $\hat{t}=2R_{\rm E}\cos \theta ~ u_{\rm T}/v_{\rm
r}$. Thus the differential rate of microlensing events, occurring per
unit timescale $\hat{t}$, is given by
\begin{eqnarray}
 \frac{\drm\Gamma}{\drm \hat{t}} &=&
\frac{\Omega_{\rm PBH}}{\Omega_{\rm DM}}
\int_0^{\ds}\!\! \drm\dl
  \int_0^\infty\!\! \drm v_{\rm r}\,
  \int_{-\pi/2}^{\pi/2}\!\!\drm\theta
  \int_0^{2\pi}\!\!\drm\alpha\, \frac{\rho_{\rm  DM}(\dl)}{M_{\rm PBH}}
  \nonumber\\
    &&\times \frac{u_{\rm T} R_{\rm E}}{\pi v_{\rm c}^2}
     \exp\left[-\frac{v_{\rm r}^2}{v_{\rm c}^2}\right]\, v_{\rm r}^2
     \cos \theta ~ \,
  \delta_{\rm D}\!\left(\hat{t}-\frac{2R_{\rm E}u_{\rm T}\cos\theta}{v_{\rm r}}\right).\hspace{1em}
\end{eqnarray}
Using the Dirac-delta function identity,
\begin{equation}
    \delta_{\rm D}\left(\hat{t}-\frac{2R_{\rm E}u_{\rm T}\cos\theta}{v_{\rm r}}\right) = \delta_{\rm D}\left(v_{\rm r}-\frac{2R_{\rm E}u_{\rm T}\cos\theta}{\hat{t}}\right)\frac{v_{\rm r}^2}{2R_{\rm E}u_{\rm T}\cos\theta},
\end{equation}
and integrating over $\alpha$ and $v_{\rm r}$, we obtain
\begin{eqnarray}
 \frac{\drm\Gamma}{\drm \hat{t}} =
  \frac{\Omega_{\rm PBH}}{\Omega_{\rm DM}}\int_0^{\ds}\!\!
  \drm\dl\int_{-\pi/2}^{\pi/2}\!\!
  \drm\theta\, 
    \frac{\rho_{\rm DM}(\dl)}{M_{\rm PBH}v_{\rm c}^2}
v_{\rm  r}^4\,  \exp\left[-\frac{v_{\rm r}^2}{v_{\rm c}^2}\right]\, ,
\end{eqnarray}
with $v_{\rm r}=2R_{\rm E}u_{\rm T}\cos\theta/\hat{t}$. One can rewrite
this equation by changing variable $\theta$ to the minimum impact
$u_{\rm min}=u_{\rm T}\sin\theta$, such that, $\mathrm{d}\theta = \mathrm{d}u_{\rm
min}/\sqrt{u_{\rm T}^2-u_{\rm min}^2}$. This results in
\begin{equation}
 \frac{\drm\Gamma}{\drm \hat{t}} =2
  \frac{\Omega_{\rm PBH}}{\Omega_{\rm DM}}
  \int_0^{\ds }\!\!\drm\dl \int_0^{u_T}\!\!
  \frac{\drm u_{\rm min}}{\sqrt{u_{\rm T}^2-u_{\rm min}^2}}
  \frac{\rho_{\rm DM}(\dl)}{M_{\rm PBH}
  v_{\rm c}^2} v_{\rm r}^4\exp\left[-\frac{v_{\rm r}^2}{v_{\rm c}^2}\right]\, ,
\label{eq:dgamma_dte}
\end{equation}
where $v_{\rm r}=2R_E \sqrt{u_{\rm T}^2-u_{\rm min}^2}/\hat{t}$. To
compute the event rate due to PBHs in both the halo regions of MW and
M31, we sum the contributions, $\mathrm{d}\Gamma=\mathrm{d}\Gamma_{\rm
MW}+\mathrm{d}\Gamma_{\rm M31}$. 
As we described above, we can express
the centric radius of each halo, $r$, entering into $v_{\rm c}(r)$, in
terms of the distance to the lensing PBH, $d$; $r=r(\dl)$.  Unless
explicitly stated, we will employ $u_{\rm T}=1$ as our default choice.
Note that the expected event number shown in Fig.~\ref{fig:eventrate}, $\mathrm{d}N_{\rm event}/\mathrm{d}\ln t_{\rm FWHM}$, is related to $\mathrm{d}\Gamma$ via
$\mathrm{d}\Gamma/\mathrm{d}\ln{t}_{\rm FWHM}=\mathrm{d}^2N_{\rm event}/\mathrm{d}\ln t_{\rm FWHM}\mathrm{d}t_{\rm obs}$, where $t_{\rm obs}$ is the unit observation time [hours].
Since our observation was done for 7 consecutive hours within one night under the similar weather conditions, $\mathrm{d}N_{\rm event}/\mathrm{d}\ln t_{\rm FWHM}=7\times \mathrm{d}\Gamma/\mathrm{d}\ln t_{\rm FWHM}$.

\begin{figure*}
\centering
\includegraphics[width=0.48\textwidth]{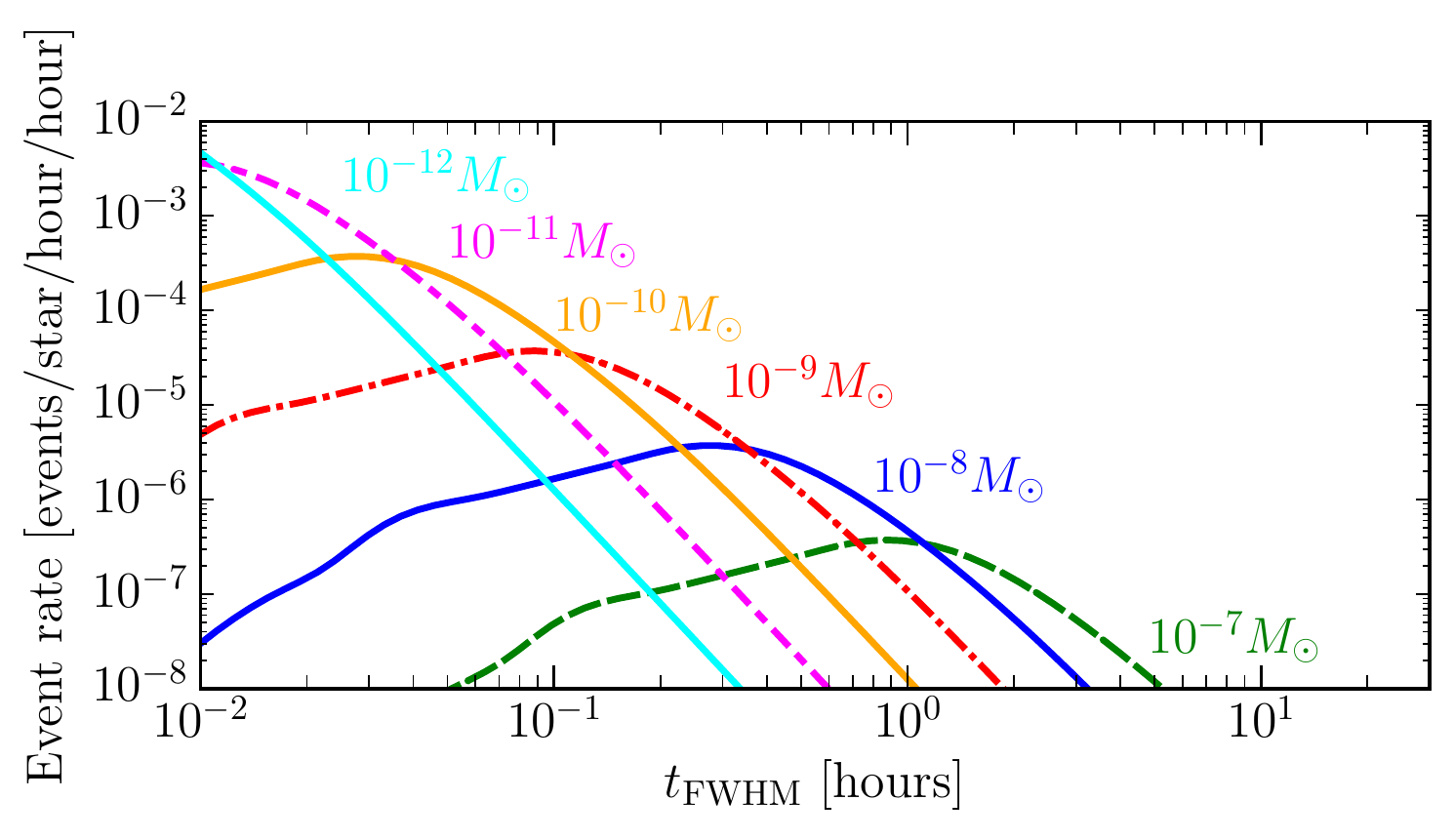}
 \includegraphics[width=0.48\textwidth]{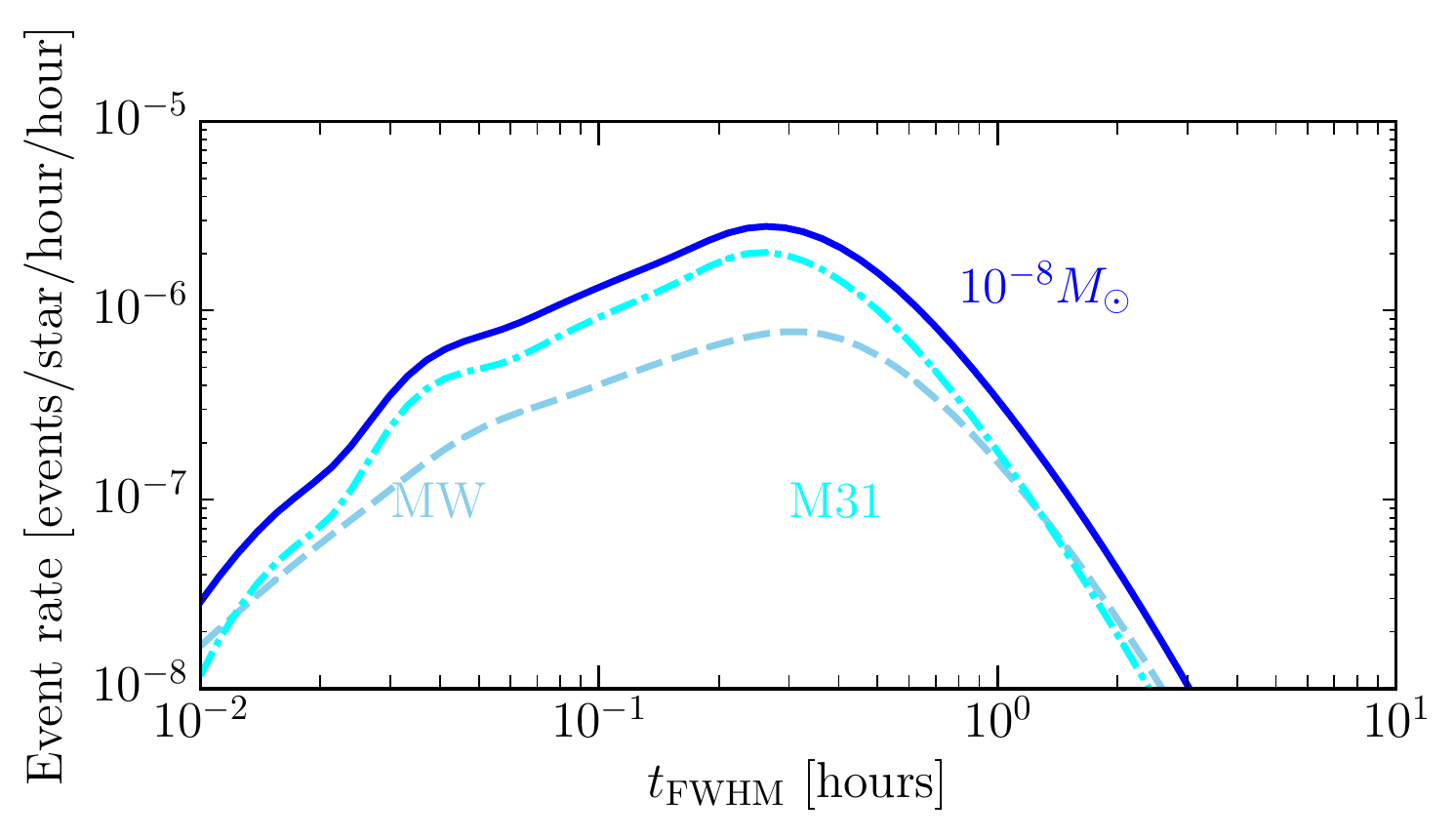}
\caption{The differential event rate of PBH microlensing for a single
 M31 star (Eq.~\ref{eq:dgamma_dte}); the rate per unit observation time
 (hour), per a single source star in M31, and per unit timescale of the
 microlensing light curve (hour) for PBHs of a given mass scale.  Here
 we assumed that all the DM in the MW and M31 halo regions is made of
 PBHs; $\Omega_{\rm PBH}/\Omega_{\rm DM}=1$. The $x$-axis is the
 full-width-half-maximum (FWHM) timescale of microlensing light curve.
 The lighter or heavier PBH has a shorter or longer timescale of
 microlensing light curve. The right panel shows the relative
 contribution to the microlensing event rate due to PBHs in either MW or
 M31 halo region, for the case of $M_{\rm PBH}=10^{-8}M_\odot$. }
 \label{fig:m31_eventrate}
\end{figure*}

Fig.~\ref{fig:m31_eventrate} shows the expected event rate for the PBH
microlensing, computed using Eq.~(\ref{eq:dgamma_dte}). Here 
we show the
event rate as a function of the full-width-half-maximum (FWHM) timescale
of the light curve, which matches our search of microlensing events from
the real HSC data. If a PBH is in the mass range
$M=[10^{-12},10^{-7}]M_\odot \simeq 2\times [10^{21},10^{26}]~{\rm g}$,
it causes the microlensing event that has a typical timescale in the
range of $[10^{-1},1]~$hour. The lighter or heavier PBHs tend to cause a
shorter or longer timescale event.  The event rate is quite high up to
$10^{-4}$
for a microlensing timescale with $[0.1,1]~$hours.
That is, if
we take about 10 hours observation and observe $10^{8}$ stars at once
for each exposure, we expect many events up to $
10^4$ events (because $10^{-4}\times 10~[{\rm
hour}]\times 0.1~[{\rm hour}]\simeq 10^4$),
assuming
that such
PBHs constitute a majority of DM in the intervening space bridging MW
and M31. The right figure shows that the PBHs in the M31 halo region
give a slightly larger contribution to the event rate, because we assumed a
larger halo mass for M31 than that of MW. 
Thus the high-cadence HSC observation of M31 is suitable for searching
for microlensing events of PBHs.

\subsection{Light curve characterization in pixel lensing regime}
\label{sec:pixellens}

As we described above, the timescale for the PBH and M31 star
microlensing system is typically several tens of minutes for a PBH with
$10^{-8}M_\odot$. However, there is an observational challenge. Since the
M31 region is such a dense star field, fluxes from multiple stars are
overlapped in each CCD pixel ($0.168^{\prime\prime}$ pixel scale for 
HSC/Subaru). In other words individual stars are not resolved even with
the Subaru angular resolution (about $0.6^{\prime\prime}$ for the seeing
size).  Hence we cannot identify which individual star in M31 is
strongly lensed by a PBH, even if it occurs. Such a microlensing of
unresolved stars falls in the ``pixel microlensing'' regime
\cite{Gould:96} (also see \cite{CalchiNovati:10} for a review).

To identify microlensing events in the pixel microlensing regime requires
elaborate data reduction techniques. In this paper, we use the image
subtraction or image difference technique first described in
Ref.~\cite{AlardLupton:98}. The image difference technique allows us to search for
variable objects including candidate stars that undergo microlensing by PBHs.
In brief, starting with the time sequenced $N_{\rm exp}$ images of M31, the
analysis proceeds as follows. (i)
We generate a reference image by co-adding
some of the best-seeing images in order to gain a higher signal-to-noise. Next
we subtract this reference image from each of the $N_{\rm exp}$ images after
carefully matching their point spread functions (PSFs) as described in
Ref.~\cite{AlardLupton:98}. (ii) We search for candidate variable objects that show
up in the difference image.  In reality, if the image subtraction is imperfect,
the difference image would contain many artifacts, as we will discuss
further.  (iii) Once secure variable objects are detected, we determine the position
(RA and dec) of each variable object in the difference image. We perform PSF
photometry for each variable candidate using the PSF center to be at the
position of the candidate in the difference image. By repeating the PSF
photometry in each difference image of the $N_{\rm exp}$ images, we can measure
the light curve of the candidate as a function of the observation time.

The light curve of a microlensing event obtained using the PSF flux in the
difference image at time $t$, obtained as described above, can be expressed as
\begin{equation}
 \Delta F(t)=F_0\left[A(t)-A(t_{\rm ref})\right],
  \label{eq:DeltaF}
\end{equation}
where $\Delta F(t)$ is the differential flux of the star at time $t$
relative to the reference image, $F_0$ is the intrinsic flux, $A(t)$ is
the lensing magnification at $t$ and $A(t_{\rm ref})$ is the
magnification at the time of the reference image, $t_{\rm ref}$. In the
above equation, $\Delta F(t)$ is a direct observable, and others ($F_0,
A(t), A(t_{\rm ref})$) are parameters that have to be modeled.

As can be seen from Eq.~(\ref{eq:cano}), the light curve for the microlensing
of a point source by a point mass can be characterized by two parameters. The
first parameter is the maximum amplification $A_0=A(u_{\rm min})$ when the
lensing PBH is closest to a source star on the sky, where $u_{\rm min}$ is the
impact parameter relative to the Einstein radius $R_E$ ($u_{\rm min}$ is
dimension-less). The second one is the timescale of the light curve, which
depends on the Einstein radius as well as the transverse velocity of the PBH
moving across the sky. For the timescale parameter we use the FWHM timescale
of the microlensing light curve, $t_{\rm FWHM}$, instead of $t_E$, defined as
\begin{eqnarray}
A\left(\frac{t_{\rm FWHM}}{2}\right)-1\equiv\frac{A_0-1}{2}.
\end{eqnarray}
Thus the light curve of microlensing can be fully modeled by the three
parameters, $F_0$, $u_{\rm min}$ and $t_{\rm FWHM}$.
In the following we will use the three parameters when performing a
fitting of the microlensing model to the observed light curve of
microlensing candidate in the image difference. Note that the use of
$t_{\rm FWHM}$, instead of $t_E$, gives slightly less degenerate
constraints on the parameters \cite{Gondolo:99}.

\section{Data Analysis and Object Selection}
\label{sec:obs2}

\subsection{Observations}
\label{sec:obsm31}

\begin{figure*}[t]
\centering
\includegraphics[width=0.6\textwidth]{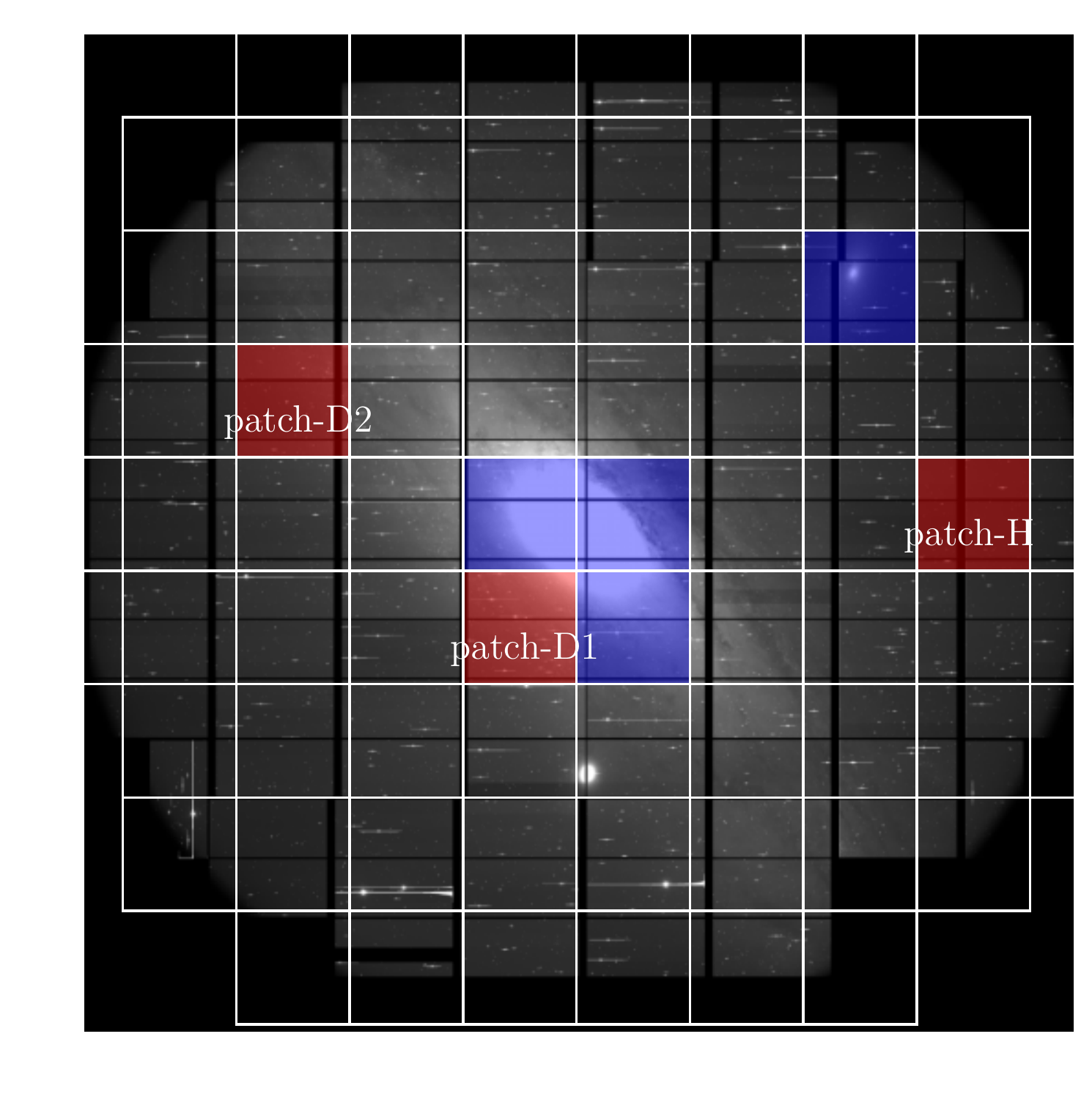}
 \caption{The background image of M31 shows configuration of 104 CCD
 chips of the Subaru/HSC camera. The white-color grids are the HSC
 ``patch'' regions. The patches labeled as ``patch-D1'', ``patch-D2''
 and ``patch-H'' are taken from representative regions of the disk
 region closer to the central bulge, the outer disk region and the halo
 region, respectively, which are often used to show example results of
 our data processing in the main text. The dark-blue regions are the
 patches we exclude from our data analysis due to too dense star fields,
 where fluxes from stars are saturated and the data are not properly
 analyzed.}
\label{fig:m31ccdpatch}
\end{figure*}
The HSC camera has 104 science detectors with a pixel scale of $0.168^{\prime
\prime}$ \cite{2018PASJ...70S...1M}.
The 1.5 degree diameter FoV of HSC enables us to
cover the entire region of M31, from the inner bulge to the outer disk
and halo regions with a single pointing. The pointing is centered at the
coordinates of the M31 central region: (RA, dec) = (00h 42m 44.420s,+41d
16m 10.1s). We do not perform any dithering between different exposures in order
to compare stars in the same CCD chip, which makes the image difference
somewhat easier. However, in reality the HSC/Subaru system has some subtle
inaccuracies in its auto-guidance and/or pointing system. This results in
variations in the pointings of different exposures, typical variations range
from a few to a few tens of pixels. 

Fig.~\ref{fig:m31ccdpatch} shows the configuration of the 104 CCD chips
relative to the image of M31 on the sky.  The white-color boxes denote
locations of HSC ``patches'', which are convenient tessellations of the HSC FoV.
The image subtraction and the search of microlensing events will be done on
a patch-by-patch basis. The patches labeled ``patch-D1'', ``patch-D2'' and
``patch-H'' denote the regions that represent inner and outer disk regions (-D1
and -D2) and a halo (-H) region, respectively. These representative regions
will be used to show how the results vary in the different regions. 

\begin{figure}[ht]
\centering
\hspace*{-1em} \includegraphics[width=0.5\textwidth]{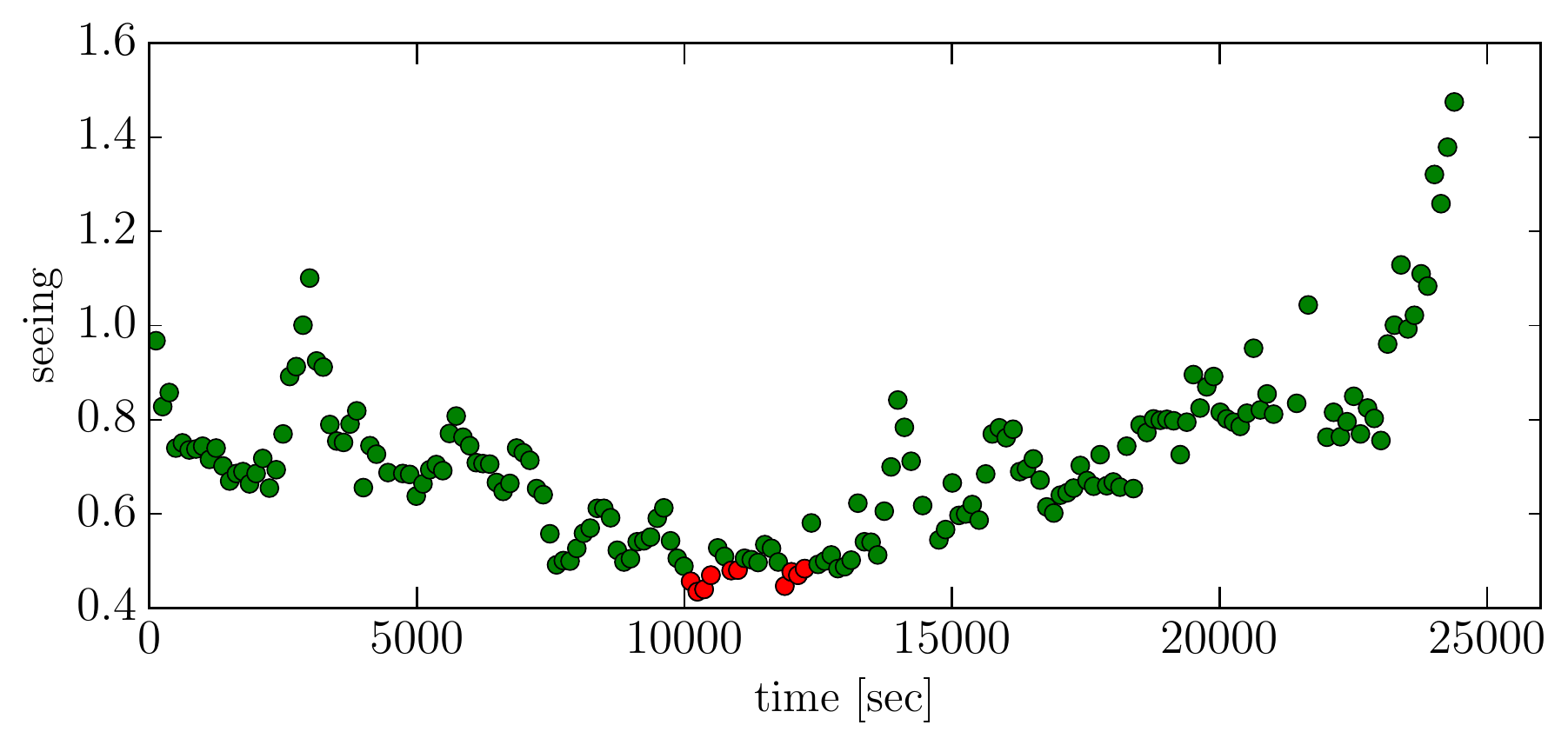}
 \caption{The PSF FWHM (seeing size) of each exposure (90~sec exposure
 each) as a function of time $t~$[sec] from the start of our
 observation. We took the images of M31 region every 2~min (90~sec
 exposure plus about 35~sec for readout), and have 188 exposures in
 total. The red points show the images of 10 best-seeing epochs ($\sim
 0.45^{\prime \prime}$) from which the
 reference image, used for the image difference, was
 constructed. \label{fig:seeing}}
\end{figure}
Our observation was conducted on November 23, 2014 which was a dark night, a
day after the new moon. In total, we acquired 194 exposures of M31 with the HSC
$r$-band filter\footnote{See
\url{http://www.naoj.org/Projects/HSC/forobservers.html} for the HSC
filter system}, for the period of about 7 hours, until the elevation of M31
fell below about $30$ degrees. We carried out the observation with a
cadence of 2~minutes, which allows us to densely sample the light curve for
each variable object. The total exposure time was 90~seconds on source and
about 35 seconds were spent for readout on average.  The weather was excellent
for most of our observation as can be seen from Fig.~\ref{fig:seeing}, which
shows how the seeing FWHM changed with time from the start of our observation.
The seeing size was better than $0.7^{\prime\prime}$ for most of the
observation period, with a best seeing FWHM of about $0.4^{\prime \prime}$
at $t\sim 10,000~$sec (2.8~hours). However, the seeing got worse than
$1^{\prime\prime}$ towards the end of our observation. We exclude 6 exposures
which had seeing FWHM worse than $1.2^{\prime\prime}$ and use the remaining 188
exposures for our science analysis.

We also use the $g$- and $r$-band data, which were taken during the
commissioning run on June 16 and 17 in 2013, respectively, in order to
obtain color information of stars as well as to test a variability of
candidates at different epochs. The $g$-band data consist of
$5\times 120~{\rm sec}$ exposures and $5\times 30~{\rm sec}$ exposures
in total, while the $r$-band data consists of $10\times 120~{\rm sec}$
exposures.

\subsection{Data reduction and Sample selection}
\label{sec:datareduce}

\subsubsection{Standard data processing}
\label{sec:stadard_dataprocess}

We performed basic standard data reduction with
the dedicated software package for HSC, {\tt hscPipe} (version
3.8.6; also see Bosch et al. \cite{Boschetal:17}), which is being developed
based on the Large Synoptic Survey Telescope software package
\cite{Ivezicetal:08,Axelrodetal:10,Juricetal:15} \footnote{Also see
\url{http://www.astro.princeton.edu/~rhl/photo-lite.pdf} for details of
the algorithm used in the pipeline.}. This pipeline performs a number of common
tasks such as bias subtraction, flat fielding with dome flats, coadding,
astrometric and photometric calibrations, as well as source detection and
measurements.

After these basic data processing steps, we subtract the background
contamination from light diffusion of atmosphere and/or unknown
scattered light. However the background subtraction is quite
challenging for the M31 region, because there is no blank region and
every CCD chip is to some extent contaminated by unresolved, diffuse
stellar light. To tackle this problem, we first divide each CCD chip
into different meshes (the default subdivision is done into 64 meshes in
each CCD chip). We then employ a higher-order polynomial fitting to
estimate a smooth background over different meshes.  We employed a
10-th order polynomial fitting for the CCD chips around the bulge
region, which are particularly dense star regions. For other CCD chips,
we use a 6-th order polynomial fitting scheme. However, we found
residual systematic effects in the background subtraction, so we
will further use additional correction for photometry of the difference
image, as we will discuss later.

For our study, accurate PSF measurements and accurate astrometric
solutions are crucial, because those allow for an accurate subtraction of
different images.  The pipeline first identifies brightest star objects
($S/N\simgt 50$) to characterize the PSF and do an initial astrometric
and photometric calibration. From this initial bright object catalog, we
select star candidates in the size and magnitude plane for PSF
estimation (see Ref.~\cite{Boschetal:17} for details). The selected stars are fed
into the PSFEx package \cite{Bertin:11} to determine the PSF as a
function of positions in each CCD chip. The functional form of the PSF
model is the native pixel basis and we use a second-order polynomial per
CCD chip for the spatial interpolation. For the determination of the astrometry, we used a
30~sec calibration image that we took at the beginning of our
observation, where bright stars are less saturated. We obtain an
astrometry solution after every 11 images, 30~sec calibration frame plus 10
time-consecutive science exposures, by matching the catalog of stars to
the Pan-STARRS1 system
\cite{Schlaflyetal:12,Tonryetal:12,Magnieretal:13}.  The HSC pipeline
provides us with a useful feature, the so-called ``hscMap'', which
defines a conversion of the celestial sphere to the flat coordinate
system, ``hscMap coordinate'', based on a tessellation of the sky.  In
Fig.~\ref{fig:m31ccdpatch} the white-color regions denote the hscMap ``patch''
regions. We perform image difference separately on each patch. Due to too many
saturated stars in the bulge region and M101, we exclude the patches, marked by
dark blue color, from the following analysis.

\subsubsection{Image subtraction and object detection}
\label{sec:imagediff}

In order to find variable objects, we employ the difference
image technique developed in Alard \& Lupton (1998; \cite{AlardLupton:98}) (also see Ref.~\cite{Alard:00}), 
which is integrated into the HSC pipeline.  To do
this, we first generated the ``reference'' image by co-adding 10
best-seeing images among the 188 exposure images, where the 10 images
are not time-consecutive (most of the 10 images are from images around
about 3 hours from the beginning of the observation, as shown in
Fig.~\ref{fig:seeing}). We use the mean of the 10 images as the
observation time of the reference image, $t_{\rm ref}$, which is needed
to model the microlensing light curve (Eq.~\ref{eq:DeltaF}).

In order to make a master catalog of variable object candidates, we
constructed 63 target images by co-adding 3 time-consecutive images from
the original 188 exposure images. A typical limiting magnitude is about
26~mag (5$\sigma$ for point sources), and even better for images where seeing
is good (see below). When subtracting the reference image from each target
image, the Alard \& Lupton algorithm uses a space-varying convolution kernel to
match the PSFs of two images. The optimal convolution kernel is derived by
minimizing the difference between convolved PSFs of two images.  A variable
object, which has a flux change between the two images, shows up in the
difference image.

Fig.~\ref{fig:ex_diffimage} shows an example  of the image subtraction performed
by the pipeline. Even for a dense star region in M31, the pipeline properly
subtracts the reference from the target image, by matching the PSFs
and astrometry. A point source which undergoes a change in its flux shows up in
the difference image, as seen in the right panel. In this case, the candidate
appears as a black-color point source meaning a negative flux, because it has a
fainter flux in the target image than in the reference image.

We detect objects in the difference image each of which is defined from
a local minimum or maximum in the difference image, where we used
$5\sigma$ for the PSF magnitude as detection threshold. The pipeline
also measures the center of each object and the size and ellipticity
from the second moments. In this process we discarded objects that have
ill-defined center, a saturated pixel(s) in the difference and/or
original image or if the objects are placed at a position within 50
pixels from the CCD edge.

\subsubsection{PSF photometry and master catalog of variable star candidates}

For each variable star candidate, we obtain PSF photometry in the difference
image to quantify the change of flux. We allow negative PSF fluxes for
candidates that have fainter flux in the target image than in the
reference image. Since the photon counts in each CCD pixel is generally
contaminated by multiple stars in most of the M31 regions, we often find
a residual coherent background (large-scale modulated background) in
each difference image, due to imperfect background subtraction in the
original image. To avoid contamination from such a residual
background, we first measure the spatially constant background from the
median of counts in $41\times 41$ pixels around each object in the
postage-stamp image, and then subtract this background from the image.
Then we perform the PSF-photometry counts in ADU units taking the PSF
center to be at the candidate center. Hereafter we sometimes refer to PSF
magnitude in the difference image as ``PSF counts''. The pipeline also
estimates noise in each pixel assuming the background limit (Poisson
noise), and gives an estimation of the noise for the PSF photometry
(see Eqs.~14 and 15 in Ref.~\cite{Great3} for the similar definition). However, the noise estimation involves a non-trivial
propagation of Poisson noise in the image difference procedures, so we
will use another estimate for the PSF photometry error in each patch, as
described below.

In the following we focus on the PSF photometry counts in ADU units in the
difference image, rather than the magnitude, because it is a direct
observable in our analysis.  However, we will also need to infer the magnitude of each
candidate; for example, to estimate the luminosity function of source stars in
each magnitude bin or to plot the light curve of variable star candidates in units
of the magnitude.  In this case we estimate the magnitude of an object in the
$i$-th target image, $m_i$, based on
\begin{equation}
m_i = -2.5\log\left(\frac{C_{{\rm diff},i}+C_{\rm
         ref}}{F_{0,i}} \right),
\label{eq:convert_m}
\end{equation}
where $C_{\rm diff},i$ is the PSF flux for the object in the difference
image of the $i$-th target image, $C_{\rm ref}$ is the PSF flux of the
reference image at the object position, and $F_{0,i}$ is the zero-point
flux in the $i$-th image. Note that the counts of the reference image
$C_{\rm ref}$ can be contaminated by fluxes from neighboring stars, so
the above magnitude might not be accurate.

\begin{figure*}[t]
\centering \includegraphics[width=7.5cm,clip]{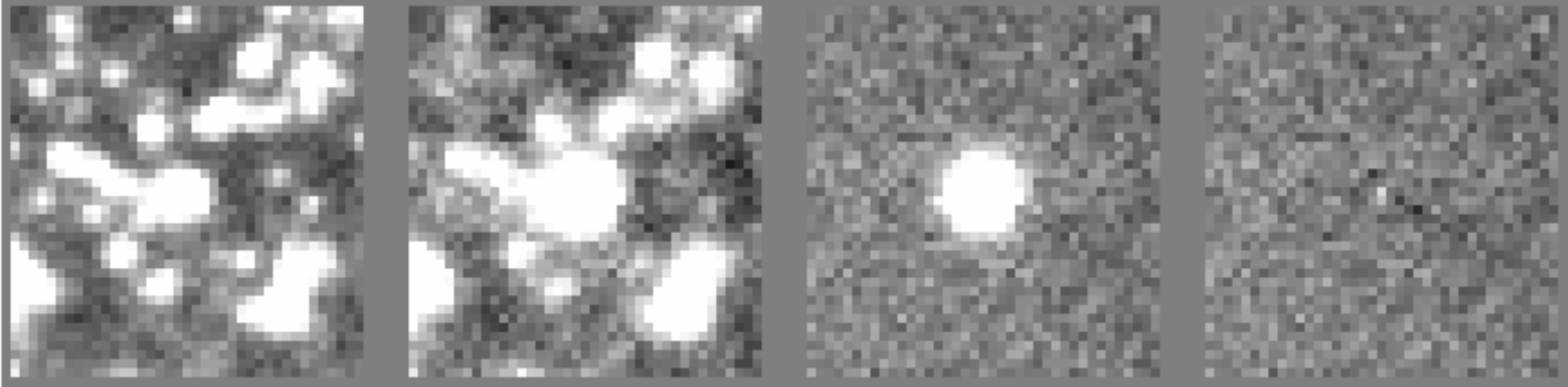}
\includegraphics[width=7.5cm,clip]{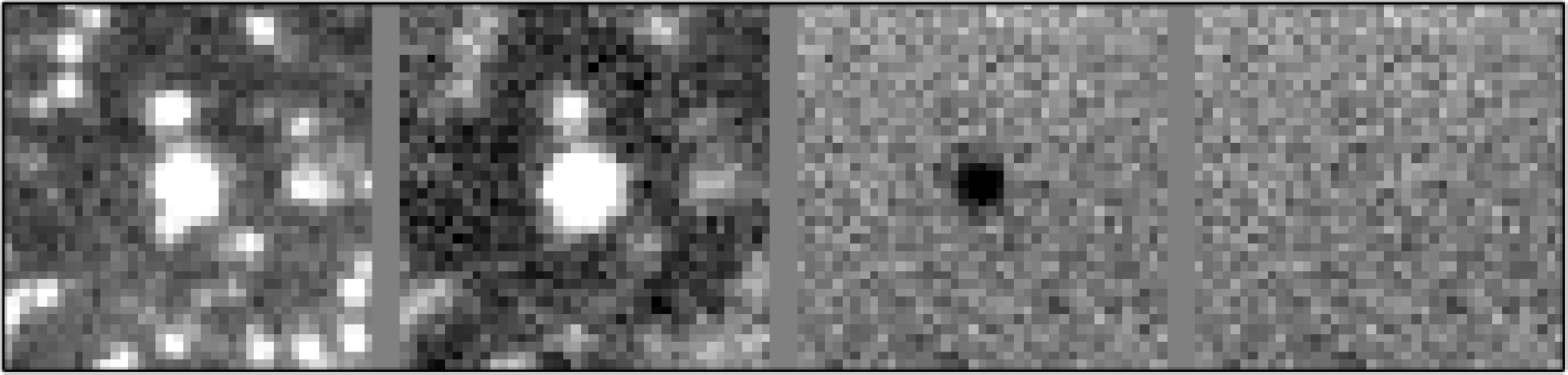}
\includegraphics[width=7.5cm,clip]{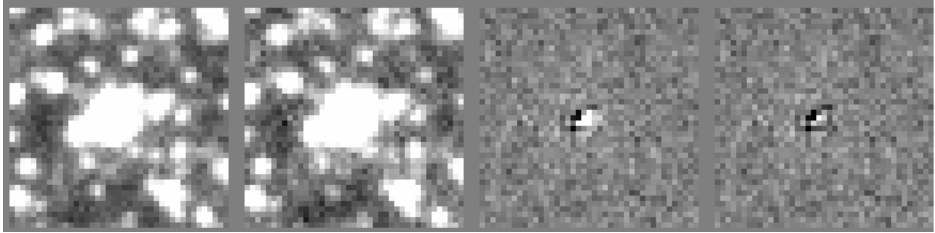}
\includegraphics[width=7.5cm,clip]{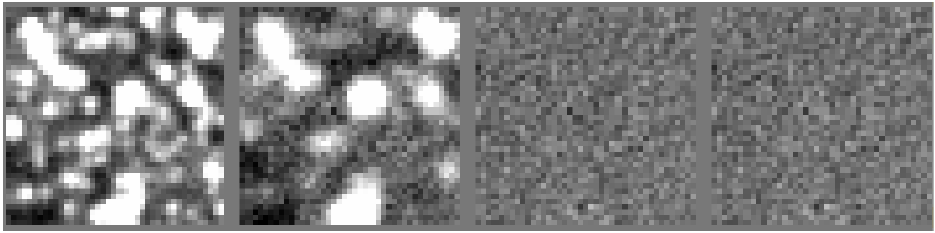}
\includegraphics[width=7.5cm,clip]{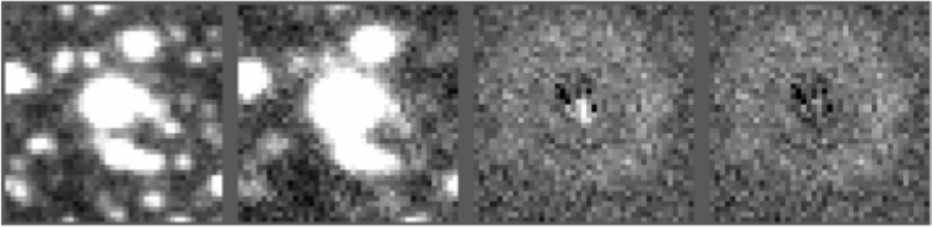}
 \includegraphics[width=7.5cm,clip]{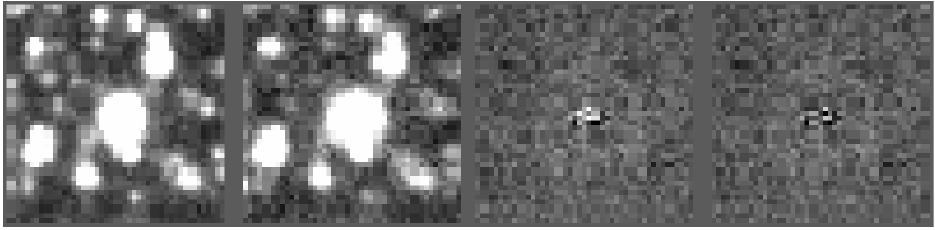}
 \caption{Examples of detected objects in the difference image, which
 pass or do not pass the selection criteria to define a master catalog
 of variable star candidates (see text for details). Each panel shows 4
 postage-stamp images: the leftmost image is the reference image (the
 coadded image of 10 best-seeing exposures), the 2nd left is the target
 image (the coadded image of 3 time-consecutive exposures), the 3rd
 image is the difference image between the reference and target images,
 and the rightmost image is the residual image after subtracting the
 best-fit PSF image from the difference image at the object position.
 The two objects in top raw are successful candidates that passed all
 the selection criteria: the left-panel object has a brighter flux in
 the target image than in the reference image, while the right-panel
 object has a fainter flux (therefore appear as a black-color image with
 negative flux).  The lower-row objects are removed from the catalog
 after the selection criteria.  The objects in the middle row are
 excluded because the object is either smaller or larger
 than the PSF size.  The left object in the bottom row is
 excluded because it has a too large ellipticity than PSF. The right
 object is excluded because of too large residual image.  }
 \label{fig:fake}
\end{figure*}
From the initial catalog constructed from the $5\sigma$ candidates from
the 63 coadded images, we prune it down to a {\it master} catalog of ``secure''
variable star candidates by applying the following criteria:
 \begin{itemize}
  \item{\it PSF magnitude threshold} -- A candidate should have a PSF
       magnitude, with a detection significance of $5\sigma$ or higher (including a negative flux), in
       any of the 63 difference images.
  \item{\it Minimum size} -- The size of the candidate should be
      greater than 0.75 times the PSF size of each difference image.
 \item{\it Maximum size} -- The size of the candidate should be smaller
      than 1.25 times the PSF size.  \item{\it Roundness} -- The candidate
          should have a round shape. We require our candidates to have an
      axis ratio greater than 0.75, as the PSF does not show extreme axis ratios.
  \item{\it PSF shape} -- We impose that the shape of an object should be
       consistent with the PSF shape. The residual image,
       obtained by
       subtracting a scaled PSF model from the candidate image in
       the difference image, should be within $3\sigma$ for the
       cumulative
      deviation over pixels inside the PSF aperture.
\end{itemize}
Fig.~\ref{fig:fake} shows examples of objects that pass or fail the above criteria.
Note that the above conditions are broad enough in order for us not to miss a
real candidate of microlensing if it exists.  We make a master catalog of
variable star candidates from objects that pass all the above conditions as well as
are detected in the image difference at least twice in the 63 difference images
at the same position within 2 pixels. These criteria result in 15,571 candidates
of variable objects, which is our master catalog of variable star candidates.

\subsubsection{Light curve measurement}

\begin{figure*}[t]
\centering \includegraphics[width=7.5cm,clip]{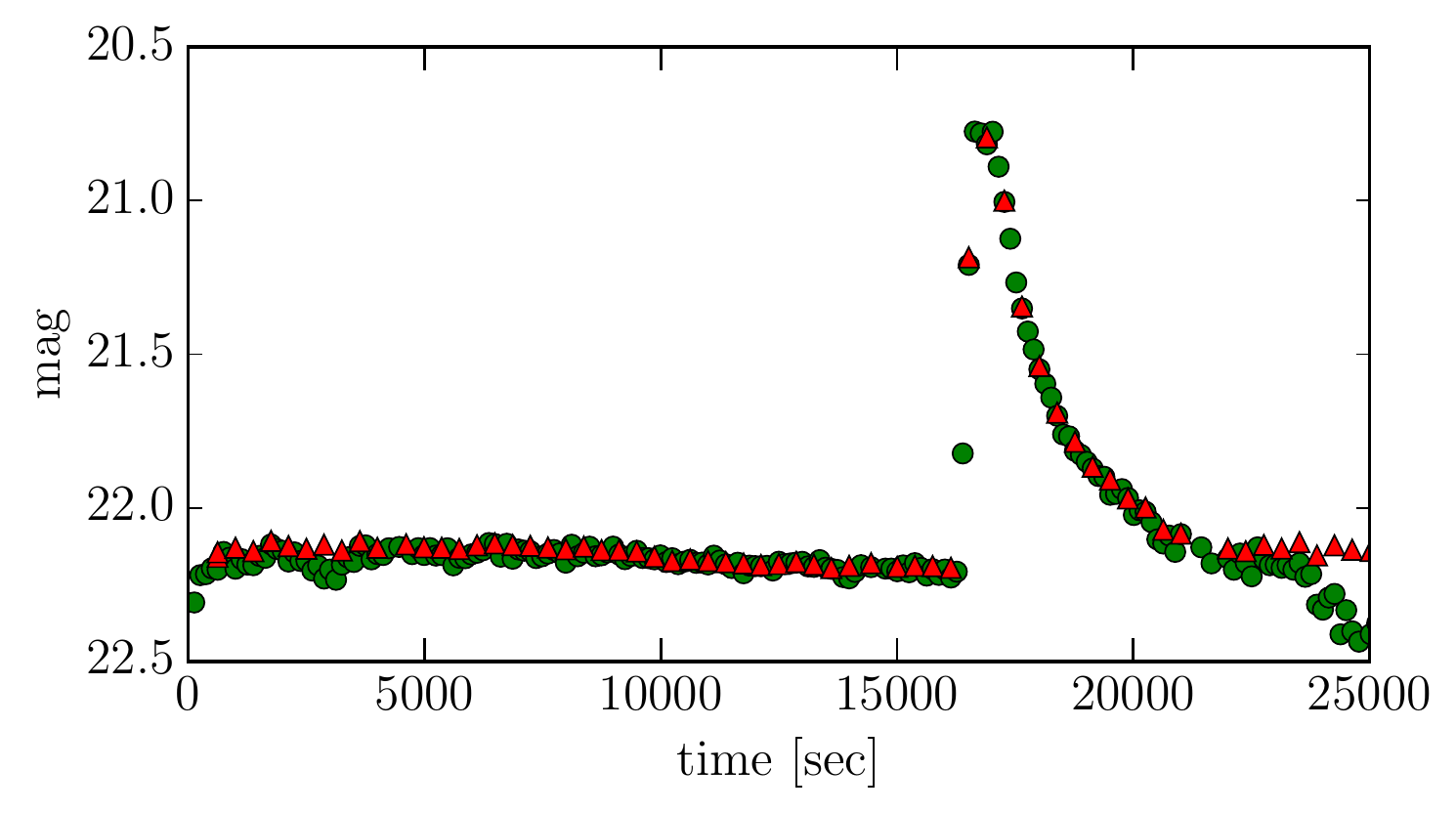}
\includegraphics[width=7.5cm,clip]{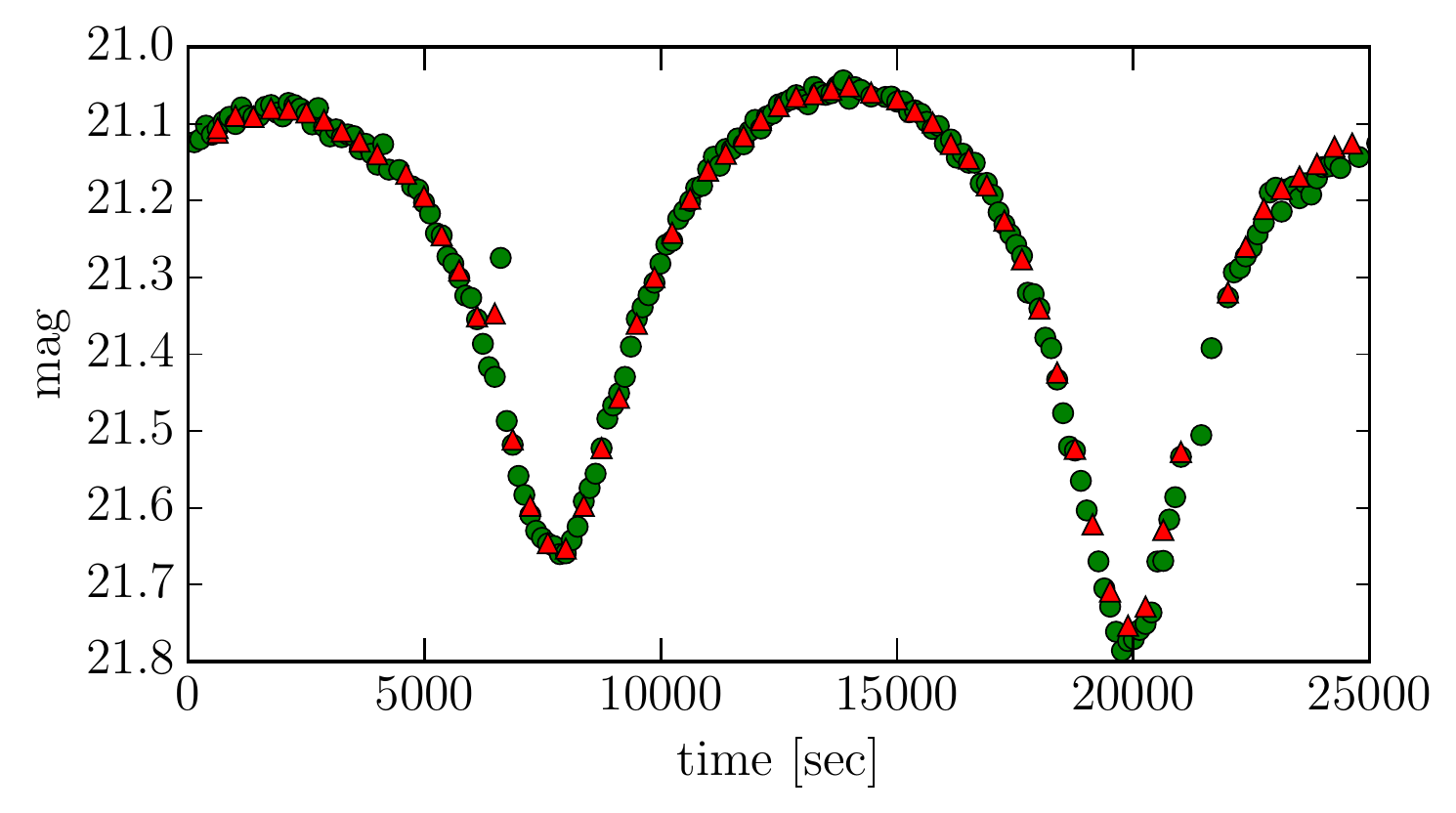}
\includegraphics[width=7.5cm,clip]{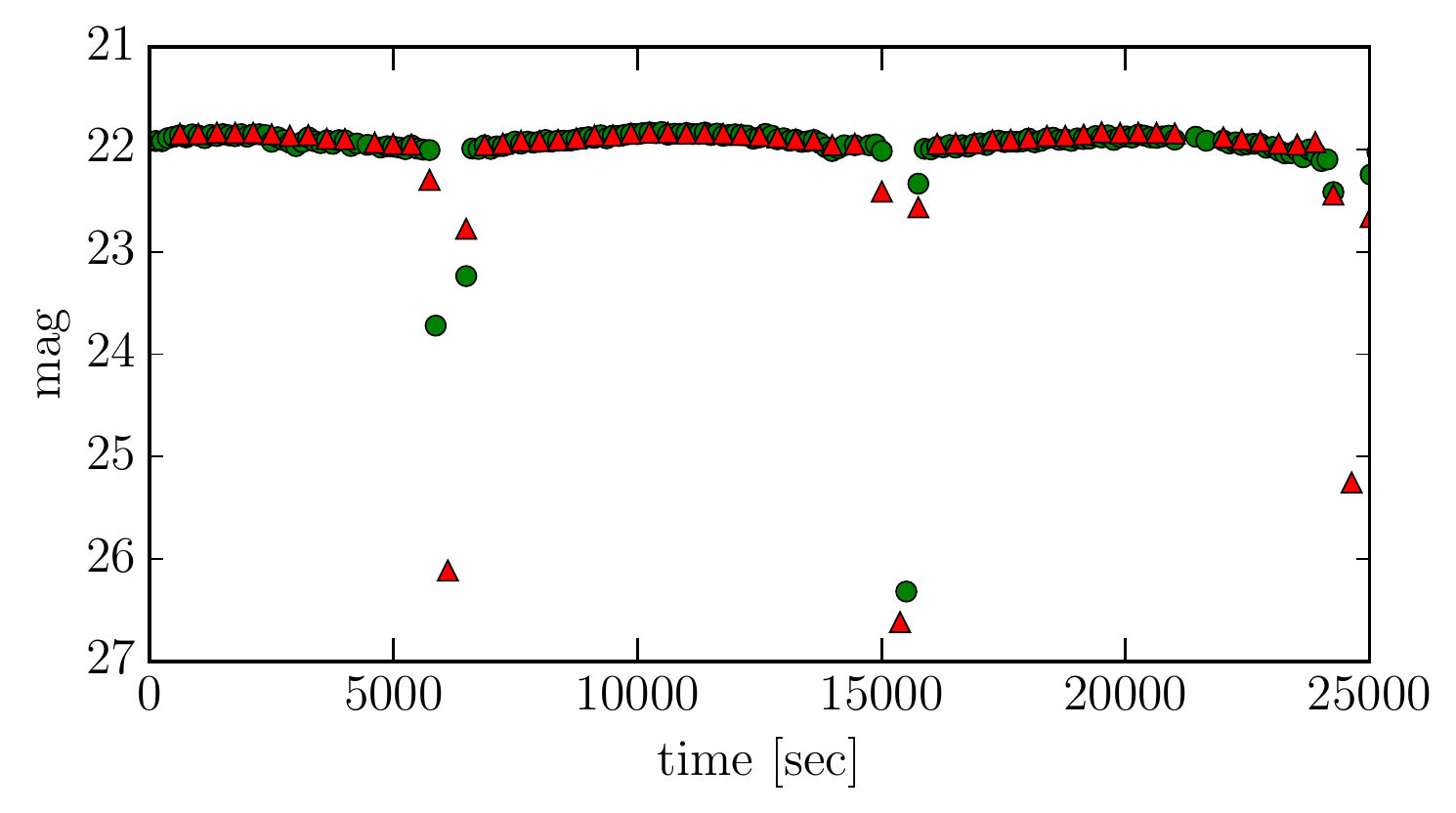}
\includegraphics[width=7.5cm,clip]{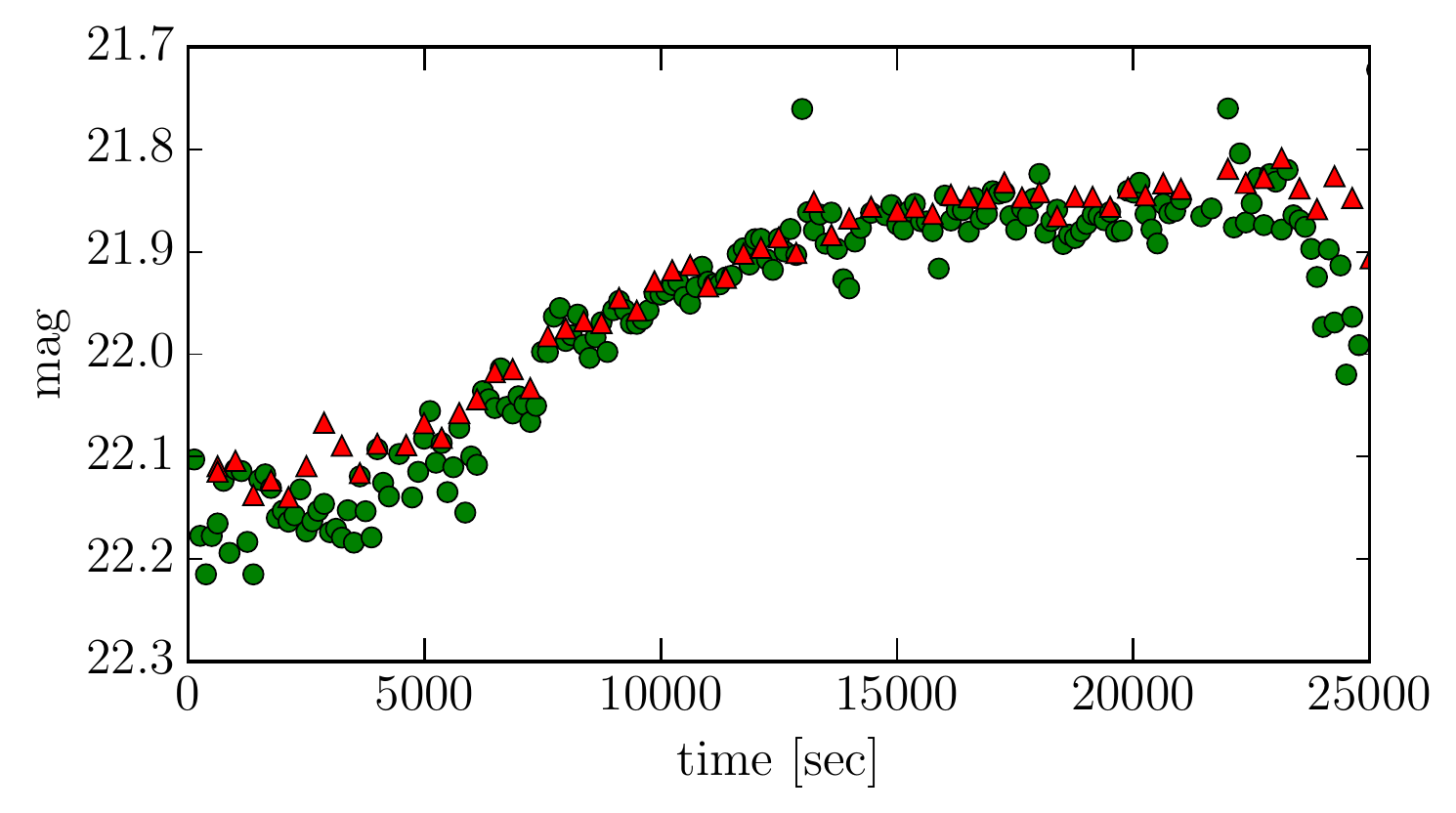}
 \caption{Examples of light curves for real variable stars identified
 in our method. The green-circle data points show the light curve
 sampled by our original data of 2~min sampling rate, while the
 red-triangle points are the light curve measured from the coadded data
 of 3 time-consecutive exposures (therefore 6~min cadence) (see text for
 details). {\it Upper left}: candidate stellar
 flare. When converting the magnitude from the counts in the difference
 image at each observation time, we used Eq.~(\ref{eq:convert_m}). Note
 that the estimated magnitude might be contaminated by fluxes of
 neighboring stars in the reference image. {\it Upper right}: 
 candidate contact binary stars. {\it Lower left}: the eclipse binary
 system, which is probably a system of white dwarf and brown dwarf,
 because one star (white dwarf) has a total eclipse over about 10~min
 duration, and then the eclipse has about 3~hours period. {\it Lower
 right}: candidate variable star, which has a longer period than
 our observation duration (7~hours).  } \label{fig:lc_real}
\end{figure*}
Once each candidate is identified, we measure the PSF counts in each of
the 63 difference images. This allows us to measure the light curve with a 6 min
resolution, as a function of time from the beginning to the end of our 7
hour long observations. In order to restore the highest time resolution of our
data, we then used each of 188 exposures and measured the PSF counts in
each of the 188 difference image that was made by subtracting the
reference image (the coadded image of 10 best-seeing exposures) from
every single exposure. Here we used the same position of candidate as used
in the 63 images. In this way we measure the light curve of the object
with 2~min time resolution. 

Fig.~\ref{fig:lc_real} shows the light curves for examples of real variable
stars. Note that we converted the PSF counts of each candidate in the
difference image to the magnitude based on Eq.~(\ref{eq:convert_m}). However,
the magnitude might be contaminated by fluxes from blended stars surrounding
the candidate star.  The figure demonstrates our ability to properly sample the light
curves with high time resolution. Thus  the difference
image technique works well and can identify variable star candidates as well as
measure their light curves.

\begin{figure*}[t]
 \centering
\hspace*{-5em}\includegraphics[width=20.cm,clip]{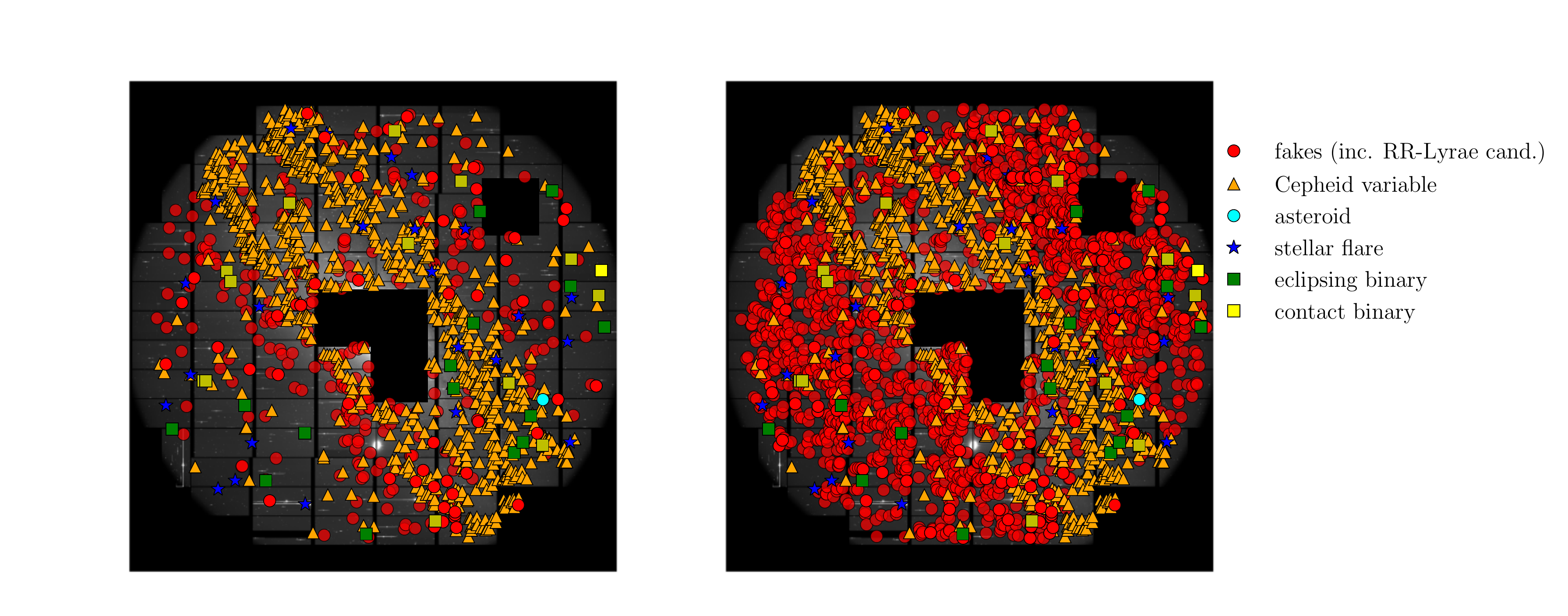}
 %
 \caption{Distribution of secure variable star candidates, 
 detected from our analysis using the image difference technique. The
 different symbols denote different types of candidates classified based
 on the shapes of their light curves. Here we exclude other non-secure
 candidates that are CCD artifacts and impostors near to the CCD edge
 or bright stars.  The left panel shows the distribution for the
 candidates with magnitudes $m_r\le 24~$mag, while
 the right panel shows the
 candidates at $m_r\le 25~$mag. The number of candidates are 1,334 and
 2,740, respectively. 
 } \label{fig:dist_candidates}
\end{figure*}
Fig.~\ref{fig:dist_candidates} shows the distribution of secure
variable star candidates detected in our analysis over the HSC
field-of-view, for candidates with magnitudes $m_r\le 24$~ and 25~mag
in the left and right panel, respectively. To estimate the magnitude of
each candidate, we used the PSF magnitude of the candidate in the
reference image. Based on the shape of the light curve for each candidate,
we visually classified the candidates in different types of variable stars;
i) stellar flares, ii) eclipsing or contact binary systems, iii) asteroids
(moving object), iv) Cepheid variables if the candidates appear to have
a longer period than our observation duration (7~hours), and v)
``impostors''. Here impostors are those candidates which show time variability only
when the seeing conditions are as good as $\simlt 0.6^{\prime\prime}$. Since
such good-seeing data is deeper as found from Figs.~\ref{fig:seeing} and
\ref{fig:nulltest}, we seem to find RR-Lyrae type variables whose
apparent magnitudes would be around $r\sim 25~$mag. When the seeing gets worse,
these stars cannot be reliably seen in the difference image. Since RR-Lyrae
stars should exist in the M31 region, we think the ``impostor'' stars are good
candidates for RR-Lyrae stars. The figure shows that our analysis successfully
enables to find variable stars across the disk and halo regions. The total
number of candidates are 1,334 and 2,740 for $m_r\le 24~$ and 25~mag,
respectively.

\section{Statistics and Selection Criteria}
\label{sec:selection}

Given the catalog of variable star candidates each of which has its
measured light curve, we now search for secure candidates of PBH
microlensing.  In this section we describe our selection criteria to
discriminate the microlensing event from other variables.

\subsection{Photometric errors of the light curve measurement}
\label{sec:noise_est}

\begin{figure}
\centering
\includegraphics[width=0.45\textwidth]{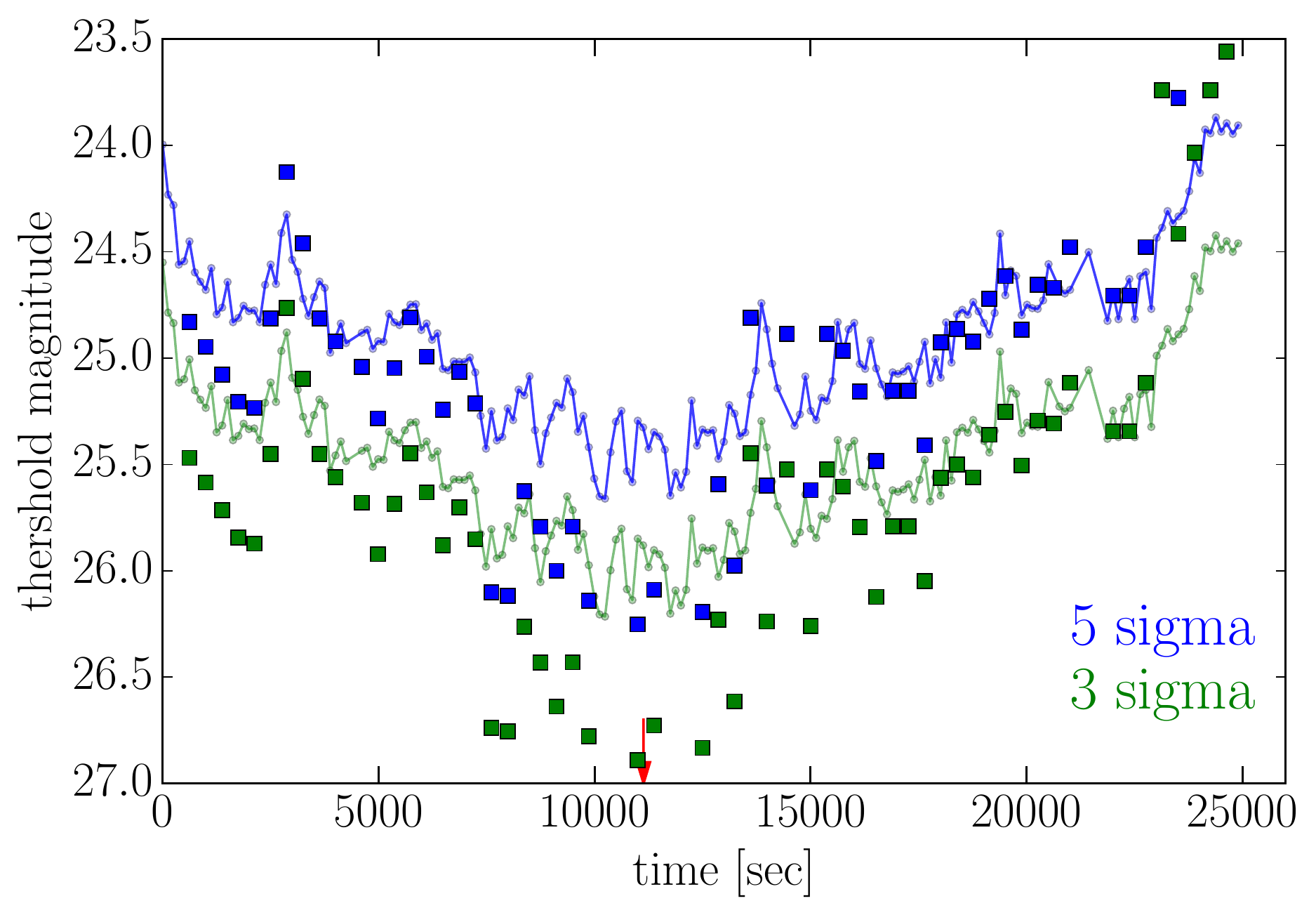}
\caption{The photometric error used for the light curve measurement in
 the difference image; we randomly select 1,000 points in the difference
 image of a given patch (here shown for the patch-D2 in
 Fig.~\ref{fig:m31ccdpatch}), measure the PSF photometry at each random
 point, and then estimate the variance of the PSF photometries (see text
 for details).  The square symbols show the 3- or 5-sigma photometric
 errors estimated from the variance when using the difference images
 constructed from the coadded images of 3 exposures, as a function of
 observation time. The circle symbols, connected by the line, are the
 results for each exposure. Although we use the photometric error in the
 ADU counts for a fitting of the microlensing model to the light curve,
 we here convert the counts to the magnitude for illustrative
 convenience.  }\label{fig:nulltest}
\end{figure}
Our primary tool to search for variable objects in the dense star
regions of M31 is the use of the image difference technique, as we have
shown. To robustly search for secure candidates of PBH microlensing
that have the expected light curve shapes, it is crucial to
properly estimate the photometry error in the light curve measurement.
However, accurate photometry for dense star regions in M31 is challenging.
To overcome this difficulty, we use the following approach to obtain a
conservative estimate of the error. The pipeline performs image subtraction on
each patch basis (as denoted by white-color square regions in
Fig.~\ref{fig:m31ccdpatch}). For a given difference image, we randomly select
1,000 points in each patch region, and then perform PSF photometry at each
random point in the same manner as that for the variable star candidates. In
selecting random points, we avoided regions corresponding to bad CCD pixels or
near the CCD chip edges. We then estimate the variance from those 1,000 PSF
magnitudes, repeat the variance estimation in the difference image for every
observation time, and use the variance as a $1\sigma$ photometry error in the
light curve measurement at the observation time. The photometric error
estimated in this way would include a contamination from various effects such
as a large-scale residual background due to an imperfect background
subtraction. We find that the photometric error is larger than the error
estimated from the pipeline at the candidate position, which is locally
estimated by propagating the Poisson noise of the counts through the image
subtraction processes.

Fig.~\ref{fig:nulltest} shows the photometric error on the light curve
measurement in the difference image, estimated based on the above
method. The shape of the photometric error appears to correlate with
the seeing conditions in Fig.~\ref{fig:seeing}. The figure shows that
most of our data reaches a depth of 26~mag or so thanks to the 8.2m
large aperture of Subaru.

\subsection{Microlensing model fit to the light curve data}

\begin{table}[htb]
\begin{center}
{\small
\caption{Definitions of Statistics \label{tab:stats_def}}
\begin{tabular}{lll}\hline\hline
{\bf Statistic} &
{\bf Definition}\\
\hline 
$\Delta C(t_i)$ & PSF-photometry counts of a candidate in the $i$-th difference
 image at the observation time $t_i$; \\
& the time sequence
   of $\Delta C(t_i)$
 forms the light curve of each candidate (188 data
 points, \\
& sampled by every 2~min).\\
$\Delta C_{{\rm coadd}}(t_i)$ & PSF-photometry counts of a candidate in the $i$-th difference
 image of 3 coadded images at $t_i$ \\
& (63 data points,
 sampled by every 6~min)\\
 $\sigma_i$ &  $1\sigma$ error of PSF-photometry in the $i$-th
 difference image
 (see text for details)\\
$\sigma_{{\rm coadd},i}$ & $1\sigma$ error in the $i$-th difference
 image of 3 coadded images at $t_i$\\
bump & sequence of 3 or more time-consecutive data points with $\Delta C_i\ge 5
 \sigma_i$ in the light curve \\
 bumplen & length (number) of time-consecutive data points with $\Delta C_i\ge 5
 \sigma_i$  \\
 mlchi2\_dof & $\chi^2$ of the light curve fit to microlensing model
 divided by the degrees of freedom \\
 mlchi2in\_dof & $\chi^2$ of the microlensing fit for data points with $t_i$
 satisfying
 $t_0-t_{\rm FWHM}^{\rm obs}\le t_i\le t_0+t^{\rm obs}_{\rm FWHM}$\\
 asymmetry $a_{\rm asy}$ &
 $(1/N_{\rm asy})\left.\sum_{t_i}\left[\Delta C(t_0-\Delta t_i)-\Delta
 C(t_0+\Delta t_i)\right]\right/[\overline{\Delta C}-\Delta C_{\rm
 min}]$ (see text for details)\\
 seeing\_corr & correlation between the light curve shape and the seeing
 variation (see text for details)
\\
\hline
\hline
\end{tabular}
}
\end{center}
\end{table}

%
\begin{table}[htb]
\begin{center}
{\small 
\caption{Selection Criteria\label{tab:cuts}}
\begin{tabular}{lll} \hline\hline
{\bf Selection Criterion} &
{\bf Purpose} &
{\bf No. of remained candidates}\\ \hline
 $\Delta C_{{\rm coadd},i}\ge 5\sigma_{{\rm coadd}, i}$  & initial
 definition of candidates  & 15,571 \\  
 bumplen$\ge 3$ & select candidates with a significant peak(s) in
 the light curve  & 11,703\\ 
 mlchi2dof$<3.5$ & select candidates whose light curve is
 reasonably well fit  & 227\\
 & by the microlensing \\
 $a_{\rm asy}<0.17$ & remove candidates that have an asymmetric light curve
 such   & 146 \\
 & as star flares & \\
 significant peak & select candidates that show a clear peak in its
 light curve   & 66 \\
 & (see text for details) &\\
 visual inspection & visually check each candidate (its light curve and images) 
 & 1\\
 seeing\_corr & remove candidates whose light curve is correlated with
 time  & 1 \\ 
 &variation of seeing & \\ \hline\hline
\end{tabular}
}
\end{center}
\end{table}

\begin{figure*}[t]
\centering \includegraphics[width=8cm,clip]{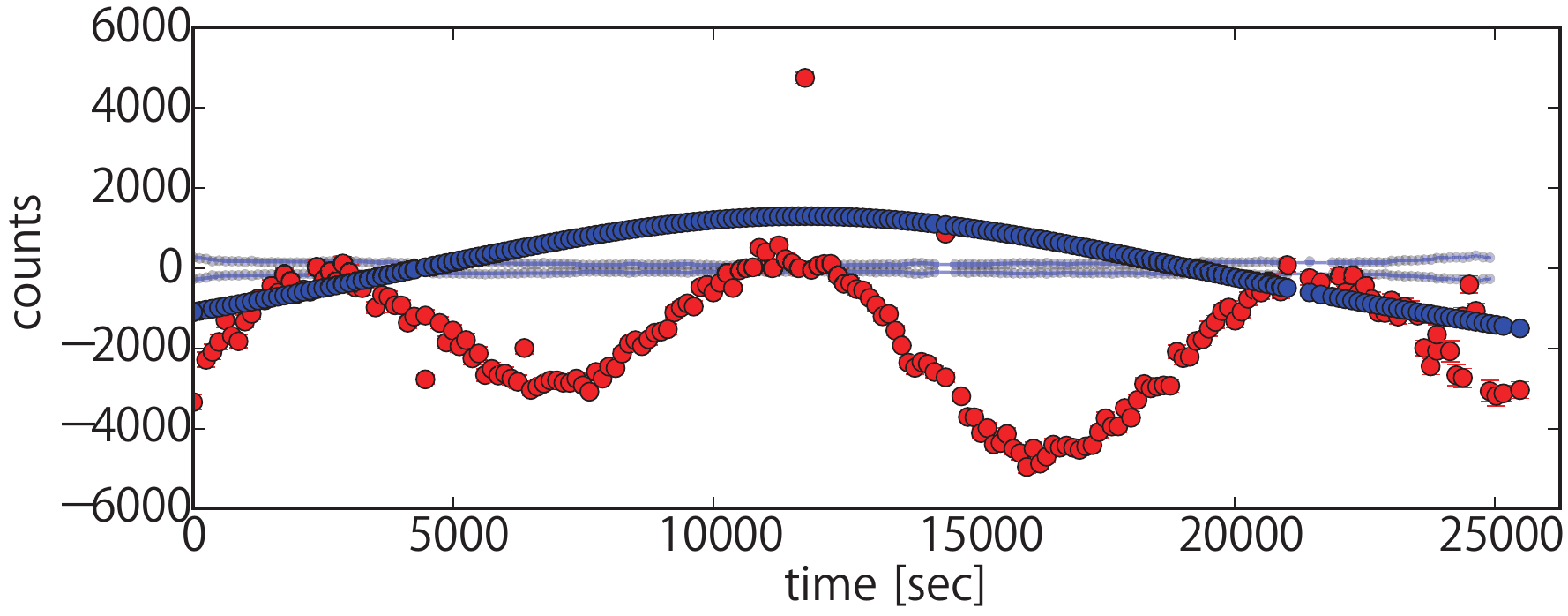}
\includegraphics[width=8cm,clip]{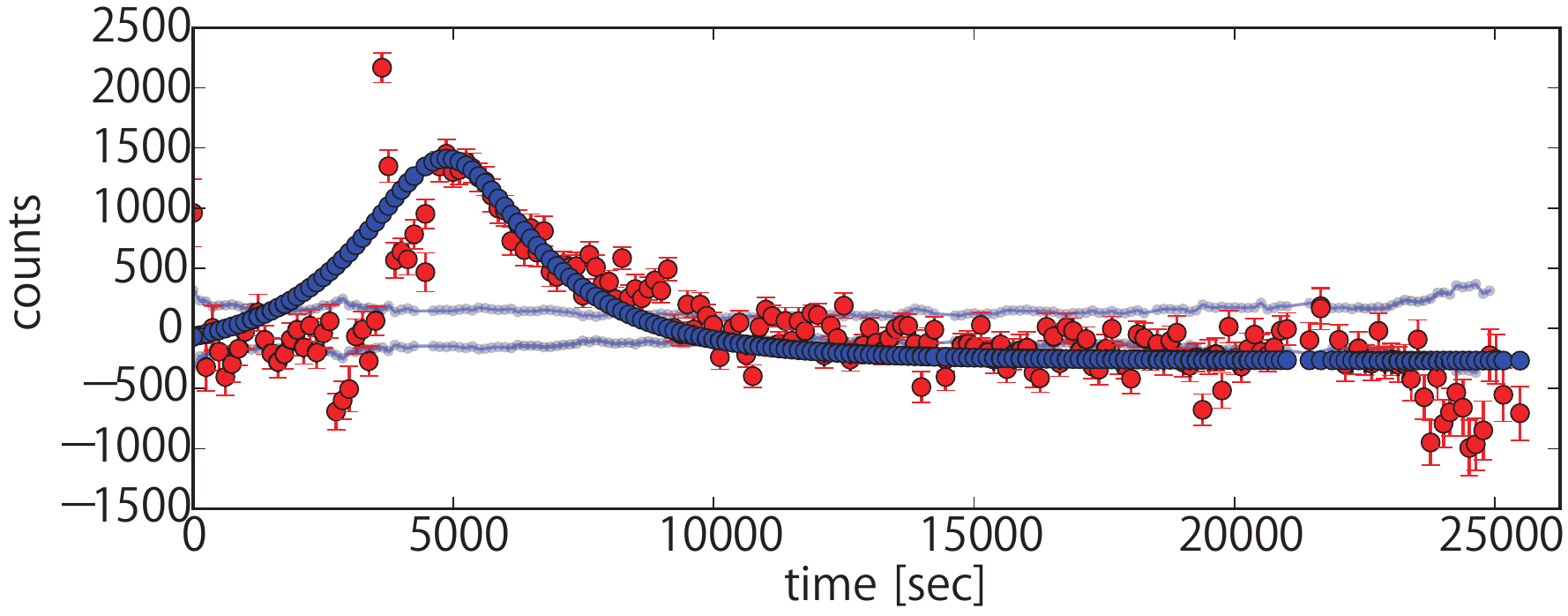}
\includegraphics[width=8cm,clip]{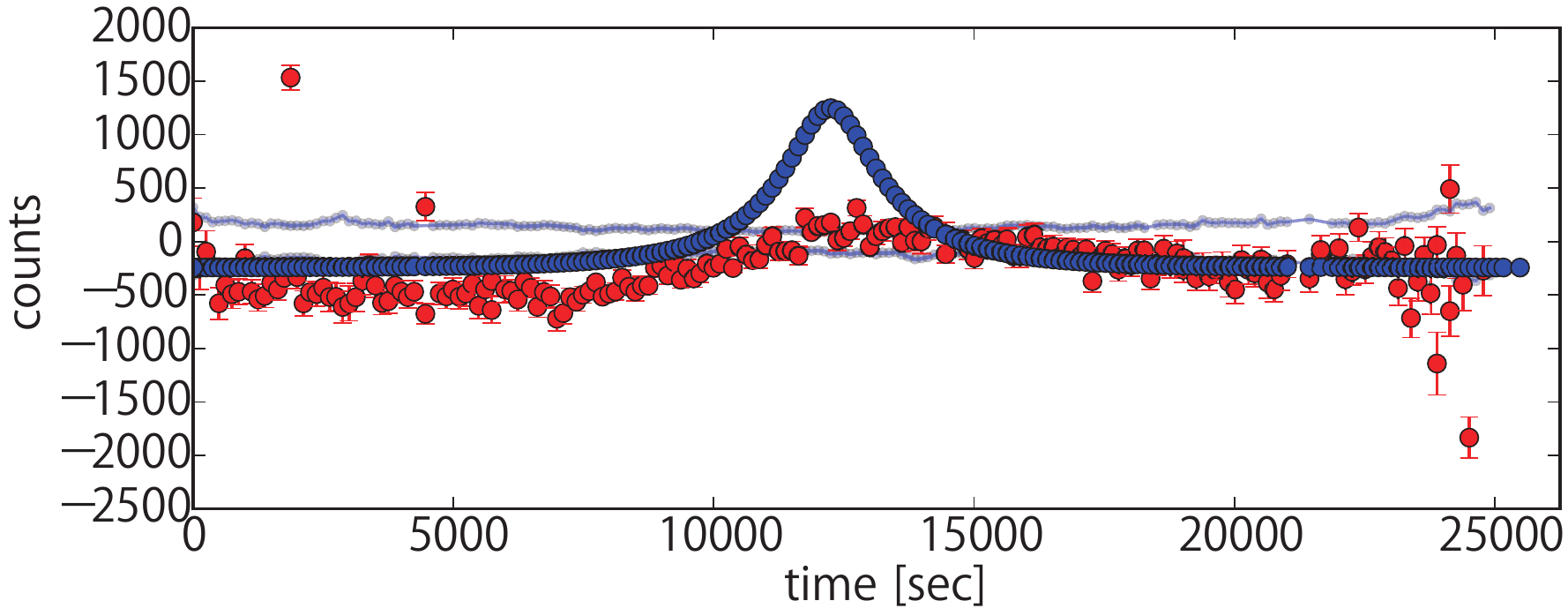}
\includegraphics[width=8cm,clip]{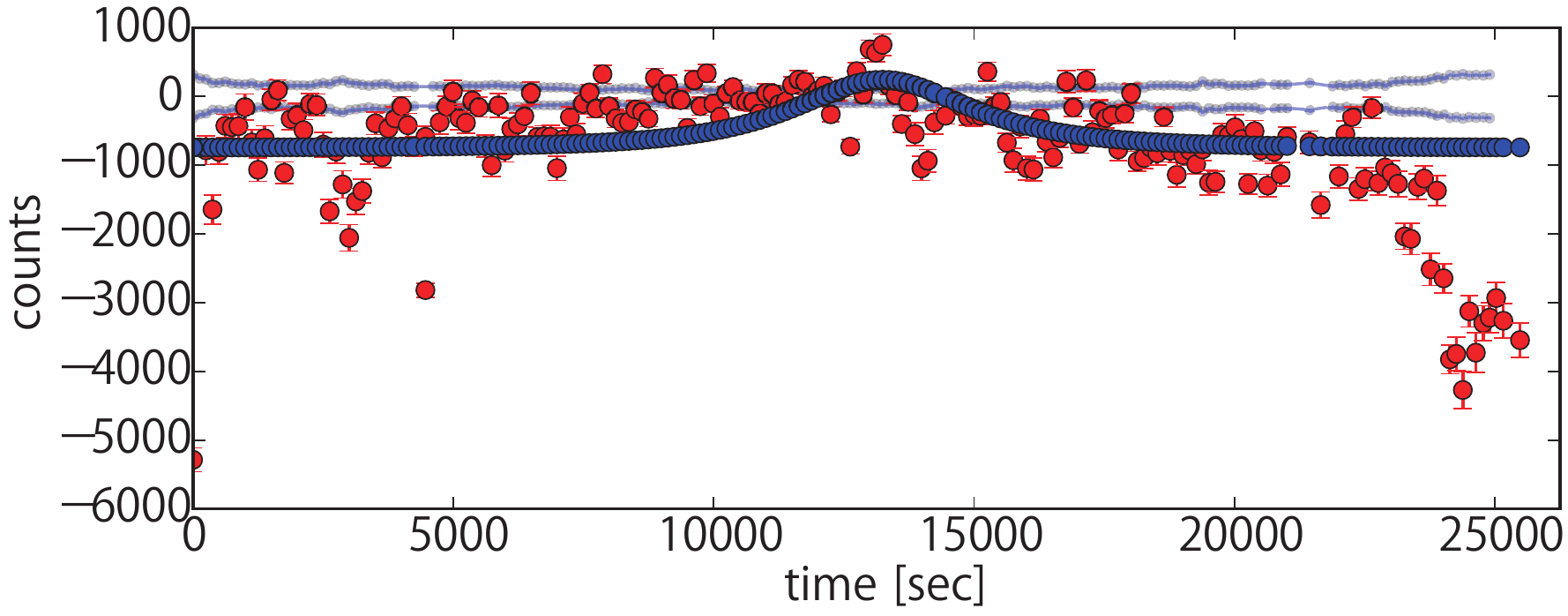}
 \caption{Example of the light curves of candidates that are rejected by
 our selection criteria for a microlensing event. The red points in each
 panel shows the PSF photometry at each observation time and consist of
 188 data points to form the light curve sampled by every 2~min in the
 difference images. The errorbar around each data point is the
 photometry error that is locally estimated by propagating the Poisson
 noise of counts through the difference image processes at the candidate
 position. The range bracketed by the two data points around zero counts
 is the $\pm 1\sigma$ photometry error that is estimated from the PSF
 photometries of 1,000 random points as shown
 Fig.~\ref{fig:nulltest}. The blue data points are the light curve for
 the best-fit microlensing model.  The upper-left panel shows an example
 of the candidates that is rejected due to a bad $\chi^2_{\rm min}$ for
 the fitting to the microlensing light curve. The upper-right panel
 shows an example of the candidates that is rejected by the asymmetric
 shape of the light curve around the peak. The lower two panels show
 examples of the candidates that do not show a prominent peak feature as
 expected for a microlensing event.
 } \label{fig:lc_rejected}
\end{figure*}
\begin{figure}
 \centering
\noindent\includegraphics[width=8cm,clip]{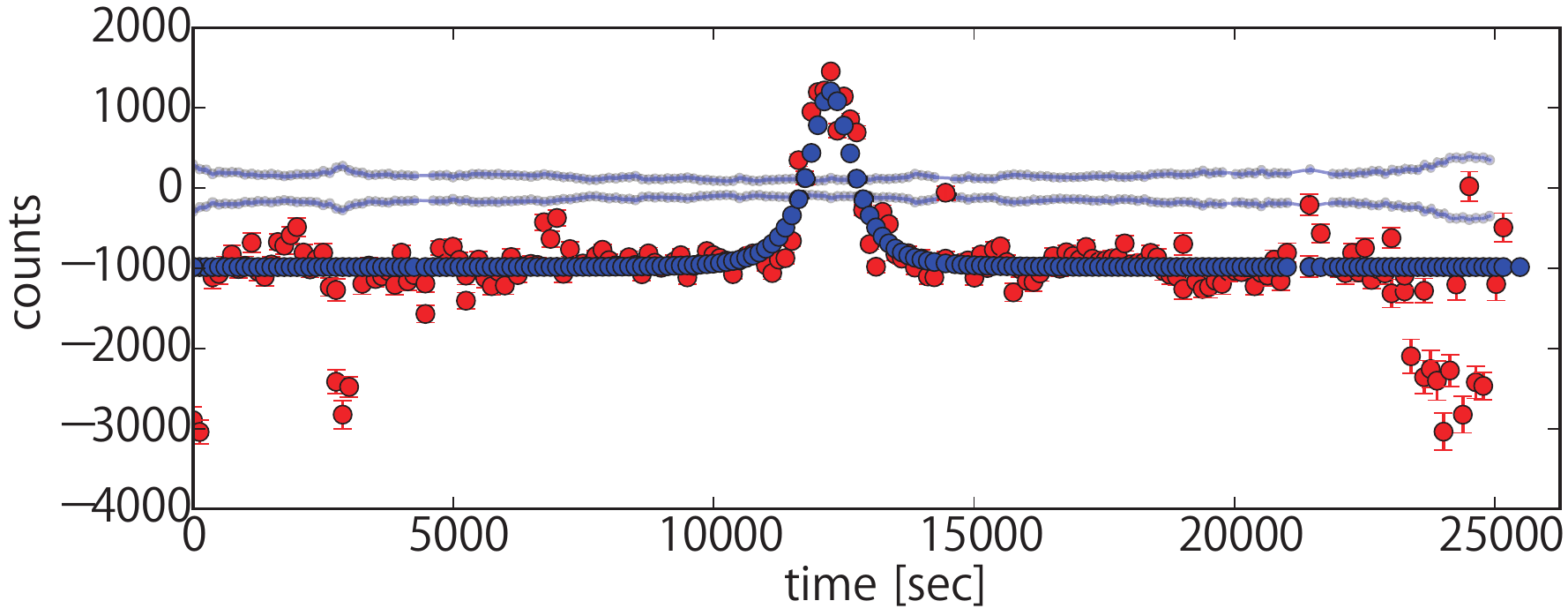}\\
\noindent \includegraphics[width=8cm,clip]{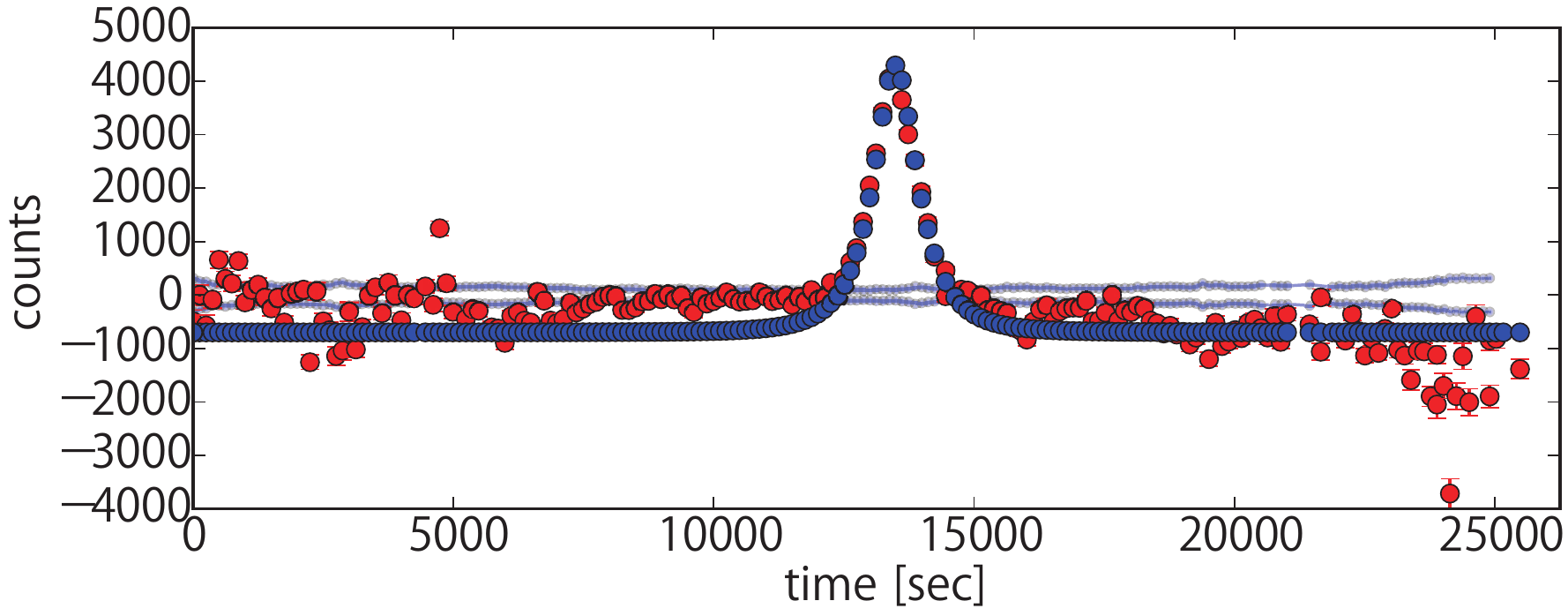}
 \includegraphics[width=3cm,clip]{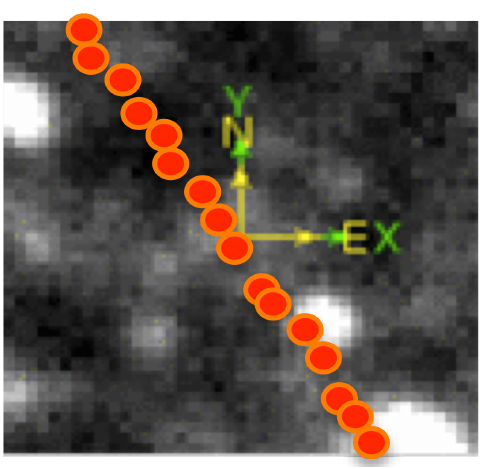}
 \caption{The upper panel shows an example of light curves for impostors that are caused by a spike-like image around a bright star. The
 light curve appears to look like a microlensing event, but it is found
 to be near a bright star. The lower panel shows the light curve for an
 asteroid that also shows a microlensing-like light curve. If the PSF
 photometry is made at the fixed position (the center in the lower-right
 image), the measured light curve looks like a microlensing event. The
 red points in the image denotes the asteroid trajectory. From our
 analysis of M31 observation, we identified one asteroid.
 \label{fig:spike_asteroid}}
\end{figure}
Here we describe our selection procedure for PBH microlensing events
from the candidates. The unique part of our study is the 
the high cadence for the light curve of each candidate, sampled by
every 2~min over about 7~hours. However the monitoring of each
light curve is limited by a duration of 7~hours.  If a microlensing
event has a longer time duration than $7~$hours, we can not identify
such a candidate. We use the statistics in Table~\ref{tab:stats_def} to
quantify the characteristics of each light curve.  Our selection procedure
for the candidates are summarized in Table~\ref{tab:cuts}. We will
describe each of the selection steps in detail.

As we described, we start with the master catalog of variable star
candidates, which contains 15,571 candidates, to search for microlensing
events.  Our level 1 requirement is that a candidate event should have a
``bump'' in its light curve, defined as 3 time-consecutive flux
changes each of which has a signal-to-noise ratio greater than $5\sigma$
in the difference image; $\Delta C_{i}\ge 5\sigma_i$, where the
subscript $i$ denotes the $i$-th difference image (at the observation
time $t_i$).  This criteria leaves us with $11,703$ candidates over all the
patches.

Next we fit the observed light curves of each candidate with a model describing
the expected microlensing light curve. As we described in
Section~\ref{sec:pixellens}, the light curve of a microlensing in the
difference image is given as
\begin{equation}
 \Delta C(t_i) = C_0 \left[A(t_i)-A(t_{\rm ref})\right],
  \label{eq:DeltaC_model}
\end{equation}
where $C_0$ is the PSF-photometry counts of an unlensed image in the
difference image, corresponding to $F_0$ in Eq.~(\ref{eq:DeltaF}), and
$A(t_i)$ and $A(t_{\rm ref})$ are lensing magnifications at the
observation time $t_i$ and the time of the reference image $t_{\rm
ref}$.  As described in Section~\ref{sec:pixellens}, the light curve in
the difference image is characterized by 3 parameters: $(u_{\rm min},
t_{\rm FWHM}, C_0)$, where $u_{\rm min}$ is the impact parameter of
closest approach between PBH and a source star in units of the Einstein
radius, and $t_{\rm FWHM}$ is the FWHM timescale of the light curve.

We identify the time of maximum magnification in the light curve and denote it by $t_{\rm 0}$.
For the model fitting, we employ the following range for the model parameters:
\begin{itemize}
 \item $0.01\le u_{\rm min}<1$, which determines the maximum
       magnification, $A_{\rm max}\equiv A(u_{\rm min})$ (see
       Eq.~\ref{eq:cano}). Thus we assume the range of maximum
       magnification to be $1.34\le A_{\rm max}\simlt 100$.
 \item $0.01\le t_{\rm FWHM}/[{\rm sec}]<25,000$. Here the lower limit
       is much shorter than the sampling rate of light curve (2~min),
       but we include such a short time-scale light curve for safety
       (see below).  The upper limit corresponds to the longest duration
       of our observation ($\sim 7~$hours).
 \item Once the parameters, $u_{\rm min}$ and $t_{\rm FWHM}$, are
       specified, the intrinsic flux can be estimated as $C_0=\Delta
       C_{\rm max}^{\rm obs}/[A_{\rm max}-A(t_{\rm ref})]$, where
       $\Delta C_{\rm max}^{\rm obs}$ is the counts of the light curve
       peak in the difference image.  In practice, the flux measurement
       is affected by measurement noise as well as the sampling
       resolution of light curve, so we allow the intrinsic flux to vary
       in the range of $0.5\times \Delta C_{\rm max}^{\rm obs}/(A_{\rm
       max}-1)\le C_0 \le 1.5\times \Delta C_{\rm max}^{\rm obs}/(A_{\rm
       max}-1)$.
\end{itemize}
The above ranges of parameters are broad enough in order for us not to
miss a real candidate of microlensing. For each candidate, we perform a
standard $\chi^2$ fit by comparing the model microlensing light curve to the
observed light curve: 
\begin{equation}
 \chi^2=\sum_{i=1}^{188}\frac{
  \left[\Delta C^{\rm obs}(t_i)-\Delta C^{\rm model}(t_i; C_0, t_{\rm
   FWHM}, u_{\rm min})\right]^2}{\sigma_i^2},
\end{equation}
where $\Delta C^{\rm model}(t_i)$ is the model light curve for
microlensing, given by Eq.~(\ref{eq:DeltaC_model}), and $\sigma_i$ is
the rms noise of PSF photometry in the $i$-th difference image,
estimated from the 1,000 random points as described above.

We compute the reduced $\chi^2$ by dividing the minimum $\chi^2$ by
the degrees of freedom (188-3=185). We discard candidates that have ${\rm
mlchi2\_dof}>3.5$. This criterion is reasonably conservative (the P-value is
$\sim 10^{-5}$). We further impose the condition that the best-fit
$t_{\rm FWHM}<14,400~$sec (4 hours), in order to remove candidates whose
light curve has a longer time variation than what we can robustly
determine. This selection removes most of Cepheid-type variables.  This
selection leaves 225 candidates. The upper-left panel of
Fig.~\ref{fig:lc_rejected} shows an example of candidates that are
removed by the condition ${\rm mlchi2\_dof} <3.5$ (i.e. ${\rm
mlchi2\_dof}>3.5$ for this candidate). This is likely to be a binary
star system.

Microlensing predicts a symmetric light curve with respect to the
maximum-magnification time $t_{0}$ ($A_{\rm max}$); the light
curve at $t_i=|t_{0}\pm\Delta t|$ should have a similar flux as
the lensing PBH should have a nearly constant velocity within the Einstein
radius. Following \cite{Griestetal:14}, we define a metric to quantify the
asymmetric shape of the light curve,
\begin{equation}
 a_{\rm asy}=\frac{1}{N_{\rm asy}}\sum_{t_i\in |t_0\pm t_{\rm FWHM}^{\rm obs}|}
  \frac{\left|\Delta C(t_0-\Delta t_i)-\Delta C(t_0+\Delta t_i)\right|}
  {\overline{\Delta C}-\Delta C_{\rm min}}.
\end{equation}
Here $t_{\rm FWHM}^{\rm obs}$ is the timescale that the observed light
curve declines to half of its maximum value. For this purpose, we take the
longer of the timescales from either side of the two half-flux points from the
maximum peak. If the expected half-flux data point is outside the
observation window of light curve, we take the other side of the light curve
to estimate $t_{\rm FWHM}^{\rm obs}$.  The summation runs over the data
points satisfying $t_i\le |t_0\pm t_{\rm FWHM}^{\rm obs}|$, 2 times the
FWHM timescale around the light curve peak. Note that, if the summation
range is outside the observation window, we take the range $|t_0\pm
(t_0-t_{\rm start})|$ or $|t_0\pm (t_{\rm end}-t_0)|$, where $t_{\rm
start}$ or $t_{\rm end}$ is the start or end time of the light curve.
$N_{\rm asy}$ is the number of data points in the above summation,
$\overline{\Delta C}$ is the average of the data points taken in the
summation, and $\Delta C_{\rm min}$ is the minimum value of the counts.

By imposing the condition $a_{\rm asy}<0.17$, we eliminate candidates
that have an asymmetric light curve, and we have confirmed that this
condition eliminates most of the star flare events from the data base.
This condition also eliminates some of the variable stars that are likely to
be Cepheids. After this cut the number of candidates is reduced to 
$146$. The upper-right panel of Fig.~\ref{fig:lc_rejected} shows an example of
the candidates that are removed by the condition $a_{\rm
asy}<0.17$.

In addition we discard candidates, if the observed light curve does not
have any significant peak; e.g., we discard candidates if ${\rm
mlchi2in\_dof}>3.5$ (see Table~\ref{tab:stats_def} for the definition)
or if the time of the light-curve peak is not well determined. The lower
panels of Fig.~\ref{fig:lc_rejected} show two examples of such rejected
candidates, which do not show a clear bump feature in the light curve as
expected for microlensing. This selection cut still leaves us with 66
candidates.

Finally we perform a visual inspection of each of the remaining
candidates. We found various impostors that are not removed by the
above automated criteria. Most of the impostors are caused by an
imperfect image subtraction; in most cases the difference image has
significant residuals near the edges of CCD chips and around bright
stars. In particular, bright stars cause a spiky residual image in
the difference image, that results in impostors that have
microlensing-like light curve if measured at a fixed position. We found
44 impostors caused by such spike-like images around bright
stars. There are 20 impostors around the CCD edges. The upper panel of
Fig.~\ref{fig:spike_asteroid} shows an example of spike-like impostors.
We were also able to identify 1 impostor caused by a moving object, an
asteroid. If the light curve is measured at the fixed position which the
asteroid is passing, it results in a light curve which mimics microlensing, as
shown in the lower panel of Fig.~\ref{fig:spike_asteroid}.

Thus our visual inspection leads us to conclude that 65 events among 66
remaining candidates are impostors and we end up with one candidate event which
passes all our cuts and visual checks. The candidate position is $({\rm RA},
{\rm dec})=(00{\rm h}~45{\rm m}~33.413{\rm s}, +41{\rm d}~07{\rm m}~53.03{\rm
s})$.

\begin{figure}[htb]
 \centering
 \includegraphics[width=0.5\textwidth]{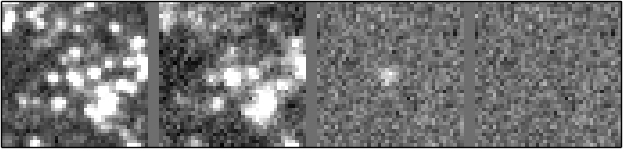}
 \includegraphics[width=0.5\textwidth]{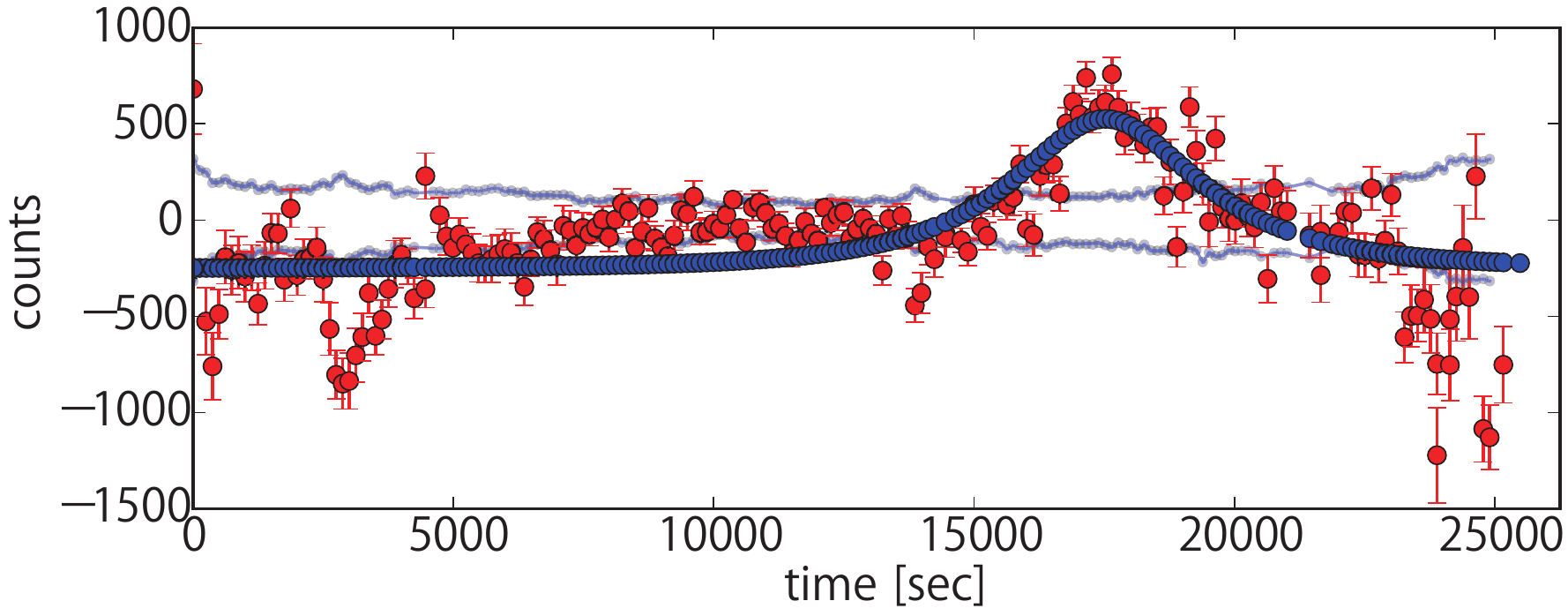}
 \caption{One remaining candidate that passed all the selection criteria
 of microlensing event. The images in the upper plot show the
 postage-stamped images around the candidate as in Fig.~\ref{fig:fake}:
 the reference image, the target image, the difference image and the
 residual image after subtracting the best-fit PSF image,
 respectively. The lower panel shows that the best-fit microlensing
 model gives a fairly good fitting to the measured light curve.
 \label{fig:one_remained2}}
\end{figure}
Fig.~\ref{fig:one_remained2} shows the images and the light curve for this
candidate of microlensing event. Although the light curve looks noisy, it is
consistent with the microlensing prediction. 
The magnitude inferred from the
reference image implies that the candidate has a magnitude of $r\sim 24.5~{\rm
mag}$.  The obvious question to consider is whether this candidate is real.
Unfortunately, the candidate is placed outside the survey regions of the
Panchromatic Hubble Andromeda Treasury (PHAT) catalog in
\cite{Williamsetal:14} (also see \cite{Dalcantonetal:12})
\footnote{\url{https://archive.stsci.edu/prepds/phat/}}, so the HST image is
not available. It is unclear if there are any variable stars that could produce
the observed light curve, with a single bump. To test the hypothesis that the
candidate is a variable star, we looked into another $r$-band data that was taken
in the commissioning run in 2013, totally different epoch from our observing
night. However, the seeing condition of the $r$-band is not good (about
$1.2^{\prime\prime}$), so it is difficult to conclude whether the star pops out
of the noise in the difference images. Similarly we looked into the $g$-band
images taken in the HSC commissioning run. However, due to the short duration
of the data itself ($\sim 15~$min), it is difficult to judge whether this
candidate has a time variability between the $g$ images. Hence we cannot draw
any convincing conclusion on the nature of this candidate. In what follows, we
derive an upper bound on the abundance of PBHs as a constituent of DM for both
cases where we include or exclude this remaining candidate.

\section{Results: Upper Bound on the Abundance of PBH Contribution to Dark Matter}
\label{sec:results}

In this section we describe how we use the results of our PBH microlensing
search to derive an upper limit on the abundance of PBHs assuming PBHs consist
of some fraction of DM in the MW and M31 halos. In order to do this, we need
three ingredients -- (1) the event rates of microlensing as we estimated in
Section~\ref{sec:mlrate}, (2) a detection efficiency for PBH microlensing
events, which quantifies the likelihood of whether a microlensing event, even if
it occurs during our observation duration, will pass all our selection
cuts, and (3) the number of source stars in M31. In this section we describe how
to estimate the latter two ingredients and then derive the upper bound result.

\subsection{Efficiency calculation: Monte Carlo simulation}
\label{sec:efficiency}

\begin{figure}[htb]
\centering
\includegraphics[width=0.6\textwidth]{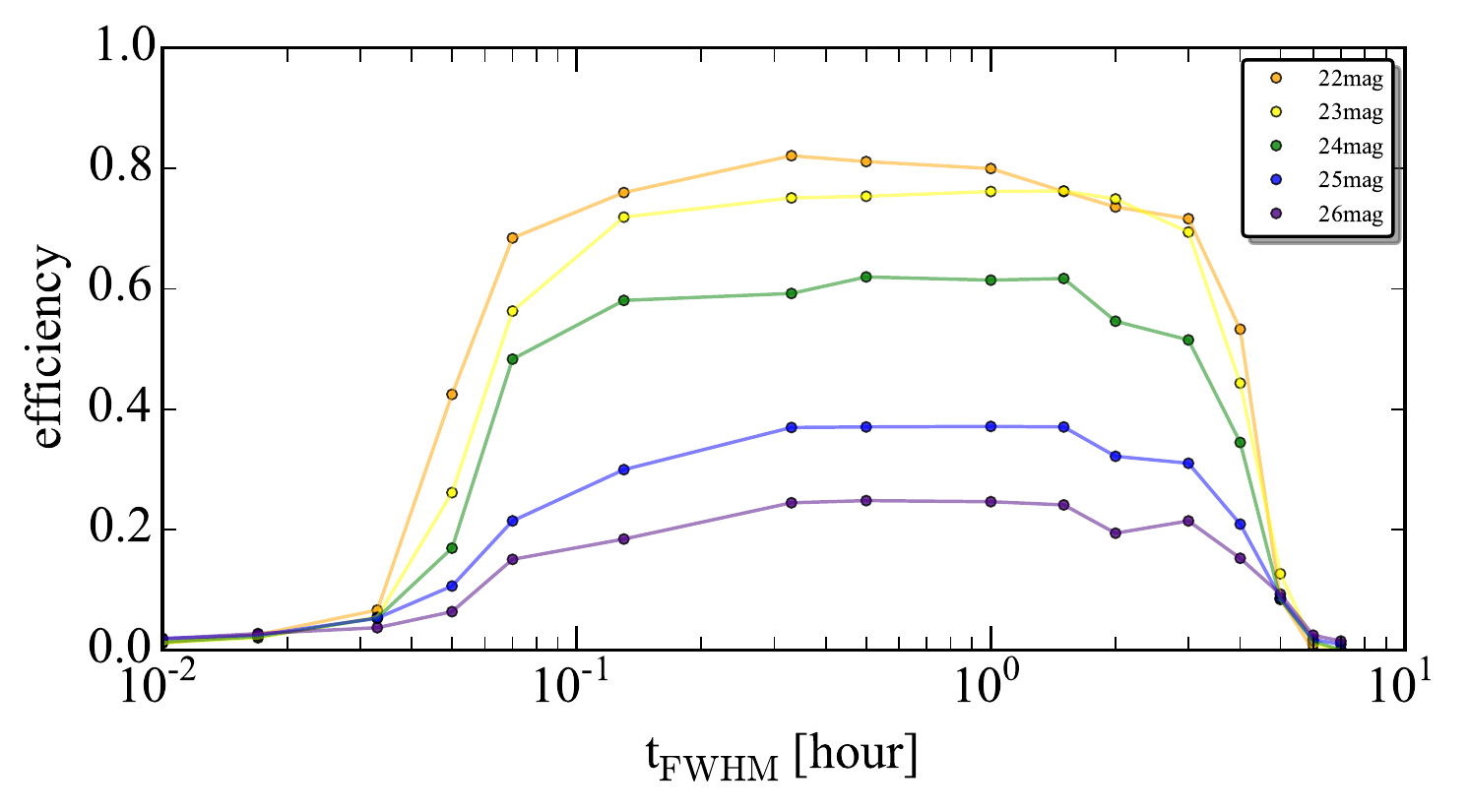}
 \caption{The detection efficiency estimated from light curve
 simulations taking into account the PSF photometry error in each of 188
 target images we used for the analysis (see text for details). Here we
 generated Monte Carlo simulations of microlensing events randomly
 varying the three parameters: the impact parameter (or maximum lensing
 magnification), the FWHM timescale of microlensing light curve
 ($x$-axis), and the observation time of the microlensing magnification
 peak,
 for source stars of a fixed magnitude as indicated by
 legend.  The detection efficiency for each source magnitude is
 estimated from 1,000 realizations.}  \label{fig:efficiencyeach}
\end{figure}
The detection efficiency of PBH microlensing events depends upon the unlensed
flux of the star in M31, $F_0$, and quantifies the fraction of microlensing
events with a given impact parameter ($u_{\rm min}$) and timescale ($t_{\rm
FWHM}$) that can be detected given our selection cuts.

To estimate the efficiency we carry out simulations of microlensing light
curves. We vary the model parameters to generate a large number of
realizations of the simulated microlensing light curves. First we
randomly select the time of maximum magnification ($t_{\rm max}$) from
the observation window, the impact parameter $u_{\rm min}\in[0, 1]$ 
and the FWHM timescale $t_{\rm FWHM}$ in the range of $0.01\le t_{\rm
FWHM}/[{\rm sec}]\le 25,000~$
to
simulate the input light curve in the difference image for a given
intrinsic flux of a source star, $F_0$ (more precisely, the intrinsic
counts $C_0$ in the difference image). Then, we add random Gaussian
noise to the light curve at each of the observation epochs $t_i$, estimated
from the $i$-th difference image in a given patch (Section~\ref{sec:noise_est}).
For each intrinsic flux, we generate 10,000 simulated light curves in each
patch region.

For each simulated light curve, we applied all of our selection cuts (see
Section~\ref{sec:selection} and Tables~\ref{tab:stats_def} and
\ref{tab:cuts}) to assess whether the simulated event passes all the
criteria. Fig.~\ref{fig:efficiencyeach} shows the estimated efficiency
for a given intrinsic flux of a star as a function of the timescale
($t_{\rm FWHM}$) of the simulated light curve, in the patch-D2 of
Fig.~\ref{fig:m31ccdpatch}.
A microlensing event for a bright star is easier to detect, if it occurs,
because even a slight magnification is enough to identify it in the difference
image. On the other hand, a fainter star needs more significant magnification
to be detected.  If the microlensing timescale is in the range of $4~{\rm
min}\simlt t_{\rm FWHM}\simlt 3~{\rm hours}$, the event can be detected by our
observation (2~min sampling rate and 7~hours observation). We interpolated the
results for different intrinsic fluxes to estimate the detection efficiency for
an arbitrary intrinsic flux. We repeated the simulations using the photometry
errors to estimate the efficiency for each patch.

\begin{figure}[htb]
\centering
\includegraphics[width=0.6\textwidth]{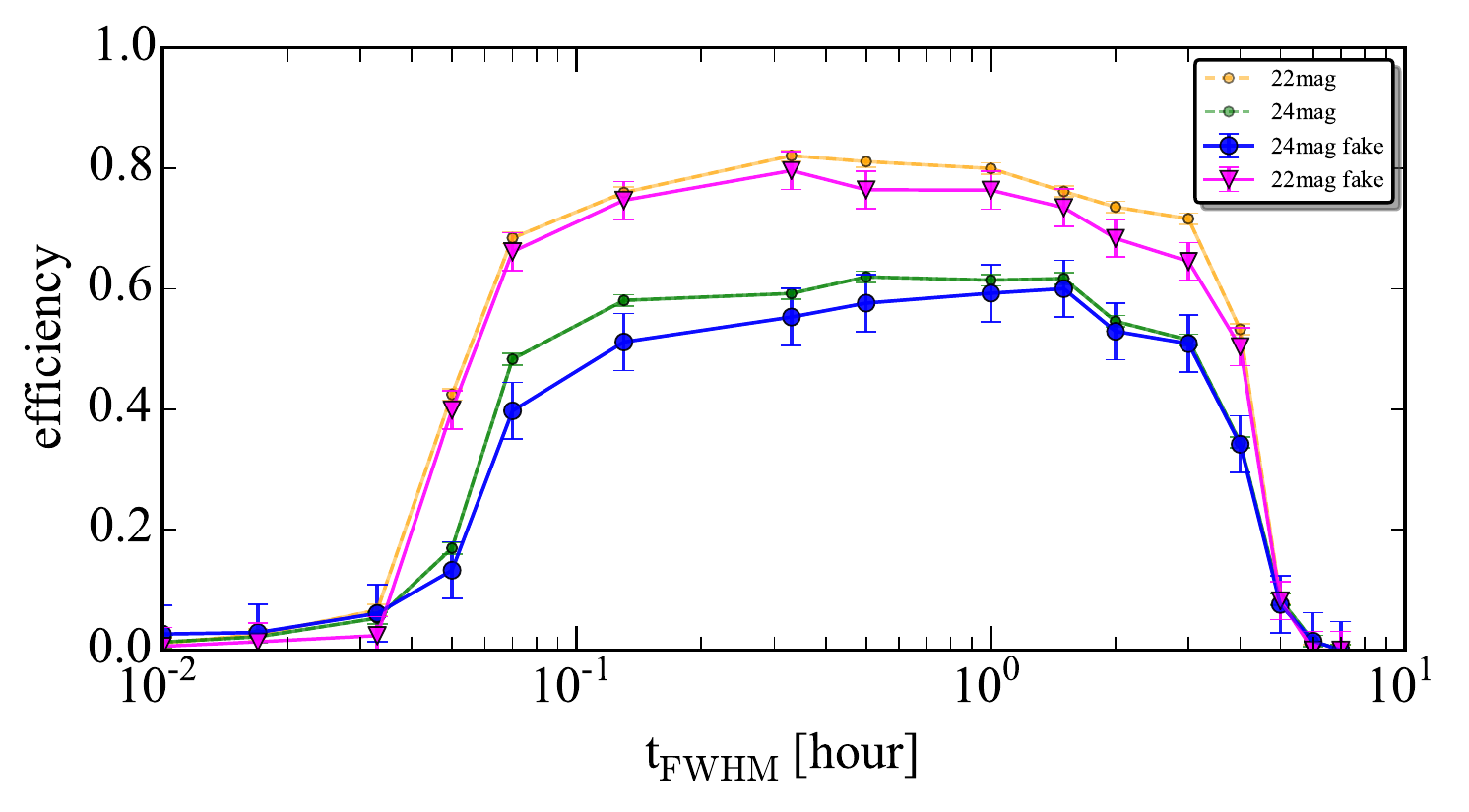}
 \caption{A justification of the detection efficiency estimation, based
 on the different method using the fake image simulations. We injected
 fake microlensing star images in individual exposures of the real HSC
 data (patch-D2 in Fig.~\ref{fig:m31ccdpatch}), re-ran the whole
 data processing, and assessed whether the fake images pass all the
 selection criteria for a microlensing event.
The small circles show the results from light curve simulations (the
same as shown in Fig.~\ref{fig:efficiencyeach}), and the large symbols
show the results from the fake image simulations, for the intrinsic
 magnitudes of 22 and 24~mag, respectively. 
 } \label{fig:fake_sim}
\end{figure}
We also performed an independent estimation of the detection efficiency. We
used fake image simulations where we injected fake microlensing star events
into individual HSC images using the software {GalSim} in Refs.~\cite{Roweetal:15,Huangetal:17},
and then re-ran the whole data
reduction procedure including image subtraction to measure the light curve. We
then assessed whether the fake microlensing event can be detected by our
selection criteria. Fig.~\ref{fig:fake_sim} compares the detection efficiency
estimated using the fake image simulations with the results of the simulated
light curves (Fig.~\ref{fig:efficiencyeach}) in the patch-D2. The figure
clearly shows that the two results fairly well agree with each other. The fake
image simulations are computationally expensive. With the results in
Fig.~\ref{fig:fake_sim}, we conclude that our estimation of the detection
efficiency using the simulated light curves are fairly accurate.

\subsection{Estimation of star counts in M31}
\label{sec:snumber}

\begin{figure}[htb]
\centering
 \includegraphics[width=0.4\textwidth]{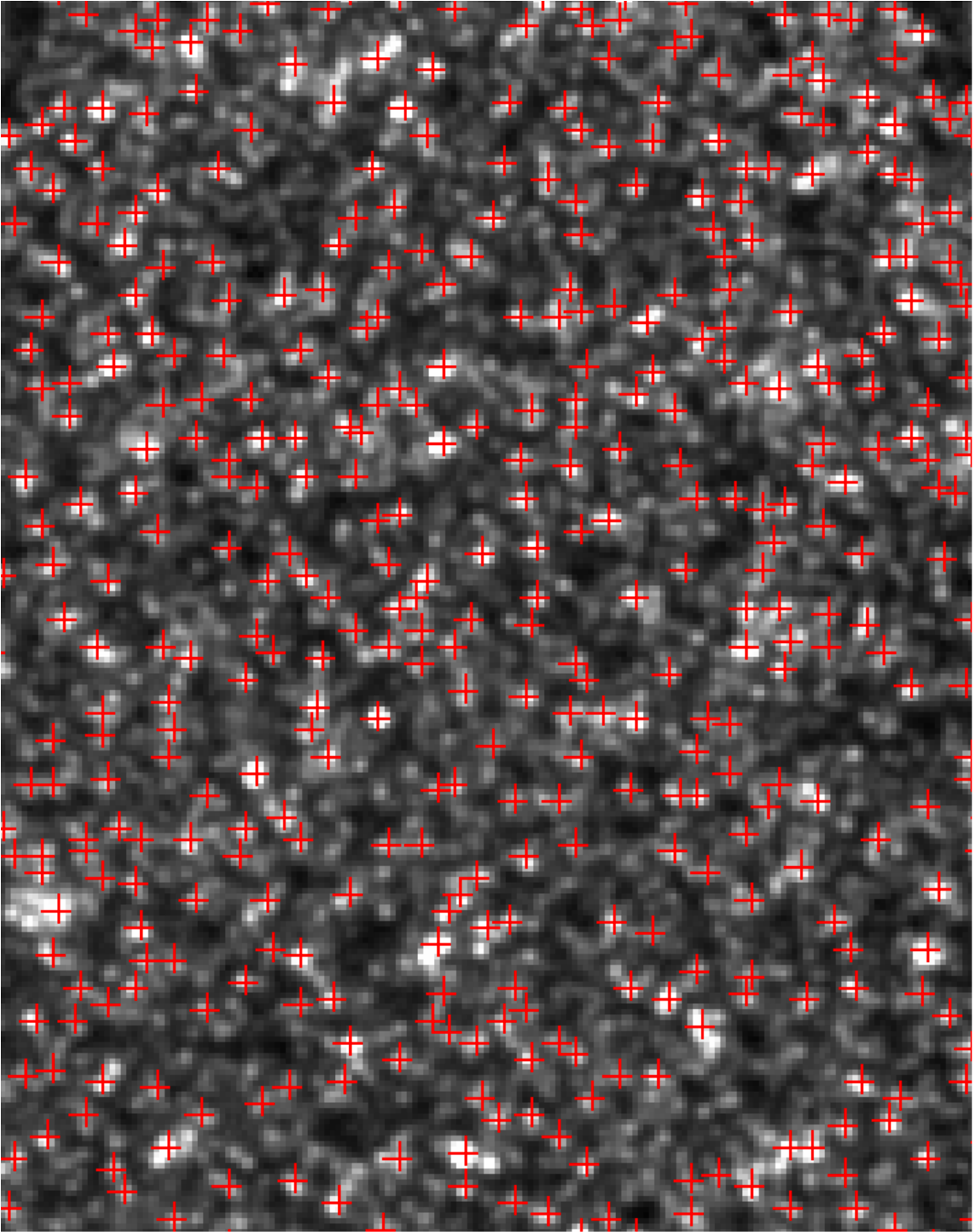}
 \caption{An example image of the distribution of peaks (cross symbols)
 identified in a small region of the reference image (the coadded image
 of 10 best-seeing exposures), which has a size of about
 $38^{\prime\prime}\times 30^{\prime\prime}$ area and is taken from the
 patch-D2 region. We measure the PSF photometry of each peak, and then
 use the number of peaks as an estimation of the number of source stars
 in each magnitude bin.  \label{fig:Npeak}}
\end{figure}
\begin{figure}[htb]
\centering
\hspace*{-3em}\includegraphics[width=0.55\textwidth]{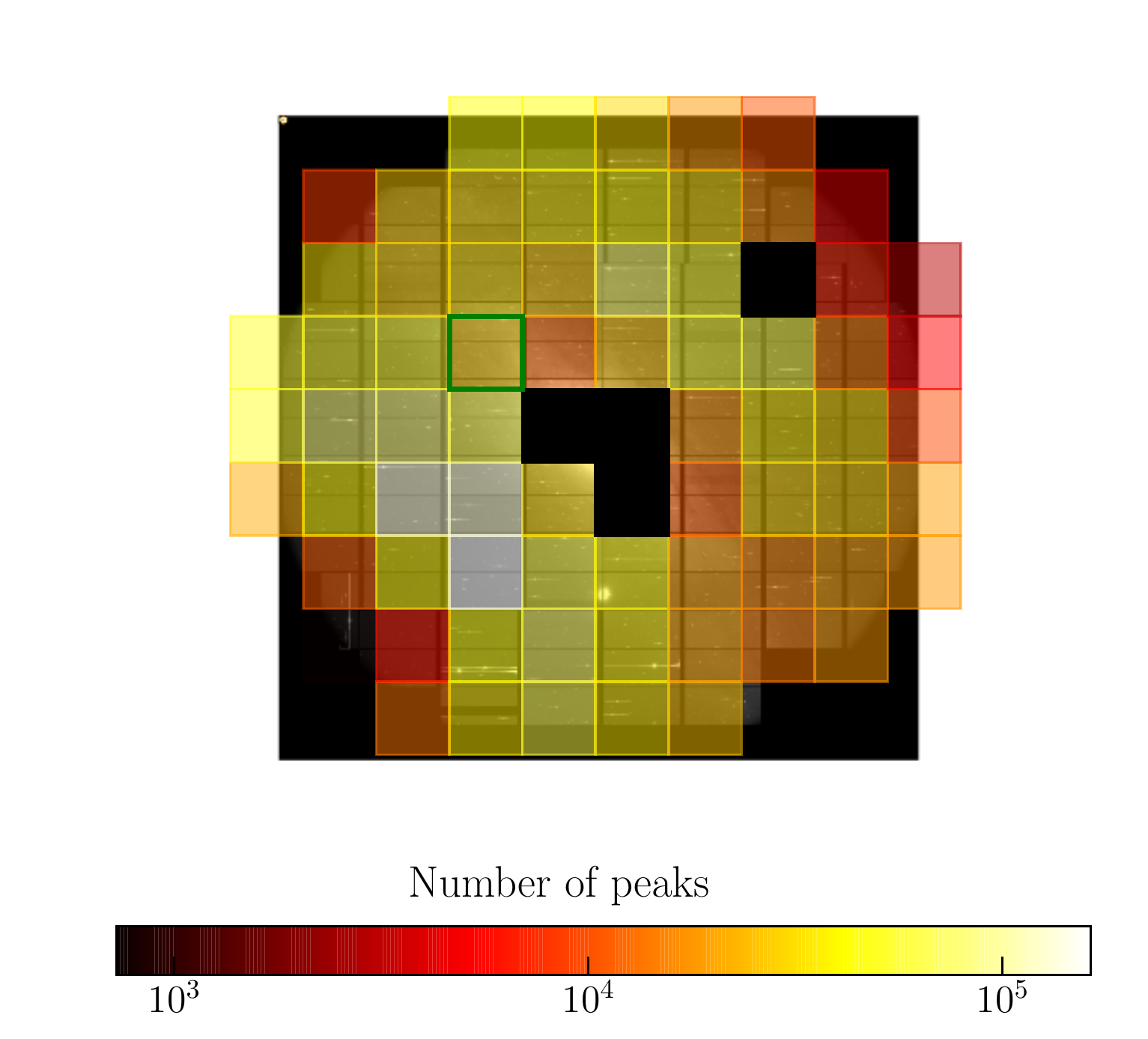}
 \caption{The color scale denotes the total number of detected peaks in
 each patch region for the HSC data. Note that the black-color patches
 are excluded from our analysis due to too crowded regions. The number
 of the peaks in a disk region tends to be smaller than that in a outer,
 halo region, because stars in a disk region are more crowded and only
 relatively brighter stars or more prominent peaks are identified.  }
 \label{fig:peakdist}
\end{figure}
\begin{figure}[htb]
\centering
 \includegraphics[width=0.3\textwidth]{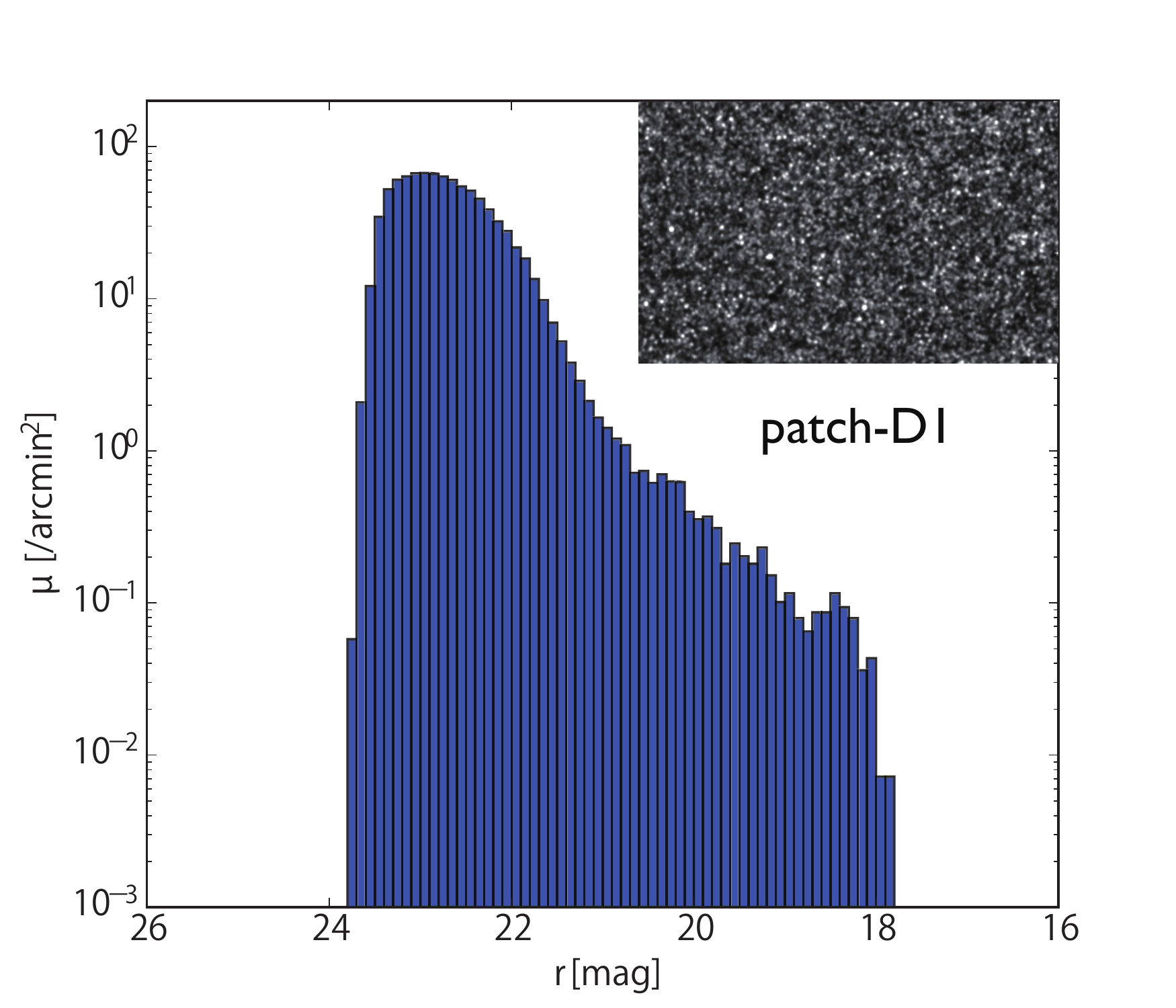}
 \includegraphics[width=0.3\textwidth]{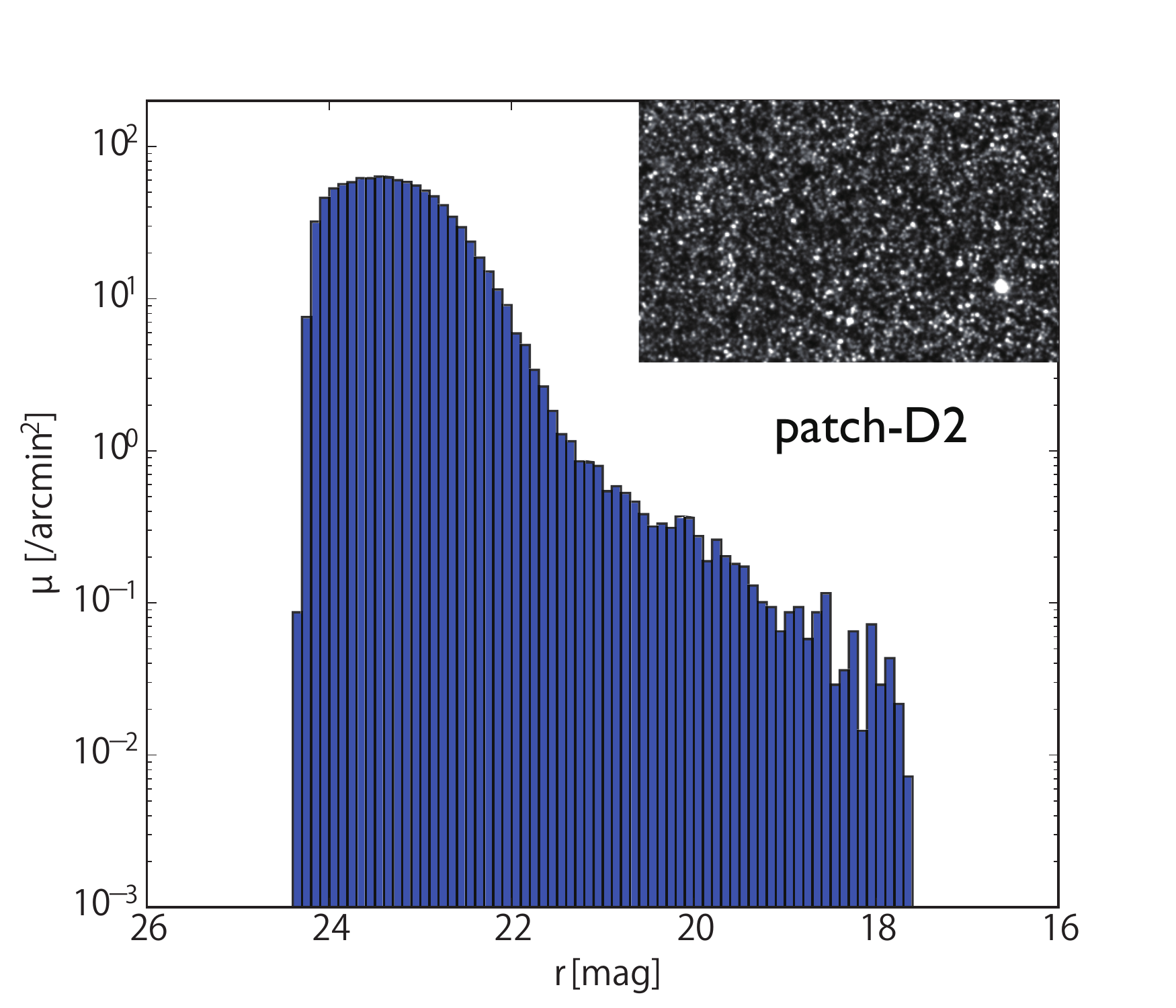}
 \includegraphics[width=0.3\textwidth]{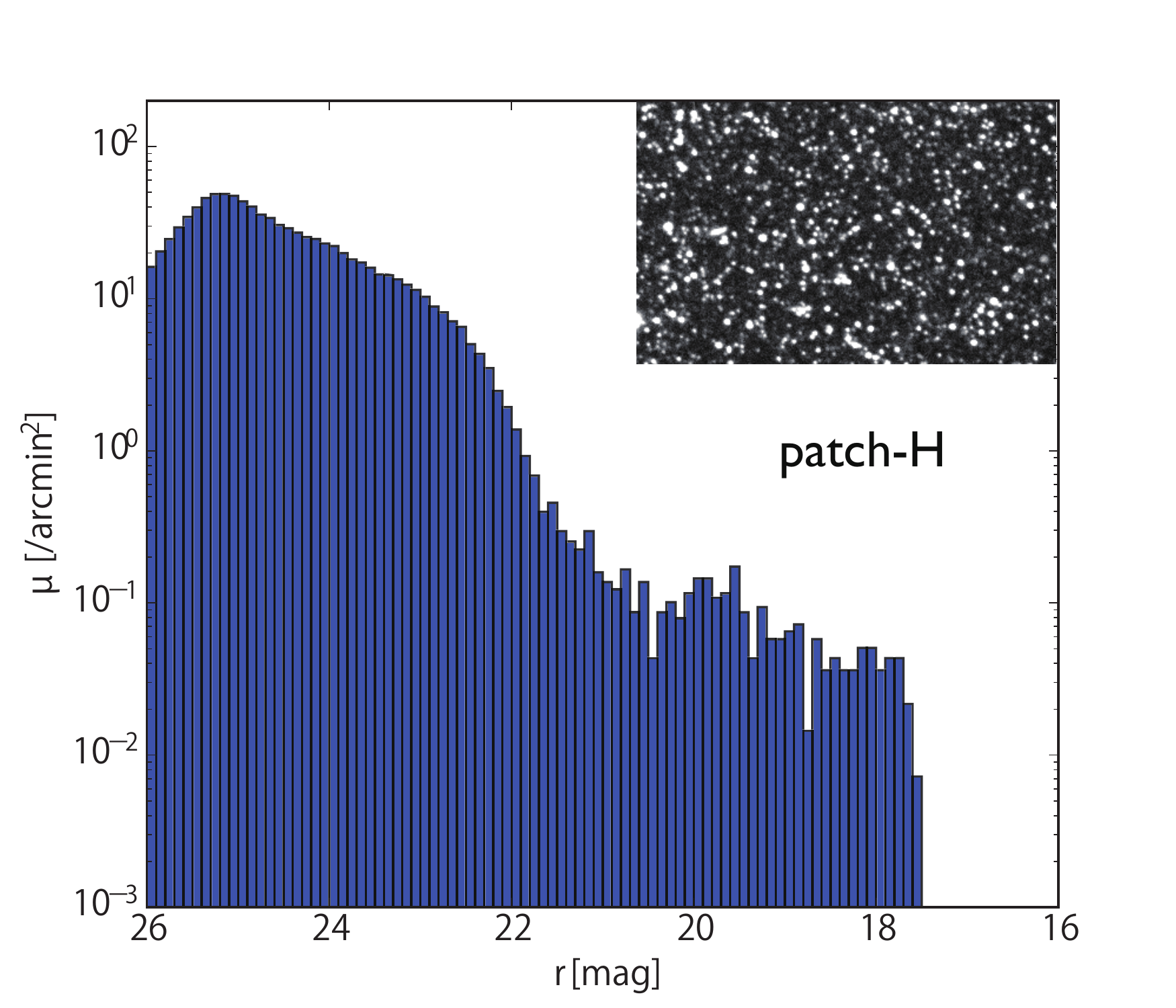}
 \caption{The peaks counts of HSC data in different regions of M31; two
 disk regions denoted as patch-D1 and patch-D2 and the halo region
 denoted as patch-H in Fig.~\ref{fig:m31ccdpatch}. The HSC data can find
 a more number of fainter peaks in the halo regions because individual
 stars are more resolved and less crowded.  \label{fig:Npeaks_d_vs_h}}
\end{figure}
The expected number of microlensing events depends on the number of
source stars in M31. However, since individual stars are not resolved in
the M31 field, it is not straightforward to estimate the number of
source stars from the HSC data. This is the largest uncertainty in our
results, so we will discuss how the results change for different
estimations of the source star counts. As a conservative estimate for the
number of source stars, we use the number of ``detected peaks'' in the
reference image of M31 data, which has the best image quality (coadding the 10
best-seeing exposures) and is used for the image subtraction.
Fig.~\ref{fig:Npeak} shows the distribution of peaks identified from the
reference image in an example region (with a size $226\times 178$~pixels
corresponding to about $38^{\prime\prime}\times 30^{\prime\prime}$), taken from
the patch-D2 region. The figure clearly shows that only relatively bright
stars, or prominent peaks, are identified, but a number of faint stars or even
bright stars in a crowded (or blended) region will be missed. Thus this
estimate of the source star counts is extremely conservative. Nevertheless this
is one of the most secure way to obtain source counts, so we will use these
counts in each patch region.

The color scale in Fig.~\ref{fig:peakdist} shows the total number of
peaks in each patch region. It can be seen that a relatively larger
number of the peaks are identified in the outer halo region of M31,
because each star can be resolved without confusion. On the other hand,
there are less number of resolved peaks in the patches corresponding to the
disk region due to crowding.
The total number of peaks identified over all the patch regions is about 6.4
million.  Fig.~\ref{fig:Npeaks_d_vs_h} shows the surface density of peaks
identified in HSC in the disk and halo regions of M31 for the three patches
marked in Fig.~\ref{fig:m31ccdpatch}. To estimate the magnitudes for the
surface density, we performed PSF photometry of each peak using its location 
as the PSF center. The figure confirms that more number of peaks are
identified in the halo region.

As another justification for the estimation of the source star counts, we
compare the number counts of peaks in the HSC image with the luminosity
function of stars in the HST PHAT catalog in Ref.~\cite{Williamsetal:14} 
(aslo see \cite{Dalcantonetal:12}), where individual stars are more
resolved thanks to the high angular resolution of the ACS/HST data.  Since
the PHAT HST data was taken with F475W and F814W filters, we need to
make color transformation of the HST photometry to infer the HSC
$r$-band magnitude. For this purpose, we first select 100 relatively bright
stars in the PHAT catalog. Then we match the HST stars with
the HSC peaks by their RA and dec positions, and compare the magnitudes
in the HST and HSC photometries. In order to derive the color
transformation, we estimated a quadratic relation between the HST and
HSC magnitudes for the matched stars in a two-dimensional space of
$(m_r^{\rm HSC}-m_{\rm F475W})$ and $(m_{\rm F475W}-m_{\rm F814W})$:
\begin{eqnarray}
 m_r^{\rm HSC}&=&m_{\rm F475W}-0.0815
 -0.385\left(m_{\rm F475W}-m_{\rm F814W}\right)\nonumber\\
&&-0.024\left(m_{\rm F475W}-m_{\rm F814W}\right)^2.
\label{eq:HSTmag_trans}
\end{eqnarray}
We then applied this color transformation to all the PHAT stars. Although the
above one-to-one color transformation is not perfect for different types of
stars, we do not think that the uncertainty largely affects our main results as
we will discuss below.

\begin{figure}[htb]
\centering
 \includegraphics[width=0.49\textwidth]{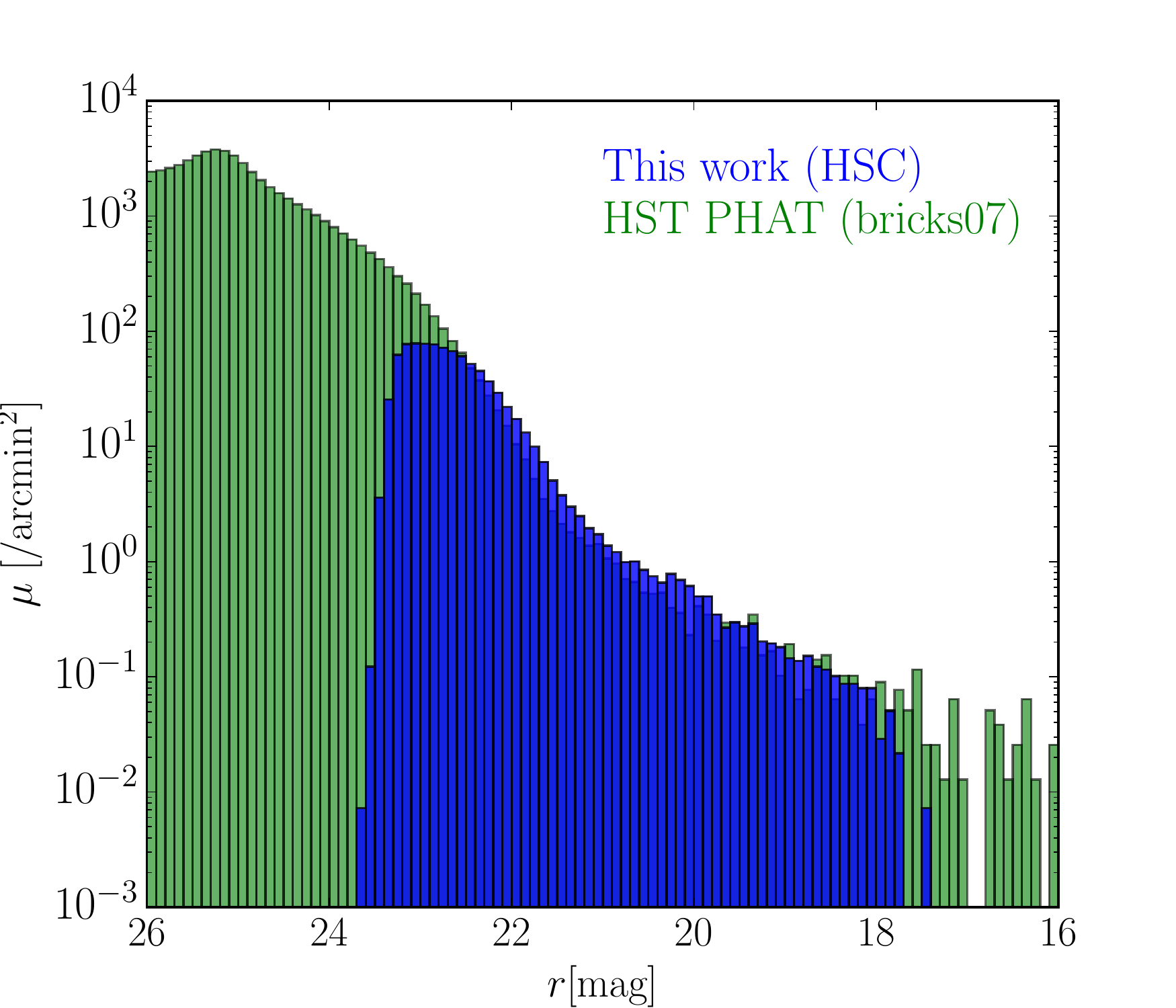}
 \includegraphics[width=0.49\textwidth]{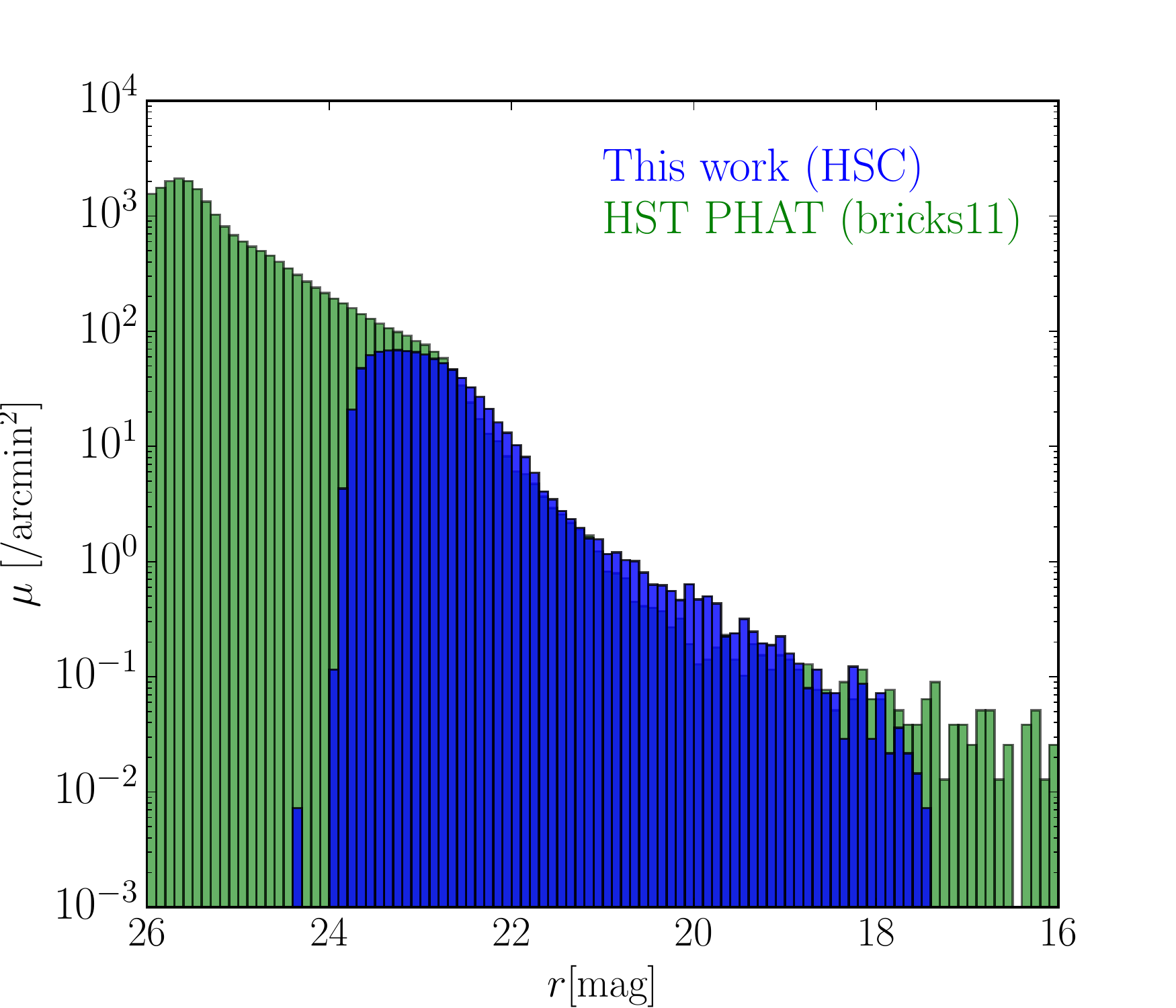}
 \caption{The green histogram shows the luminosity function of M31 stars
 in the HST PHAT catalog, while the blue histogram shows that of the
 peaks in the HSC image. We converted the magnitudes of HST stars to the
 HSC $r$-band magnitudes using Eq.~(\ref{eq:HSTmag_trans}).  The
 comparison is done using the PHAT catalog in the two regions of
 ``bricks07'' (or B7) and ``bricks11'' (B11) in Fig.~1 of
 \cite{Dalcantonetal:12}, which are contained in the patch right next
 to or one-upper to the patch-D2 in the HSC data (see
 Fig.~\ref{fig:m31ccdpatch}). These regions are in a disk region of
 M31. The luminosity function of HSC peaks fairly well reproduces the
 HST result down to $r\sim 23~$mag, but clearly misses fainter stars.
  The PHAT luminosity functions in the two regions appear to be in a
 similar shape.  } \label{fig:HSTHSC}
\end{figure}
Fig.~\ref{fig:HSTHSC} compares the surface density of stars in the HST
PHAT catalog with that of the HSC peaks, as a function of magnitudes, in
the overlapping regions between our M31 data and HST PHAT.  These
regions correspond to ``bricks07'' and ``bricks11''.
The figure clearly shows that the HSC peak counts fairly well reproduces
the HST results down to $r\sim 23~$mag. Since the HSC photometry of each
peak should be contaminated by fluxes of neighboring stars, we would
expect a systematic error in the PSF photometry, which causes a
horizontal shift in the surface density of peaks (the HSC photometry is
expected to over-estimate the magnitude). Even with this contamination,
the agreement looks promising. However, it is clear that the HSC peak
counts clearly misses the fainter stars, which can be potential source
stars for PBH microlensing. The surface density of HST stars in
different regions look similar.

The data overlap between HSC and PHAT covers the disk region only
partially.  Nevertheless, as an estimate of our star counts,
we infer the underlying luminosity function of stars in the disk region
from the HST PHAT catalog based on the number counts of HSC peaks at
$m_r=23~$mag in each patch of the disk regions, assuming that the
luminosity function of HST stars is universal in the disk regions.  For
the halo regions, we use the HSC peak counts.  In this 
estimate of source stars, we find about $8.7\times 10^7$ stars down to
$m_r=26~$mag over the entire region of M31, which is a factor of 14 more
number of stars than that of HSC peaks. However, the source stars
extrapolated from the HST data are faint, and will suffer from lower
detection efficiency.  Therefore, the final constraints do not improve a
lot from these improved star counts.

One might worry about a possible contamination of dust extinction to the
number counts of source stars. However our estimation of the source star
counts is based on the HSC photometry that is already affected by dust
extinction. Hence, we do not think that dust extinction largely affects
the following results.

\section{Discussion}
\label{sec:discussion}

\begin{figure}[htb]
\centering
 \includegraphics[width=9cm,clip]{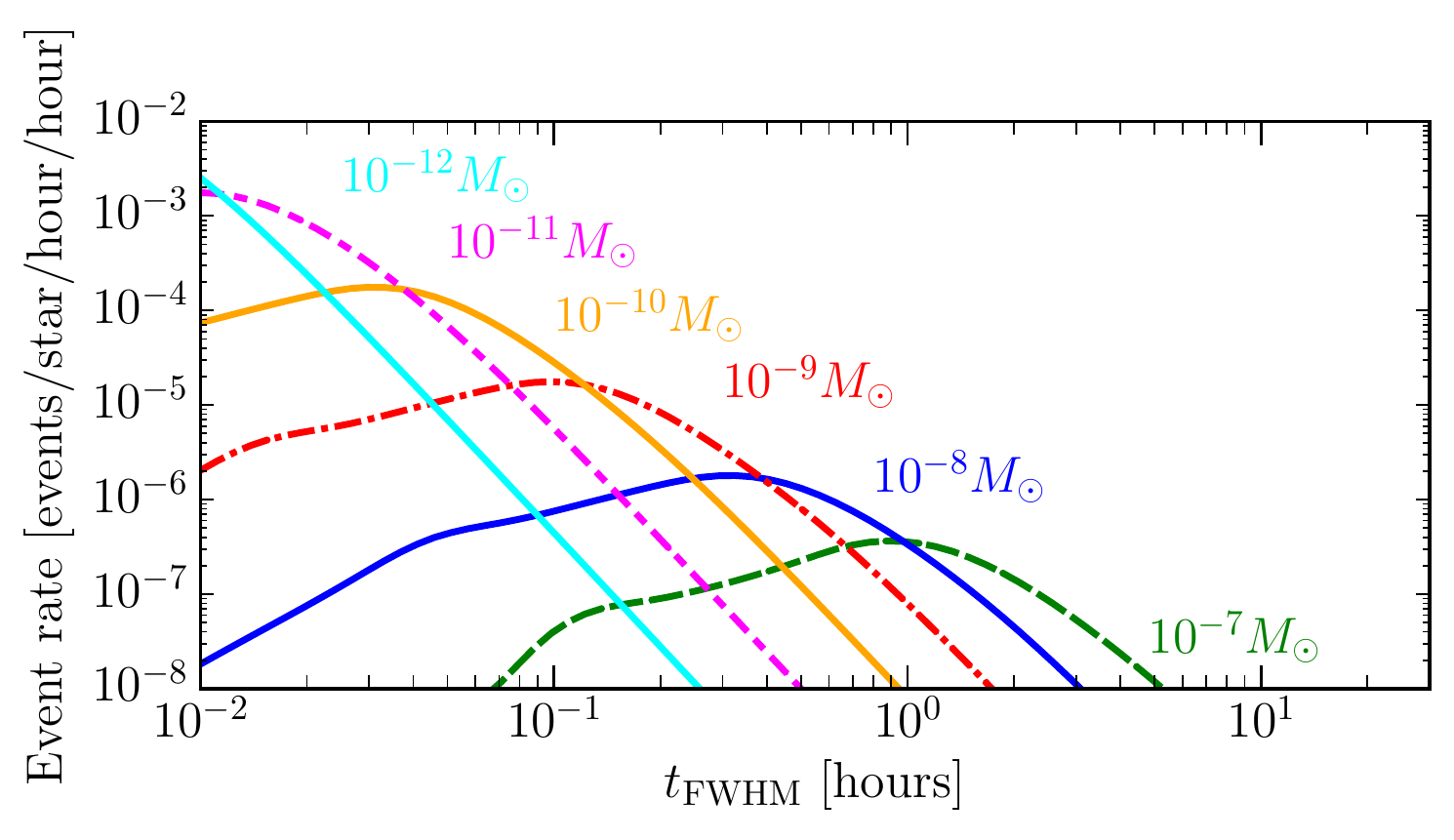}
 \includegraphics[width=9cm,clip]{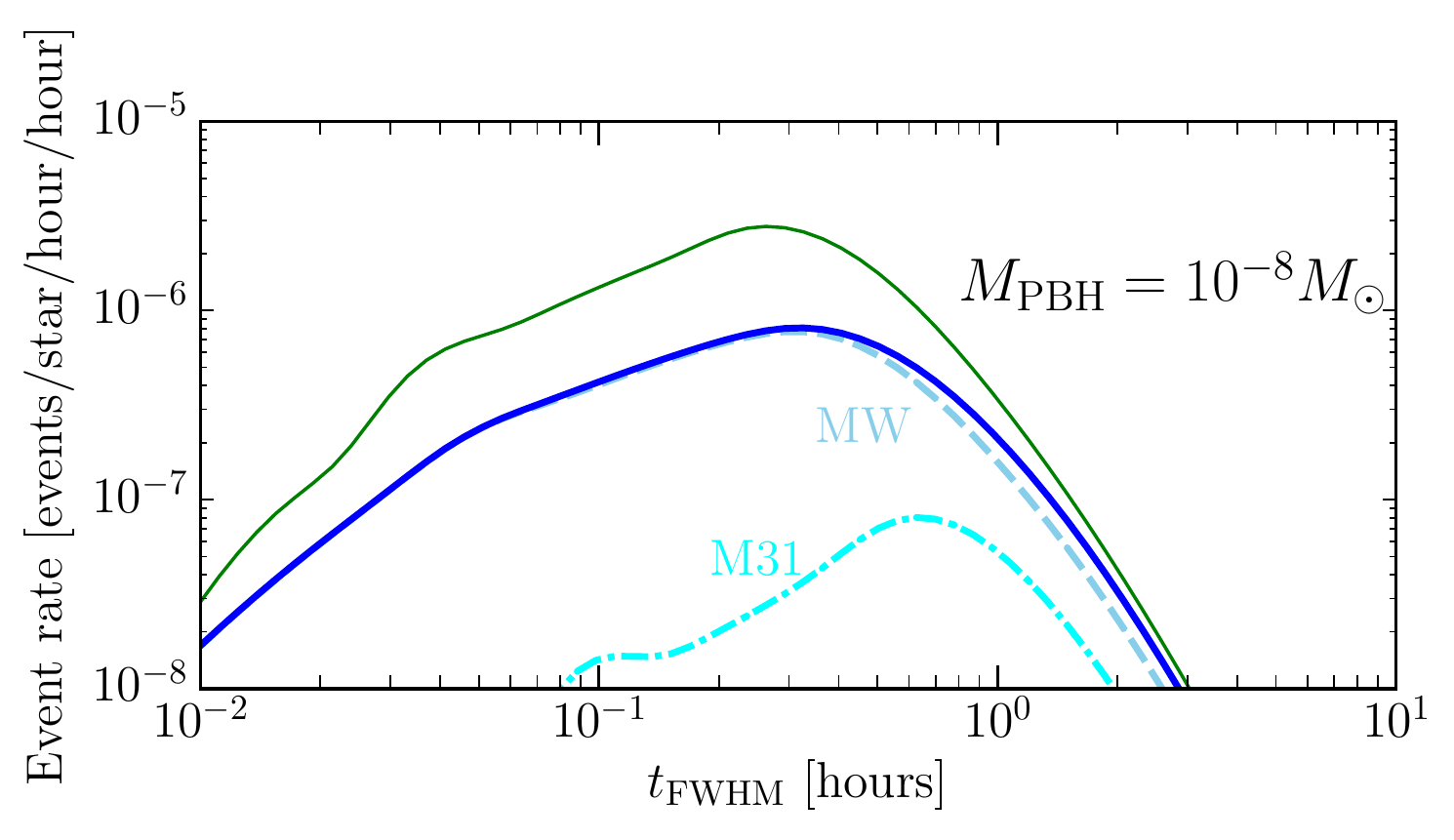}
 \caption{The event rate of PBH microlensing for a single star in M31
 when taking into account the effect of finite source size. Given the
 fact that the HSC data (down to $r\simeq 26$~mag) is sufficiently deep
 to reach main-sequence stars in M31, rather than red-giant branch
 stars, we assume a solar radius for source star size. The finite source
 size effect lowers the event rate compared to
 Fig.~\ref{fig:m31_eventrate}. The lower panel shows the relative
 contribution of PBHs in the MW or M31 halo region to the event rate for
 PBHs with $M_{\rm PBH}=10^{-8}M_\odot$. The upper thin solid curve is
 the result for a point source, the same as in the right panel of
 Fig.~\ref{fig:m31_eventrate}. The microlensing events in M31 are mainly
 from nearby PBHs to a source star at distance within a few tens of kpc
 (see Fig.~\ref{fig:tau}), so the finite source size effect is more
 significant for such PBHs due to their relatively small Einstein radii.
 \label{fig:eventrate_finitesource}}
\end{figure}
\begin{figure}[htb]
\centering
\includegraphics[width=0.95\textwidth]{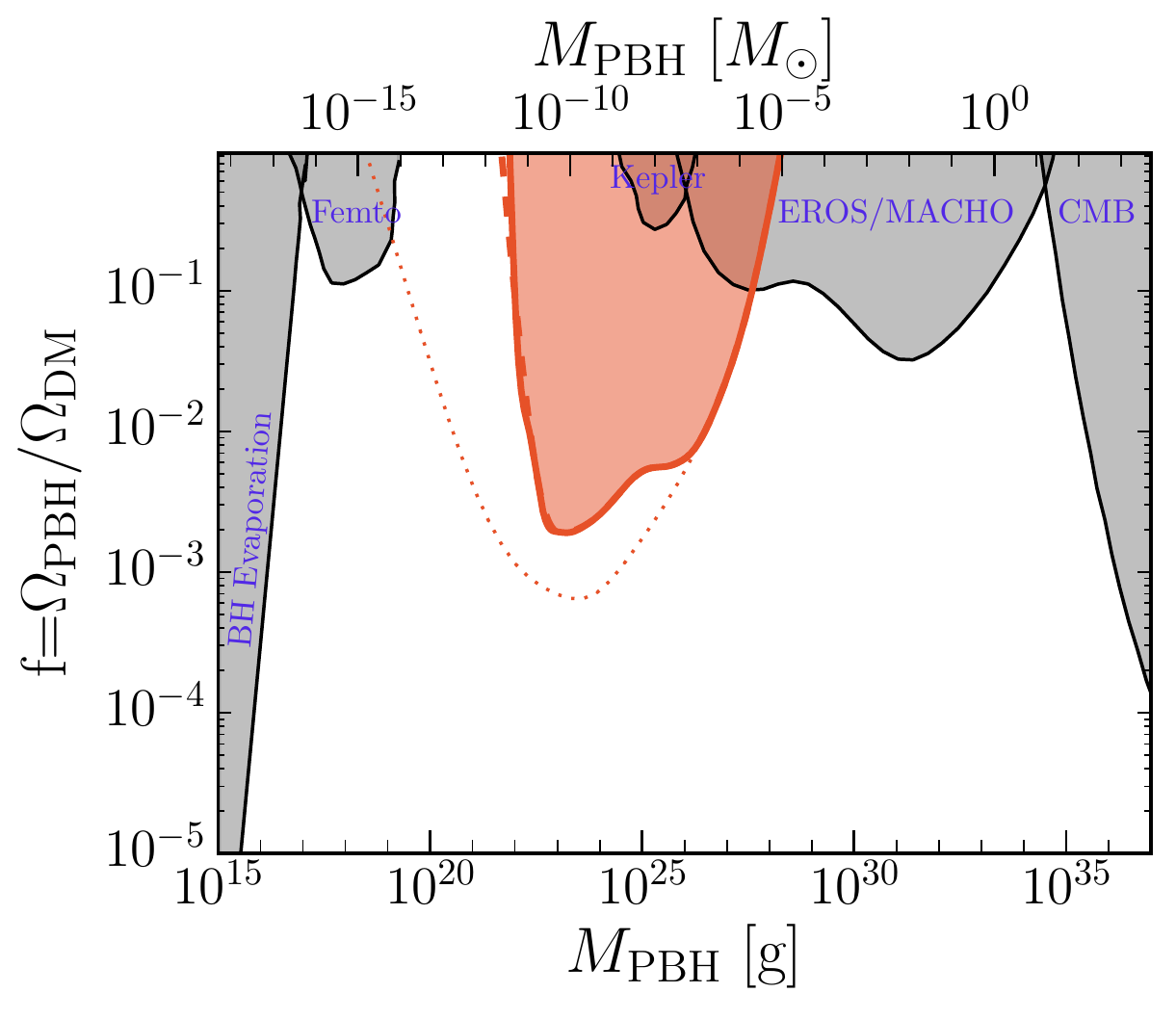}
 \caption{The solid curve shows the 95\% C.L. upper bound when 
ignoring the wave optics effect or equivalently taking 
 into account only the effect of finite source size on the event rate of microlensing, 
 assuming a solar radius for stars in M31. For comparison, the dotted curve shows the result for a point source, i.e. 
 when ignoring both effects of the finite source size and the wave optics, while
 the dashed curve shows
 the results including both the effects, which is our default result shown 
 in Fig.~\ref{fig:upper_bound}.  
 \label{fig:upperbound_finitesourcesize}}
\end{figure}

Although our results for the upper bounds in Fig.~\ref{fig:upper_bound}
are promising, we employed several assumptions. In this section, we
discuss the impacts of our assumptions.

One uncertainty in our bounds comes from the number counts of
source stars in M31, which is a result of blending of stars in the HSC data due
to overcrowding, especially in the disk regions of M31. 
If we use the number of HSC peaks for the counts of source stars, 
$6.4\times 10^6$ 
 instead of $8.7\times 10^7$, 
the counts extrapolated from the HST luminosity function, 
the upper bounds in Fig.~\ref{fig:upper_bound} are weakened by a factor of 10. 
Nevertheless the upper bounds are quite tight, and very meaningful. 
However, we again stress that the use of HSC peak counts is extremely conservative, so 
we believe that our fiducial method using the HST-extrapolated counts of source stars
is reasonable. 

Another uncertainty in our analysis is the effect of finite source size. As can
be found by comparing Eqs.~(\ref{eq:thetaE}) and (\ref{eq:theta_star}), the
angular size of the source star can be greater than the Einstein radius if PBHs
are  close to M31 or if PBHs are in the small mass
range such of $M_{\rm PBH}\simlt 10^{-10}M_\odot$ (assuming solar radius for
the star), all of which result in a smaller Einstein radius. Compared to the
distance modulus for M31 is $\mu\simeq 24.4~$mag, our HSC depth is deep enough
($r\simeq 26~$mag) to reach main sequence stars whose absolute magnitudes
$M_r\simeq 1.5$~mag. According to Figs.~23 and 24 in
\cite{Dalcantonetal:12}, most such faint stars at $r\sim 25$--26~mag would be
either main sequence stars (probably A or F-type stars) or subgiant stars. In
either case such stars have radii similar to the Sun within a factor of 2 or so
\footnote{\url{http://cas.sdss.org/dr4/en/proj/advanced/hr/radius1.asp}}\cite{Northetal:07}.
The shallower data such as the work by \cite{deJongetal:06} probes the
microlensing events only for much brighter stars such as red giant
branch (RGB) stars. RGB stars have much greater radius than that of main
sequence stars, where the finite source size effect is more
significant. Here we employ a solar radius ($R_\odot\simeq 6.96\times
10^{10}$~cm) for all source stars for simplicity, assuming that the
upper bound is mainly from the microlensing for main sequence stars,
rather than for RGB stars \cite{Dalcantonetal:12}.  We followed
\cite{WittMao:94} to re-estimate the event rates of PBHs microlensing
taking into account the finite source size effect.
Fig.~\ref{fig:eventrate_finitesource} shows that the finite source size
effect lowers the event rate, compared to
Fig.~\ref{fig:m31_eventrate}. In particular the effect is greater for
PBHs of smaller mass scales and in the M31 halo region.

For the results in Fig.~\ref{fig:upper_bound}, we also took into account the effect of 
wave optics. Since the Schwarzschild radii for light PBHs with $M\simlt 10^{-10}M_\odot$ become comparable with or smaller than the wavelength of the HSC $r$-band filter, the wave effect lowers the maximum magnification of the microlensing light curve 
\cite{Gould:92,Nakamura:98,TakahashiNakamura:03}. 
The dashed curve in Fig.~\ref{fig:upperbound_finitesourcesize} shows the upper bounds
when ignoring the wave effect or equivalently when including only the finite source size effect. 
As can be found the finite source size effect is a dominant effect compared to the wave effect, and the upper bounds are not largely different in the two cases. This confirms the recent study where the finite source size effect
is more important 
for femtolensing of gamma ray bursts
\cite{2018arXiv180711495K}.
These finite source size and wave optics effects for microlensing searches 
need to be further carefully studied, and this is our future work. 

Theory for PBH formation, via the nature of primordial fluctuations or the nonlinear collapse mechanism, predicts 
that PBHs generally have a mass spectrum, rather than the monochromatic spectrum. 
To compare models with non-monochromatic spectrum, our observed number of
events should be compared to the events predicted using Eq.~\ref{eq:expN}
further integrated over the PBH mass spectrum, i.e.,
\begin{equation}
N_{\rm exp}\!\left(\frac{\Omega_{\rm PBH}}{\Omega_{\rm DM}}\right)  = \frac{\Omega_{\rm PBH}}{\Omega_{\rm DM}}
\int {\rm d}M_{\rm PBH} \int_0^{t_{\rm obs}}\!\!\frac{\mathrm{d}t_{\rm FWHM}}{t_{\rm FWHM}}~ 
 \int\!\!\mathrm{d}m_{r}~
\frac{\mathrm{d}N_{\rm event}}{\mathrm{d}{\ln t}_{\rm FWHM}}
\frac{\mathrm{d}N_{s}}{\mathrm{d}m_{r}}\epsilon({t}_{\rm FWHM}
,m_r)P(M_{\rm PBH}) ,
\label{eq:expN_extended}
\end{equation}
where $P(M_{\rm PBH})$ is a mass spectrum of PBHs, normalized so as to satisfy $\int_0^{\infty}\!\mathrm{d}M_{\rm PBH}~P(M_{\rm PBH})=1$. Then one can use our constraints to constrain the overall PBH mass fraction to DM, $\Omega_{\rm PBH}/\Omega_{\rm DM}$, following the method in Refs.~\cite{Carretal:16,Green:16,Inomataetal:17,Carretal:17}).

\section*{Data availability} The catalog of variability star candidates including the candidates shown in this paper is available from the corresponding author upon reasonable request.


\let\oldthebibliography=\thebibliography
\let\oldendthebibliography=\endthebibliography
\renewenvironment{thebibliography}[1]{%
  \oldthebibliography{#1}%
  \setcounter{enumiv}{47}%
}{\oldendthebibliography}


\end{document}